\documentclass[prc,preprint,showpacs,showkeys,nofootinbib,tightenlines]{revtex4}

\usepackage{graphicx,color}  %
\usepackage{amsmath}

\newcommand{\beq}{\begin{equation}}
\newcommand{\eeq}{\end{equation}}
\newcommand{\bea}{\begin{eqnarray}}
\newcommand{\eea}{\end{eqnarray}}
\newcommand{\ben}{\begin{eqnarray*}}
\newcommand{\een}{\end{eqnarray*}}

\newcommand{\boldtau}{\mbox{\boldmath $\tau$}}
\newcommand{\boldpi}{\mbox{\boldmath $\pi$}}
\newcommand{\boldzeta}{\mbox{\boldmath $\zeta$}}
\newcommand{\boldtheta}{\mbox{\boldmath $\theta$}}
\newcommand{\boldS}{\mbox{\boldmath $S$}}
\newcommand{\boldP}{\mbox{\boldmath $P$}}
\newcommand{\boldV}{\mbox{\boldmath $V$}}
\newcommand{\boldt}{\mbox{\boldmath $t$}}

\newcommand{\boldx}{\mbox{\boldmath $x$}}

\def\lixo#1{}

\newcommand{\slashT}{\slash \hspace{-0.4em}T}
\newcommand{\slashchi}{\slash \hspace{-0.4em}\chi}

\def\slashchar#1{\setbox0=\hbox{$#1$}           
  \dimen0=\wd0                                    
  \setbox1=\hbox{/} \dimen1=\wd1                  
  \ifdim\dimen0>\dimen1                           
    \rlap{\hbox to \dimen0{\hfil/\hfil}}            
    #1                                             
  \else                                          
    \rlap{\hbox to \dimen1{\hfil$#1$\hfil}}        
    /                                           
 \fi}                                           %
\begin{document}


\bigskip


\title{The Effective Chiral Lagrangian
{}From the Theta Term}

\author{E. Mereghetti}\email{emanuele@physics.arizona.edu}
\affiliation{Department of Physics, University of Arizona,
          Tucson, AZ\ 85721}
\author{W.H. Hockings}\email{whockings@bmc.edu}
\affiliation{Department of Mathematics and Natural Sciences, 
Blue Mountain College, Blue Mountain, MS 38610}
\author{U. van Kolck}\email{vankolck@physics.arizona.edu}
\affiliation{Department of Physics, University of Arizona,
          Tucson, AZ\ 85721}

\begin{abstract}
We construct the 
effective chiral Lagrangian involving hadronic and electromagnetic 
interactions originating from the QCD $\bar{\theta}$ term.
We impose vacuum alignment at both quark and hadronic levels,
including field redefinitions to eliminate pion tadpoles.
We show that leading
time-reversal-violating (TV) hadronic interactions are related
to isospin-violating interactions that can in principle be determined
from charge-symmetry-breaking experiments.
We discuss the complications that arise from TV electromagnetic interactions. 
Some implications of the expected sizes of 
various pion-nucleon TV interactions are presented,
and the pion-nucleon form factor is used as an example.
\end{abstract}

\smallskip

\maketitle  

\section{Introduction}
\label{sec:intro}

Time-reversal ($T$) and 
$CP$ violation have been a subject of intense interest for 
nearly half a century.  
The Standard Model (SM) 
with three families has a natural source of $CP$ violation
in the form of
a complex phase in the Cabibbo-Kobayashi-Maskawa (CKM) quark-mixing matrix.
However, this violation is small in the sense
that it comes \cite{Jarlskog} in a combination of CKM parameters 
$J_{CP}\simeq 3\cdot 10^{-5}\ll 1$.
Moreover, this mechanism appears to be insufficient for electroweak 
baryogenesis \cite{genesis}.
As a consequence, it has been hoped that 
the study of $T$ violation
will offer a window into new physics.
In this paper we study systematically the effects on hadronic and
electromagnetic interactions of a source
of $T$ violation yet to be detected, the QCD $\bar{\theta}$ term.

$CP$ violation has been observed in kaon and B-meson systems 
at a level consistent with SM expectations \cite{pdg}. 
On the other hand, electric dipole moments (EDMs)
signal $T$ violation as well, but 
they are relatively insensitive to the CKM phase
because they involve flavor-diagonal $CP$ violation.
Indeed, in the SM with $\bar{\theta}=0$, the neutron EDM, for example, is
expected to be very small, $d_n \sim 10^{-32}\, e$ cm \cite{PRrev}. 
In contrast, the present experimental bound is
$|d_n| < 2.9 \cdot 10^{-26} \, e$ cm \cite{currentbound}, 
and plans exist to decrease this limit by one or
two orders of magnitude using ultracold neutrons at 
SNS \cite{SNS} and ILL+PSI \cite{ILL}.
A less strict bound on the proton EDM,
$|d_p|< 7.9 \cdot 10^{-25} \, e$ cm,
can be extracted from the EDM of the $^{199}$Hg atom  \cite{mercury}
using a calculation of the contribution of the nuclear Schiff moment
\cite{dmitriev}.
In addition,
there exist exciting plans to probe the deuteron EDM in a storage ring
at the level of $|d_d|\sim 10^{-29} \, e$ cm \cite{DeuteronEDM}.
Hadronic and nuclear EDMs are thus sensitive
to non-CKM sources of $T$ violation in the strong interactions.
(For a review of both experimental and 
theoretical issues, see, for example, Refs. \cite{KLbook, PRrev}.) 

A natural question that arises, if the proposed experiments
do measure a non-vanishing hadronic or nuclear EDM, is whether we can identify
the dominant mechanism(s) of $T$ violation.
In the following we would like to present a step in the direction of answering
this question. 
Calculating hadronic and nuclear properties directly from QCD 
has proven difficult
to say the least.  Nevertheless, at low momenta $Q\sim m_\pi \ll M_{QCD}$, 
where $m_\pi$ is the pion mass and $M_{QCD}\sim 1$ GeV is the typical
mass scale in QCD, these properties can be described
in terms of an effective field theory (EFT) involving
nucleons, pions and delta isobars, known
as chiral perturbation theory ($\chi$PT) 
\cite{Weinberg:1978kz,Gasser:1983yg}. 
In the EFT, all interactions are allowed which transform
under Lorentz, parity, time-reversal, and chiral symmetry 
in the same way as do terms in the QCD Lagrangian.
Long-range effects due to the light pions are separated from 
short-range effects due to all higher-energy degrees of freedom.  
Observables are systematically expanded in powers of $Q/M_{QCD}$ 
(times functions of $Q/m_{\pi}$).
$\chi$PT has been successfully
applied to a variety of hadronic and nuclear systems.
(For reviews, see for example Refs. \cite{Wbook, ulfreview, paulo}.)  
We want to use EFT to analyze $T$ violation in a way
similar to what has been done for parity violation \cite{PV}.

We will present here an extension of chiral EFT to
include $T$ violation from the lowest-dimension QCD operator,
the $\bar{\theta}$ term.
The basic idea \cite{Hockings,HvK}
is that $T$ violation is accompanied at the quark level
by a specific form of chiral-symmetry breaking, and thus
the interactions among low-energy hadrons and photons
break chiral symmetry in the same way.
We construct here the $T$-violating Lagrangian
governing the low-energy interactions of pions and nucleons.
(Some of these interactions have already been considered in
Refs. \cite{Hockings,HvK}.)
The extension to delta isobars is straightforward.
We plan in future papers to apply the same method to 
more nucleons and other sources
of $T$ violation. 
Since various sources
of $T$ violation have distinct
chiral-symmetry transformation properties, they
will generate different interactions at the hadronic level \cite{Hockings}.
This, in turn, leads to different relationships among observables.

Since we are interested in low-lying hadronic and nuclear systems,
we limit ourselves to two quark flavors, when the chiral symmetry is 
$SU(2)\times SU(2)$. We extend 
to higher orders in the chiral expansion the pioneering
work of Ref. \cite{CDVW79}.
We do not assume that the strange-quark mass makes
a good expansion parameter. 
With such an assumption more stringent (approximate) relations among
observables exist \cite{su3}.
On the other hand, focusing on $SU(2)\times SU(2)$ will make
some intrinsic aspects of the connection between $T$ violation and
chiral-symmetry breaking more obvious. 

As we are going to see, $T$ violation from the $\bar{\theta}$ term
is intimately connected to isospin violation from the quark mass
difference, which is more specifically charge-symmetry breaking (CSB).
As a consequence, hadronic $T$-violating operators in the chiral
Lagrangian are related in lowest orders to CSB operators
and in higher orders to more general isospin violation,
until the connection becomes ineffective due to unpaired operators.
In the EFT context, isospin-violating operators were first
constructed systematically in Refs. \cite{vanKolck,isoviolphen},
and we follow the same method here.
These interactions have been partially studied in 
nuclear forces \cite{isoviolOPE,isoviolNN,isoviolNNN} 
and reactions \cite{CSBd,CSBa,CSBexp,CSBth}.
Similar work on pion-nucleon scattering can be found in 
Refs. \cite{oldpanzer,panzer}.
We discuss the extent to which this connection can be used
to constrain this specific source of $T$ violation.
This connection is known in specific circumstances 
\cite{CDVW79, su3, Hockings, Savage},
but does not seem to be widely appreciated.

An important limitation we discuss 
comes from electromagnetic interactions between quarks,
which contain both a chiral invariant and CSB. 
However, even when the link to CSB interactions is ineffective,
the chiral Lagrangian suggests a hierarchy of $T$-violating interactions 
stemming from naive dimensional analysis. For example,
while much of the discussion of $T$ violation in nuclear physics
(see, for example, Ref. \cite{nuclear})
employs various forms of non-derivative pion-nucleon couplings
on the same footing \cite{TVpiN}, they appear in the EFT 
at different orders in powers of $Q/M_{QCD}$.
The $T$-violating pion-nucleon form factor,
previously considered in Ref. \cite{TVpiNNFF}, is used here
to illustrate this hierarchy.

In establishing these results, one needs to be mindful that
$T$ violation can lead to the disappearance
of a neutral pion into the vacuum. Such tadpoles reflect vacuum
misalignment. This problem has been solved at the quark level
long ago by Baluni \cite{Baluni}, building on earlier
work \cite{Dashen,Nuyts}.
Here we reformulate it at the EFT level using field redefinitions,
extending results from Ref. \cite{Hockings}.
We discuss not only the leading-order effect considered by Baluni,
but also small tadpoles appearing at higher orders.

This paper is organized as follows. 
In Sec. \ref{secChLag} we review the procedure to construct the most general 
chiral Lagrangian in the presence of chiral symmetry breaking, 
using technology detailed in Apps. \ref{appA} and \ref{appB}.
In Sec. \ref{VA} we discuss the issue of vacuum alignment, reviewing Baluni's 
solution at the quark level 
and presenting its counterpart at the hadronic level. 
(We relegate to Apps. \ref{appC} and \ref{appD} other aspects
of vacuum alignment, including 
how to work with a misaligned Lagrangian by resumming pion tadpoles.) 
In Sec. \ref{thetainteractions} we list the $T$-violating EFT operators 
in the purely hadronic sector of the chiral Lagrangian.
(Some high-order, but nevertheless interesting, 
interactions are presented 
in App. \ref{appE}.
Constraints from Lorentz invariance are discussed in App. \ref{rel}.)
We include the effects of the electromagnetic interaction 
to $T$ violation in Sec. \ref{EMinteractions}. 
($T$-conserving electromagnetic interactions of the same order,
relevant for the link with isospin violation, are shown in 
App. \ref{TconsEM}.)
In Sec. \ref{Discussion} we discuss some of the implications
of these EFT operators, in particular the link
with $T$-conserving operators and the role of residual tadpoles.
In Sec. \ref{FF} we calculate the $T$-violating pion-nucleon form factor
up to second order in our expansion after eliminating tadpoles.
(Modifications needed when tadpoles are kept are shown in 
App. \ref{piNFFwtad}.)
We draw our conclusions in Sec. \ref{Conclusion}.

\section{Framework}
\label{secChLag}

We are interested in possible $T$-violating hadronic processes at 
low energies stemming from the QCD $\bar\theta$ term.
This section, which has as its main goal to set the notation
and framework used in the rest of the paper, introduces
the well-known $\bar\theta$ term and briefly reviews
the method of building the low-energy chiral Lagrangian.

\subsection{$T$ violation from the $\bar\theta$ term}

Well below the electroweak scale, strong interactions
can be described by the most general Lagrangian
with Lorentz, color, and electromagnetic gauge invariance
among left-handed ($q_L$) and right-handed ($q_R$)
quarks, gluons ($G_{\mu}$) and photons ($A_{\mu}$).
The lowest-dimension operators are included in the
QCD Lagrangian,
\begin{eqnarray}
\mathcal L &=&  -\frac{1}{2} {\rm Tr} G^{\mu\nu} G_{\mu\nu}   
-\frac{\theta}{32\pi^2}\varepsilon^{\mu \nu \alpha \beta}
 {\rm Tr} G_{\mu \nu} G_{\alpha \beta}   
-\frac{1}{4} F^{\mu\nu} F_{\mu\nu}  \nonumber\\
&&+ \bar q_L i \slashchar D \,q_L + \bar q_R i \slashchar D \,q_R 
+ e\bar q_L \slashchar A Q \,q_L + e\bar q_R \slashchar A Q \,q_R 
- \bar{q}_{R} M q_{L}  - \bar{q}_{L} M^{*} q_{R}, 
\label{eq:QCDt.1}
\end{eqnarray}
where $G_{\mu \nu}$ and $F_{\mu \nu}$ are the gluon and
photon field strengths, respectively; 
$D_\mu$ is the color-gauge covariant derivative;
$M$ and $Q$ are the quark mass and charge matrices;
$e$ is the electron charge;
and $\theta$ is a real parameter \cite{theta}.

The field $q=q_L+q_R$ represents a multiplet of fields, 
of dimension equal to the number of 
quark flavors we consider. We work for simplicity
with two light flavors, $u$ and $d$, so 
\begin{equation}
 q = \left(\begin{array}{c}
           u \\ d
           \end{array} \right)
\label{eq:q=ud}
\end{equation}
is an isospin doublet. Objects in isospin space
can be written in terms of the identity
and the Pauli matrices $\boldtau$,
for example
\begin{equation}\label{eq:QCDt.0.2}
Q= \frac{1}{3} \left(\begin{array}{cc} 
                     2 & 0 \\
                     0 & -1
                     \end{array}\right)
 = \frac{1}{6}+\frac{\tau_3}{2}.
\end{equation}
The most general form of the diagonal mass matrix is
\begin{equation}
 M = e^{i \rho} \left(\begin{array}{cc} 
                        m_u & 0 \\
                        0 & m_d
                       \end{array}\right)
=e^{i \rho} \bar{m} \left(1- \varepsilon \tau_3\right),
\label{eq:QCDt.2}
\end{equation}
with real parameters $\rho$ and $m_{u,d}$, or alternatively 
\begin{equation}
\bar{m}=\frac{m_u+m_d}{2}
\end{equation}
and
\begin{equation}
\varepsilon=\frac{m_d-m_u}{m_u+m_d}.
\end{equation}
An important role is played
by rotations in isospin space 
belonging to the chiral group $SU_L(2)\times SU_R(2)\sim SO(4)$,
\begin{equation}
q \rightarrow \exp \left[i\boldtheta_V \cdot \boldt 
+ i\boldtheta_A \cdot \boldx  \right] q,
\label{chiraltrans}
\end{equation}
where $\boldtheta_{V,A}$ are real parameters and
\begin{equation}
\boldt=\boldtau/2, \qquad \boldx=\gamma_5 \boldtau/2,
\end{equation}
the group generators.

The $\theta$ term is a total derivative, but it contributes
to physical processes through 
extended, spacetime-dependent field configurations 
known as instantons \cite{acabeca}.
The $\theta$ term can be eliminated from the Lagrangian by performing 
transformations on the quark field. The most general transformation 
that leaves $M$ diagonal is a combination of a chiral transformation
(\ref{chiraltrans}) with $\boldtheta_{V}=(0,0,\beta)$ 
and $\boldtheta_{A}=(0,0,\alpha)$, and two $U(1)$ transformations,
\begin{equation}
q \rightarrow \exp\left[i \theta_V^0 + i \theta_A^0 \gamma_5 \right] q,
\label{abeliantrans}
\end{equation}
with arbitrary parameters $\theta_{V,A}^0$.
The axial $U(1)$ transformation has an anomaly \cite{fujikawa}
and induces a transformation 
in the integration measure in the path integral that is equivalent to 
a modification of the $\theta$ term in the QCD Lagrangian.
With the choice $\theta_A^0=-\theta/4$, the $\theta$ term can be eliminated
and
the QCD Lagrangian can be written as
\begin{equation}
 \mathcal L =   - \frac{1}{2} {\rm Tr} G^{\mu \nu} G_{\mu \nu}   
- \frac{1}{4} F^{\mu \nu} F_{\mu \nu}  \nonumber\\
+ \bar q_L i \slashchar D  q_L 
+ \bar q_R i \slashchar D  q_R 
+ {\mathcal L}_{e}
+ {\mathcal L}_{\alpha},
\label{eq:QCDt.7}
\end{equation}
where 
\begin{equation}
{\mathcal L}_{e} =e A_{\mu}\bar q \gamma^{\mu} Q q
= e A_{\mu} \left(\frac{1}{6} I^\mu + T^\mu_{34} \right)
\label{eq:rot.16}
\end{equation}
is the electromagnetic interaction,
and 
\begin{eqnarray}
{\mathcal L}_{\alpha} &=&- \bar{q}_R e^{i \frac{\bar{\theta}}{2}} 
                 \left(\begin{array}{c c}
                       m_u e^{i\alpha} & 0\\
                       0               & m_d e^{-i\alpha}
                       \end{array}\right)q_L + \textrm{H.c.} 
\nonumber \\ 
&=& -\bar{m} \cos \alpha \cos \frac{\bar{\theta}}{2}
\left\{\left[ 1+\varepsilon \tan \alpha \tan \frac{\bar \theta}{2} \right] S_4 
-\left[\varepsilon +\tan \alpha \tan \frac{\bar \theta}{2}\right] P_3 
\right. \nonumber \\ 
&& \left. \qquad\qquad\qquad\quad
+\left[\varepsilon \tan \alpha - \tan \frac{\bar \theta}{2}\right] P_4
+\left[\tan \alpha - \varepsilon \tan \frac{\bar \theta}{2}\right] S_3
\right\}
\label{eq:QCDt.8}
\end{eqnarray}
is a family of $CP$-violating mass terms labeled by the angle $\alpha$
and parametrized by $\bar \theta = 2 \rho - \theta$.
We have introduced two $SO(4)$ vectors, a Lorentz scalar
\begin{equation}
S = \left(\begin{array}{c} 
            -2i\bar q \gamma^5 \boldt q\\ 
            \bar q q
           \end{array}
           \right)
\label{Sdef}
\end{equation}
and a Lorentz pseudoscalar
\begin{equation}
P = \left(\begin{array}{c} 
            2\bar q \boldt q\\ 
            i\bar q \gamma^5 q
           \end{array}
           \right),
\label{Pdef}
\end{equation}
and two Lorentz vectors, an $SO(4)$ scalar,
\begin{equation}
I^\mu=\bar q \gamma^{\mu} q,
\label{Idef}
\end{equation}
and an $SO(4)$ antisymmetric tensor,
\begin{equation}
T^\mu = \left( \begin{array}{c c} 
  \varepsilon_{ijk}\bar q \gamma^{\mu}\gamma^5 t_k q 
& \bar q \gamma^{\mu} t_i q \\
  - \bar q \gamma^{\mu} t_j q 
& 0 
\end{array}\right).
\label{Tdef}
\end{equation}

\subsection{Chiral Lagrangian}

The low-energy EFT that describes interactions among pions and nucleons 
(and delta isobars, since $m_\Delta -m_N\sim 2m_{\pi}$) 
at low momentum $Q\sim m_\pi\ll M_{QCD}$
is $\chi$PT. 
At such momenta 
we can resolve pion propagation, but not details
of its structure. 
Pions must explicitly be accounted for in the theory, while
other mesons can be integrated out. 
The special role of the pion is a consequence of the
approximate invariance of the QCD Lagrangian under chiral symmetry.
Because it is not manifest in the spectrum, which only exhibits
approximate isospin symmetry,
chiral symmetry must be spontaneously broken down to its 
diagonal subgroup $SU_{L+R}(2)\sim SO(3)$.
{}From Goldstone's theorem, one expects to find in the spectrum
massless Goldstone bosons that live on the ``chiral circle''
$S^3\sim SO(4)/SO(3)$.
There are, of course, infinite ways to parametrize
the chiral circle. Here we use stereographic coordinates, whose 
dimensionless fields we denote by an isovector field $\boldzeta$.
We can identify these degrees of freedom with canonically normalized pion
fields 
$\boldpi= F_\pi \boldzeta$, where
$F_\pi\simeq 186$ MeV, called the pion decay constant, is the diameter
of the chiral circle. 
Such fields transform in a complicated way under chiral symmetry.
However, a pion covariant derivative 
can be defined by 
\begin{equation}
D_\mu \boldpi =D^{-1} \partial_\mu \boldpi,
\label{picovder}
\end{equation}
with 
\begin{equation}
D=1+ \frac{\boldpi^2}{F_\pi^2},
\label{D}
\end{equation}
which transforms under chiral transformations
as under an isospin transformation, but with a field-dependent parameter.
Similarly, we can use a nucleon field $N$ that transforms in an analogous way,
a nucleon covariant derivative 
\begin{equation}
{\mathcal D}_\mu N
=\left(\partial_\mu 
 +\frac{2i}{F_\pi^2} \boldt\cdot \boldpi \times D_\mu\boldpi\right)N, 
\label{nucovder}
\end{equation}
{\it etc}. 
At $Q\sim m_{\pi}\ll m_N$, nucleons 
are essentially 
non-relativistic; as such, the only coordinate with which
their fields vary rapidly is $v\cdot x$, where $v$ is the velocity.
For simplicity, we employ
a heavy-nucleon field from which this fast variation has
been removed \cite{heavybaryon}. 
This simplifies the gamma matrix algebra, since only the spin
$S^\mu$ remains.
(This procedure can be easily generalized to include
a heavy-delta field.)
Details about our choice of fields are found in App. \ref{appA}.

The first step in describing QCD at low energies is
to construct the most general Lagrangian 
that transforms under the symmetries of QCD in the same way as QCD itself.  
Along with this, one needs a power-counting scheme so that interactions 
can be ordered according to the expected size of their contributions.  
The 
Lagrangian contains an infinite number of terms
that we group using an integer ``chiral index'' $\Delta$
and the (even) number of fermion fields $f$:
\begin{equation}
 {\cal L}=\sum_{\Delta=0}^{\infty}\sum_{f/2}
{\cal L}^{(\Delta)}_{f}.
\label{ChiralL}
\end{equation}
Power counting for the case $f/2 \ge 2$ is subtle \cite{nogga},
so here we limit ourselves to $f/2\le 1$.

The technology for constructing the Lagrangian is well known, see, 
for example, Ref. \cite{Wbook}.
When we neglect ${\mathcal L}_{e}$, Eq. (\ref{eq:rot.16}),
and ${\mathcal L}_{\alpha}$, Eq. (\ref{eq:QCDt.8}),
the EFT Lagrangian includes all interactions made out of 
$D_\mu\boldpi$, $N$, and their covariant derivatives
that are chiral invariant.
In this case, $f\le 2$ interactions have index
\begin{equation}
\Delta = d+f/2-2 \ge 0,
\label{Delta}
\end{equation}
in terms of 
the number  $d$ of derivatives 
(and powers of $m_\Delta -m_N$).
For example, in leading order the chiral-invariant Lagrangians are
\begin{equation}
{\mathcal L}^{(0)}_{\chi, f=0}= 
\frac{1}{2} D_{\mu} \boldpi \cdot D^{\mu} \boldpi 
\label{eq:QCD.16prime}
\end{equation}
and (omitting delta isobars)
\begin{equation}
{\mathcal L}^{(0)}_{\chi, f=2}= 
\bar{N}\left( iv\cdot {\mathcal D}
         -\frac{4 g_A}{F_\pi}S^{\mu} \boldt\cdot D_{\mu} \boldpi\right) N,
\label{eq:QCD.16}
\end{equation}
where $g_A\simeq 1.267$ is the pion-nucleon axial-vector coupling. 
Note that at this order the nucleon is static;
kinetic corrections have relative size ${\cal O}(Q/M_{QCD})$
and appear in ${\mathcal L}^{(1)}_{\chi, f=2}$.

On the other hand,
${\mathcal L}_{\alpha}$, Eq. \eqref{eq:QCDt.8},
and ${\mathcal L}_{e}$, Eq. \eqref{eq:rot.16},
break chiral
symmetry: ${\mathcal L}_{\alpha}$ as third and fourth components of 
the vectors (\ref{Sdef}) and (\ref{Pdef}),
and ${\mathcal L}_{e}$ as the third and fourth components of 
the antisymmetric tensor (\ref{Tdef}).
In the EFT they generate interactions, now involving $\boldpi$ directly
and $A_\mu$, 
that transform as these vectors and tensors, and their tensor products.
These terms are proportional to 
powers of $m_{u,d}$ and $e$.
Chiral symmetry breaking with our effective fields is
reviewed in App. \ref{appB}.

In the rest of this paper we discuss the construction
of the $T$-violating interactions stemming from Eq. (\ref{eq:QCDt.8}),
and also how they are power counted.
It is clear that $T$ violation is intimately linked with the quark masses
and the explicit breaking of chiral symmetry, in particular
isospin violation.
Explicit chiral symmetry breaking 
in the form of isospin violation is also present in the electromagnetic 
terms from Eq. (\ref{eq:rot.16}).
They generate two classes of interactions.
In one class, hadrons 
interact with soft photons (those with momenta below $M_{QCD}$)
in a gauge-invariant way.
We can minimally couple charged pions and nucleons 
to the photon by modifying their covariant derivatives,
\begin{eqnarray}\label{eq:minimal.1}
  (D_{\mu} \boldpi)_a & \rightarrow & (D_{\mu,\, \rm{em}} \boldpi)_a = 
\frac{1}{D} \left(\partial_{\mu}\delta_{ab}- eA_{\mu}\varepsilon_{3ab}\right) 
\pi_b 
\nonumber \\
 \mathcal D_{\mu} N   &\rightarrow & \mathcal D_{\mu, \, \rm{em}} N = 
\left[\partial_{\mu}+\frac{2i}{F_\pi^2}\boldt \cdot
\left(\boldpi\times D_{\mu, \rm{em}}\boldpi \right) 
- i e A_{\mu} \left(\frac{1}{2} + t_3 \right)\right]N. 
\end{eqnarray}
(For simplicity, in the text that follows
we drop the subscript ``${\rm{em}}$'' in covariant derivatives.) 
In addition, we can couple the photon through the field strength $F_{\mu\nu}$.
The other class of interactions consists of
purely hadronic
interactions from the exchange of hard photons 
(momenta above $M_{QCD}$),
which can be integrated out, giving rise 
to operators with no explicit photon fields.
The first class of interactions is very important because of EDMs;
the second class competes with interactions from Eq. (\ref{eq:QCDt.8}).
We thus construct the low-energy interactions from Eq. (\ref{eq:rot.16}) 
as well.

The index $\Delta$ defined in Eq. \eqref{Delta} can be generalized to label 
electromagnetic operators. If the operator contains soft photons, the 
definition of $d$ is enlarged to count also the number of photon fields, 
which, having dimension one, require compensating powers of $M_{\rm{QCD}}$ 
in their coefficients. Operators generated by the integration of hard photons 
are proportional to powers of $e^2$.
Typically, an extra inverse power of $4\pi^2$ appears in a loop, leading 
to a factor of $\alpha_{\rm{em}}/\pi$. 
Since the numerical value of $\alpha_{\rm{em}}/\pi$ is very close to 
$\varepsilon \, m^3_{\pi}/M^3_{QCD}$ (using $M_{QCD} \sim m_{\rho}$, 
the mass of the rho meson), we can still use $\Delta$ to label this class of 
operators, provided that each power of $\alpha_{\rm{em}}/\pi$ increases the 
chiral index by $3$ \cite{vanKolck}.

\section{Vacuum Alignment}\label{VA}

The $S_3$ term in Eq. (\ref{eq:QCDt.8}) is actually unphysical because it 
gives rise to terms in the low-energy effective Lagrangian that make the 
vacuum unstable under small fluctuations.  
At leading order a term would arise that is linear in the 
pion fields ({\it i.e.} $\pi_{3}$).  The vacuum would then be unstable because 
it could always produce mesons to lower its energy.
The problem of such leading-order tadpoles is discussed below
in Subsec. \ref{need}.

There are two approaches to removing these spurious terms. 
One approach \cite{Dashen} is to impose, at quark level, 
the condition that $T$-violating interactions should not cause 
vacuum instability.
This has been done \cite{Baluni} to first-order in symmetry-breaking
interactions, and we review this argument in Subsec. \ref{nobalony}.
Then we derive in Sec. \ref{thetainteractions}
the corresponding low-energy EFT, which will not contain terms 
which cause vacuum instability.

The other approach is to derive the low-energy EFT without putting 
any conditions on how the resulting interactions affect the vacuum, 
then employ field redefinitions on the fields at the hadronic level 
to eliminate terms that affect the stability of the vacuum.  
It is the second approach that we follow in Subsec. \ref{noparma}.
For most of this section we neglect ${\mathcal L}_{e}$,
Eq. (\ref{eq:rot.16}),
but we consider the field redefinitions in the presence of 
electromagnetism in Subsec. \ref{nosalami}. 
Such an alternative procedure will help to understand what kind of interactions
can be removed from the chiral Lagrangian.
As we will see in
Secs. \ref{thetainteractions} and \ref{EMinteractions}, 
further pion tadpoles appear in subleading orders but pose no problems.
They can also be eliminated with a field redefinition of the form
discussed in this section.

\subsection{The need for vacuum alignment}
\label{need}

To illustrate the importance of vacuum alignment for the construction of 
the EFT, let us suppose we do not align the vacuum at the quark
level.
One can construct the low-energy interactions induced by Eq. (\ref{eq:QCDt.8})
following the method sketched in App. \ref{appB}. 
Among the various types of terms are the ones that are linear
in the symmetry-breaking parameters. These are the infinitely
many operators that
transform as third and fourth components of $S$ and $P$ type vectors:
\begin{equation}
\mathcal L_\alpha =  \sum_n \,  \left\{
C_{4n} S_{4 n}[\boldpi,N] + C_{3n} S_{3 n}[\boldpi,N]\right\}
+ \sum_n \,  \left\{D_{3n} P_{3 n}[\boldpi,N] 
+ D_{4n} P_{4 n}[\boldpi,N] \right\} +\ldots, 
\label{eq:align.3}
\end{equation}
where $n$ runs over all $S[\boldpi,N]$ and $P[\boldpi,N]$ 
that can be obtained using Eq. \eqref{eq:ex.7},
and ``$\ldots$'' stand for higher-rank tensors.
The coefficient $C_{\alpha n}$ or $D_{\alpha n}$
of each term depends on details of the QCD dynamics,
and cannot at present be determined. However,
chiral symmetry fixes the ratio of coefficients of components
of the same object, which is given by
Eq. (\ref{eq:QCDt.8}). 
Thus 
\begin{equation}
 \begin{split}
\frac{C_{3n}}{C_{4n}} & = 
\frac{\tan \alpha- \varepsilon \tan \frac{\bar \theta}{2}}
      {1+\varepsilon \tan \alpha \tan \frac{\bar \theta}{2}}
=  \tan \left[\alpha - \arctan\left(\varepsilon\tan \frac{\bar\theta}{2}\right)
\right],\\
\frac{D_{4n}}{D_{3n}} & = 
-\frac{\varepsilon \tan \alpha - \tan \frac{\bar \theta}{2}}
      {\varepsilon + \tan \alpha \tan \frac{\bar \theta}{2}} 
= - \tan\left[\alpha - \arctan\left(\frac{1}{\varepsilon} 
              \tan\frac{\bar\theta}{2}\right)\right].
  \end{split}
\label{eq:align.4}
\end{equation}

The simplest symmetry breaking operator comes
from $S[0,N] = ({\mathbf 0} \;\, v_0)^T$,
in which case a piece in Eq. (\ref{eq:align.3}) is
\begin{equation}
\mathcal L_{m_\pi^2} =  \frac{{\tilde m}_\pi^2 F_\pi^2}{4}
-\frac{{\tilde m}_\pi^2}{2D}\boldpi^2
+\frac{g{\tilde m}_\pi^2 F_{\pi}}{2D}\pi_3,
\label{eq:rot.20}
\end{equation}
where the bare pion mass is
\begin{equation}
 \tilde m^2_{\pi} = \frac{4v_0}{F^2_{\pi}}{\bar m} 
   \cos\alpha \cos\frac{\bar\theta}{2} 
   \left[1 + \varepsilon \tan\alpha \tan\frac{\bar\theta}{2}\right]
\label{eq:mass}
\end{equation}
and the coupling of the neutral pion to the vacuum is
\begin{equation}
g =
\tan \left[\alpha - \arctan\left(\varepsilon\tan \frac{\bar\theta}{2}\right)
\right].
\label{eq:g}
\end{equation}

\begin{figure}
\centering
\includegraphics[width=2.5cm]{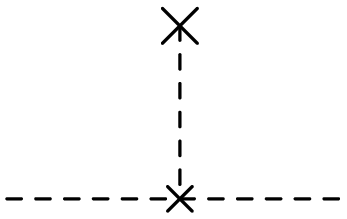} \hspace{0.5cm}
\includegraphics[width=2.5cm]{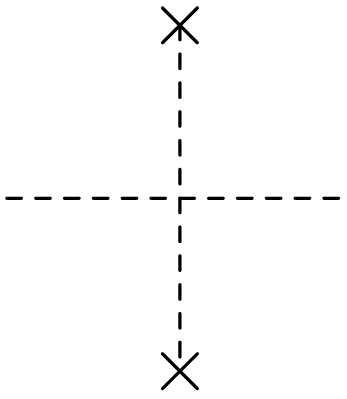} 
 
\vspace{0.5cm}
\includegraphics[width=2.5cm]{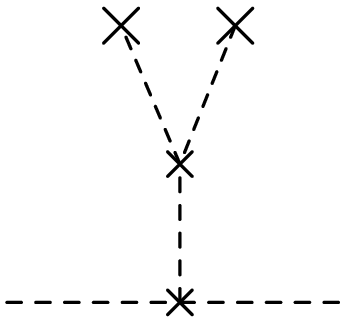} \hspace{0.5cm}
\includegraphics[width=2.5cm]{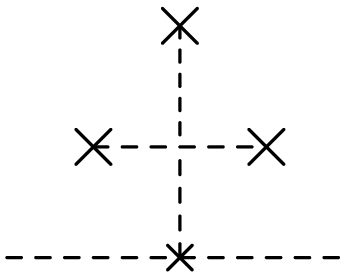}\hspace{0.5cm}
\includegraphics[width=2.5cm]{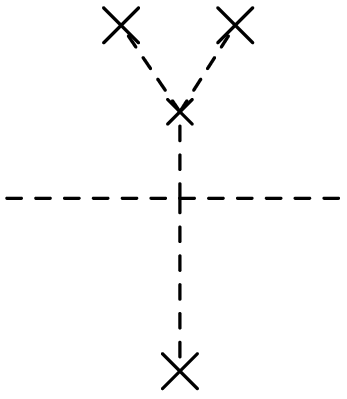} \hspace{0.5cm}
\includegraphics[width=2.5cm]{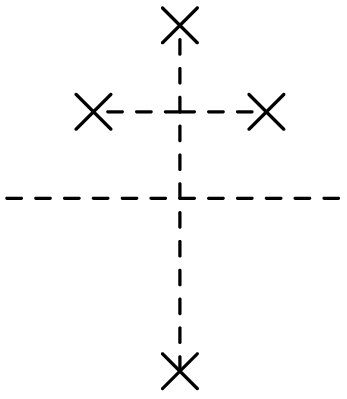} 

\vspace{0.5cm}
\includegraphics[width=2.5cm]{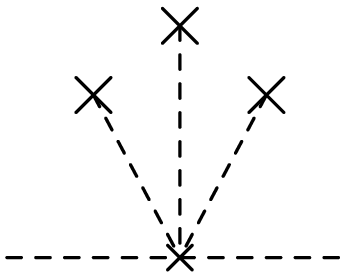} \hspace{0.5cm}
\includegraphics[width=2.5cm]{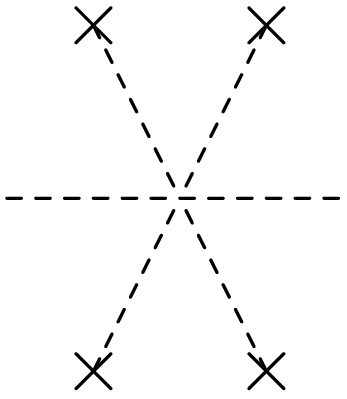} \hspace{0.5cm}
 
\vspace{0.5cm}
\includegraphics[width=3.75cm]{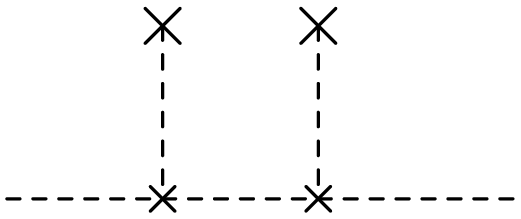}\hspace{0.5cm}
\includegraphics[width=3.75cm]{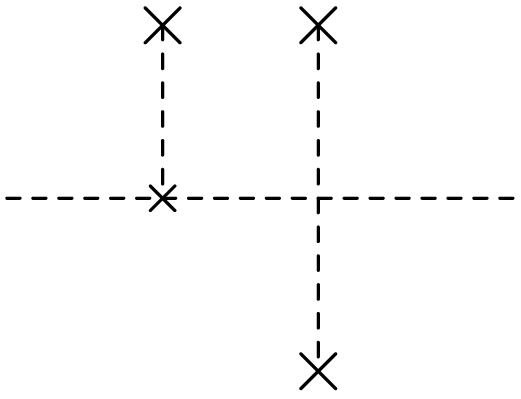} 
\includegraphics[width=3.75cm]{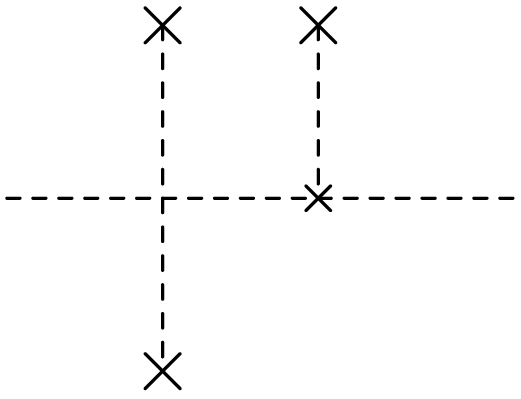} 
\includegraphics[width=3.75cm]{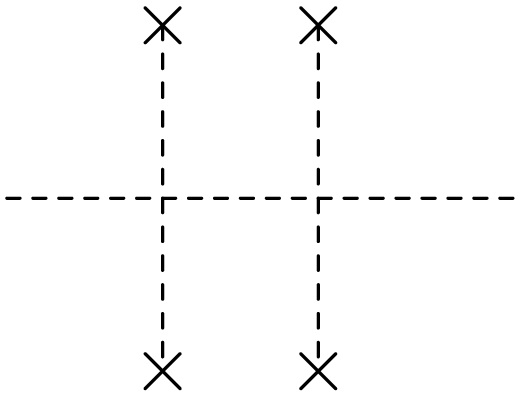} 
\caption{Contributions of order $g^2$ (first line) and $g^4$ 
(next three lines) to the pion two-point Green's function. 
A dashed line stands for a pion propagator, Eq. \eqref{prop1}.
A cross denotes a vertex coming from the
third term in Eq. \eqref{eq:rot.20}.
Other vertices arise from Eq. \eqref{eq:QCD.16prime} and 
the second term in Eq. \eqref{eq:rot.20}.} \label{Fig.2a}
\end{figure}

The first term in Eq. \eqref{eq:rot.20} is a constant that is irrelevant 
for our purposes.
The second term is a mass term, which together with the pion kinetic term
in Eq. \eqref{eq:QCD.16prime} generates a pion propagator 
of conventional form,
\begin{equation}
\frac{i\delta_{ab}}{p^2 - \tilde m^2_{\pi} + i \varepsilon},
\label{prop1}
\end{equation}
when the pion momentum is $p$.
Due to the non-linear realization of chiral symmetry, both
this term and the pion kinetic term
in Eq. \eqref{eq:QCD.16prime} generate also pion self-interactions.
The third term in Eq. \eqref{eq:rot.20}
is $T$-violating and allows neutral pions to disappear 
into the vacuum. It generates both tadpoles and
interactions among an odd number of pions. 
Together, these two effects change pion propagation,
since the full pion propagator includes now an arbitrary number of $\pi_3$s
that disappear into vacuum. Examples are illustrated in Fig. \ref{Fig.2a}, 
where we draw all the diagrams that contribute to the pion propagator up 
to order $g^4$.

The physical pion mass $m^2_{\pi}=m^2_{\pi}(\tilde{m}^2_{\pi},g)$ is given 
by the pole of the two-point Green's function. 
The difficulty is that the contributions of all the diagrams in 
Fig. \ref{Fig.2a} to the two-point 
Green's function are comparable to the one of the propagator.
Indeed, the first two diagrams in Fig. \ref{Fig.2a} give a contribution 
of order $g^2\tilde m^2_{\pi}/(p^2 - \tilde m^2_{\pi})^2$,
while the other diagrams in Fig. \ref{Fig.2a} scale as
$g^4\tilde m^2_{\pi}/(p^2 - \tilde m^2_{\pi})^2$ or 
$g^4\tilde m^4_{\pi}/(p^2 - \tilde m^2_{\pi})^3$,
where we take
$p \sim \tilde m_{\pi}$. 
These translate into contributions of relative order
$g^2$ and $g^4$, respectively, to the pion mass.
Since $g$ depends on $\alpha$ and is {\it a priori} not small, 
these diagrams have the same power counting as 
the propagator (\ref{prop1}):
to calculate the two-point Green's function at tree level we need to sum 
all the 
diagrams of the type in Fig. \ref{Fig.2a} with an arbitrary number of tadpoles.
That is, the pion two-point function 
in the presence of explicit chiral symmetry breaking in the form 
\eqref{eq:align.3} cannot be calculated in perturbation theory. 

The example of the pion 
mass can be extended to other 
observables, for example pion-pion or pion-nucleon scattering cross sections: 
at any order in 
$Q/M_{QCD}$, an infinite number of
diagrams in which zero-momentum neutral pions disappear into the vacuum 
contribute to the physical process. 
When explicit and spontaneous symmetry breaking 
are badly misaligned, explicit symmetry breaking is not just a perturbation.
In App. \ref{appC} we show this in a simple 
example.

The resummation of pion tadpoles can be performed explicitly in diagrams.
We show in App. \ref{appD} how to do so
in the case of the pion two-point Green's function at tree level.
Although calculations can be carried out with arbitrary $\alpha$,
it is unpractical to do so
for all quantities and at every order.
We are thus led to impose at least approximate vacuum alignment.

\subsection{Alignment at quark level}
\label{nobalony}

Explicit symmetry-breaking terms provide a 
preferred direction for spontaneous symmetry breaking \cite{Dashen}. 
The construction of the effective Lagrangian only relies on the fact that 
the symmetry group is broken to one of its subgroups, for example,
$SO(4)$ broken to $SO(3)$.
However, in the absence of explicit symmetry-breaking terms, 
there is no way to say which particular subgroup it is broken to. 
We choose the $SO(3)$ subgroup of rotations in the three-dimensional 
space orthogonal to the vector $n = ({\mathbf 0} \;\, 1)^T$, 
but any other choice of $n$ would 
be equivalent. 
Explicit symmetry-breaking terms force a particular choice of vacuum, 
``aligned'' with the breaking terms. 

Here we consider alignment in first order in chiral-symmetry-breaking 
parameters, as originally done by Baluni \cite{Baluni}.
The chiral-symmetry-breaking Lagrangian \eqref{eq:QCDt.8} generates
at tree level an effective potential
\begin{equation}
V_1=c_4 S_4 + d_3 P_3 + d_4 P_4 + c_3 S_3. 
\end{equation}
The vacuum alignment condition (see Eq. \eqref{align}) is
\begin{equation}
\begin{split}
& \sum_{\alpha = 1}^4 \left(\mathcal T^a \bar S \right)_\alpha
  \frac{\partial V_1}{\partial S_\alpha} 
+ \sum_{\alpha = 1}^4 \left(\mathcal T^a \bar P \right)_\alpha
  \frac{\partial V_1}{\partial P_\alpha}  = 0, \\
& \sum_{\alpha = 1}^4 \left(\mathcal X^a \bar S \right)_\alpha 
  \frac{\partial V_1}{\partial S_\alpha} 
+ \sum_{\alpha = 1}^4 \left(\mathcal X^a \bar P \right)_\alpha 
  \frac{\partial V_1}{\partial P_\alpha}  = 0,
\end{split}
\label{eq:ex.2}
\end{equation}
where the bar 
means we are considering the vacuum expectation value,
and $\mathcal T^{a}$ and $\mathcal X^a$  are the $SO(4)$ generators.
Using the explicit expression \eqref{defgen} of the generators,
the vacuum alignment condition \eqref{eq:ex.2} reads
\begin{equation}
 \begin{split}
d_3 {\bar P}_1 + c_3 {\bar S}_1 & = 0,\qquad d_3 {\bar P}_2 + c_3 {\bar S}_2 
  = 0, \\
d_4 {\bar P}_1 + c_4 {\bar S}_1 & = 0,\qquad d_4 {\bar P}_2 + c_4 {\bar S}_2  
  = 0,
\end{split}
\label{eq:QCDt.10prime}
\end{equation}
and
\begin{equation}
c_4 {\bar S}_3 + d_4 {\bar P}_3  - d_3 {\bar P}_4 - c_3 {\bar S}_4 = 0.
\label{eq:QCDt.10}
\end{equation}
Assuming that the vacuum does not break isospin \cite{vf1}
and parity \cite{vf2},
\begin{equation}
\bar S = \left(\begin{array}{c} 
           \mathbf{0}\\ 
           v
           \end{array}
           \right),
\qquad
\bar P = \left(\begin{array}{c} 
           \mathbf{0}\\ 
           0
           \end{array}
           \right),
\end{equation}
with $v\ne 0$ a real number, which we can choose to be positive.
Plugging in this guess for the vacuum, Eq. \eqref{eq:QCDt.10} becomes
\begin{equation}
  c_3 v = 0,
\label{eq:QCDt.11}
\end{equation}
which is satisfied only if the coefficient of the third component 
of the $S$ vector in Eq. \eqref{eq:QCDt.8} 
vanishes, $c_3 = 0$.
We can rephrase this result by saying that $T$-violating terms can be 
implemented as small perturbations in the usual chiral Lagrangian if 
the freedom to choose the parameter $\alpha$ in Eq. \eqref{eq:QCDt.8} 
is used to make the $T$-violating interaction
an isospin singlet \cite{Dashen,Nuyts,Baluni}.
Explicitly, the condition $c_3 = 0$ is
\begin{equation}
\tan \alpha  = \varepsilon \tan \frac{\bar \theta}{2}.
\label{eq:QCDt.13}
\end{equation}
This choice automatically kills all coefficients
$C_{3n}$ (see Eq. \eqref{eq:align.4}), and in particular the strength $g$ 
of the pion tadpole (see Eq. \eqref{eq:g}).

Substituting Eq. \eqref{eq:QCDt.13} into Eq. \eqref{eq:QCDt.8},
we obtain
\begin{eqnarray}
{\mathcal L}_{m} =
-\bar{m}\, r(\bar \theta)\, S_4
+\varepsilon \bar{m}\, r^{-1}(\bar \theta) \, P_3 
+ m_{\star}\, \sin \bar\theta \, r^{-1}(\bar \theta) \, P_4,
\label{eq:QCDt.8f}
\end{eqnarray}
where we introduced the standard parameter
\begin{equation}
m_{\star} = \frac{m_u m_d}{m_u + m_d}
=\frac{\bar m}{2} \left(1-\varepsilon^2\right)
\end{equation}
and the function 
\begin{equation}
r(\bar \theta) = \left(\frac{1+\varepsilon^2 \tan^2\frac{\bar \theta}{2}}
                            {1+\tan^2\frac{\bar \theta}{2}}\right)^{1/2}.
\end{equation}

The last term in Eq. \eqref{eq:QCDt.8f} is $T$-violating.
As it is well known, this source of $T$ violation
is small for $\bar{\theta}$ near $0$
or near $\pi$. 
If $|\bar\theta|\ll 1$, then $r(\bar \theta)=1+{\mathcal O}(\bar{\theta}^2)$
and \cite{Baluni,CDVW79}
\begin{equation}
 \mathcal L_m =  
-\bar m \,S_4
+ \varepsilon \bar m \, P_3 
+  m_{\star}\bar{\theta}\, P_4
+ {\mathcal O}\left(\bar{\theta}^2\right).
\label{eq:QCDt.15}
\end{equation}
On the other hand, for $|\bar\theta-\pi|\ll 1$, 
$r(\bar \theta)= |\varepsilon| +{\mathcal O}((\bar{\theta}-\pi)^2)$ 
and \cite{CDVW79}
\begin{equation}
 \mathcal L_m =  
-\bar m |\varepsilon| \,S_4
+ \frac{\varepsilon}{|\varepsilon|}\bar m \, P_3 
+  \frac{m_{\star}}{|\varepsilon|} \left(\pi- \bar{\theta}\right) \, P_4
+ {\mathcal O}\left((\pi- \bar{\theta})^2\right).
\label{eq:QCDt.15prime}
\end{equation}

\subsection{Alignment at hadronic level}
\label{noparma}

In the previous section we exploited the freedom in the choice of the 
parameter $\alpha$ in Eq. \eqref{eq:QCDt.8} to write the $T$-violating term 
in the QCD Lagrangian in a way compatible with the usual choice of the vacuum, 
which respects parity and isospin symmetry.   
In this section we follow a different approach:
we start from the EFT Lagrangian \eqref{eq:align.3}
that reflects Eq. \eqref{eq:QCDt.8} before alignment,
and we look for a rotation within the EFT 
that enforces the vacuum alignment condition \eqref{eq:QCDt.11}. 

We define a new field $\boldzeta'$ for the pion through
\begin{equation}
\zeta_i = \frac{1}{d'} 
\left\{\zeta_i'
-\delta_{i3}\left[2C\zeta_3'+S\left(1-\boldzeta'^2\right)\right]\right\},
\label{eq:rot.1}
\end{equation}
where
\begin{equation}
d'= 1-C\left(1-\boldzeta'^2\right)+2S \zeta_3',
\end{equation}
and 
\begin{equation}
C = \frac{1}{2} (1 - \cos \varphi), \qquad S = \frac{1}{2} \sin \varphi,
\end{equation}
in terms of an angle $\varphi$.
Although this transformation is complicated,
the pion covariant derivative simply rotates,
\begin{equation}
D_{\mu} \pi_{i} = \sum_j O'_{ij} D^{\prime}_{\mu} \pi^{\prime}_j 
\label{eq:rot.6}
\end{equation}
with a matrix
\begin{equation}
O'_{ij}=\delta_{ij} - \frac{2}{d'} 
\left\{C\left[\left(\boldzeta^{\prime 2}-\zeta^{\prime 2}_3\right)\delta_{ij} 
   -\varepsilon_{3ik}\zeta^{\prime}_k \varepsilon_{3jl}\zeta^{\prime}_l \right]
       +\left(C\zeta^{\prime}_3+S\right)
        \left(\zeta^{\prime}_i\delta_{3j}-\zeta^{\prime}_j\delta_{3i}\right)
\right\}
\label{eq:rot.6prime}
\end{equation}
that is orthogonal,
\begin{equation}
\sum_l O'_{i l} O'_{j l} = \delta_{i j}.
\label{eq:rot.5}
\end{equation}

Analogously, we define a new field $N'$ for the nucleon via
\begin{equation}
 N = U' N^{\prime},
\label{eq:rot.2}
\end{equation}
with a matrix 
\begin{equation}
U'=\frac{1}{\sqrt{d'}} 
 \left[\sqrt{1-C} + \sqrt{C} \left(\zeta^{\prime}_3 
       + 2i \varepsilon_{3 j k}\zeta^{\prime}_j t_{k}\right)\right]
\label{eq:rot.2prime}
\end{equation}
that is unitary,
\begin{equation}
U'^{\dagger} U'   =   1 .
\label{eq:rot.5prime}
\end{equation}
One can show that the covariant derivative of the nucleon is indeed 
covariant under 
this field redefinition,
\begin{equation}
 \mathcal D_{\mu} N = U' \mathcal D^{\prime}_{\mu}N^{\prime}.
\label{eq:rot.7}
\end{equation} 
As a consequence, 
nucleon bilinears change under this field redefinition as under isospin;
for example,
\begin{equation}
\begin{split}
\bar N N  & = \bar N^{\prime} N^{\prime},\\
\bar N t_i N & = \sum_j O'_{ij} \bar N^{\prime} t_{j} N^{\prime}.
\end{split}
\label{eq:rot.4}
\end{equation}

More generally then,
a generic pionless isoscalar and isovector operator, constructed with nucleon 
fields, their covariant derivatives and covariant derivatives of the pion, 
transforms under \eqref{eq:rot.2} like
\begin{equation}\label{eq:rot.9}
\begin{split}
V_4 [0, N]& = V^{\prime}_4[0, N'],\\
V_i [0, N]&= \sum_j O'_{ij} V^{\prime}_{j}[0, N'].
\end{split}
\end{equation}
The chiral-invariant part of the Lagrangian, for example
Eqs. \eqref{eq:QCD.16prime} and \eqref{eq:QCD.16}, 
is built out of isoscalar combinations of chiral-covariant objects.
The properties \eqref{eq:rot.5} and \eqref{eq:rot.5prime}
thus ensure that the  chiral-invariant Lagrangian  
is invariant under the field redefinitions 
\eqref{eq:rot.1} and \eqref{eq:rot.2}.

This is not true of the chiral-variant interactions \eqref{eq:align.3}.
After the redefinitions \eqref{eq:rot.1} and \eqref{eq:rot.2},
\begin{eqnarray}
\mathcal L_\alpha 
&=&\sum_n \,  \left\{
C^{\,\prime}_{4n} S_{4 n}[\boldpi',N'] 
+ C^{\,\prime}_{3n} S_{3 n}[\boldpi',N']
\right\}
+ \sum_n \,  \left\{D^{\,\prime}_{3n} P_{3 n}[\boldpi',N'] 
+ D^{\,\prime}_{4n} P_{4 n}[\boldpi',N'] \right\} 
\nonumber\\
&&
+\ldots, 
\label{eq:align.3rotated}
\end{eqnarray}
with 
\begin{equation}
 \begin{split}
C^{\,\prime}_{3n} & = \left( 1-2C \right) C_{3n} + 2S C_{4n}, 
\qquad \; C^{\,\prime}_{4n}  = \left( 1-2C \right) C_{4n} -2S C_{3n},
\\
D^{\,\prime}_{3n}  & = \left( 1-2C \right) D_{3n} +2S D_{4n},
\qquad D^{\,\prime}_{4n} = \left( 1-2C \right) D_{4n} -2S D_{3n}.
\end{split}
\label{eq:align.3rotatedprime}
\end{equation}
{\it All} $S_{3n}$ can be eliminated from the Lagrangian
by choosing
\begin{equation}\label{eq:rot.12twoprimes}
\tan \varphi  = \frac{2S}{1-2C} = - \frac{C_{3n}}{C_{4n}}
= -\tan 
 \left[\alpha -\arctan\left(\varepsilon\tan \frac{\bar\theta}{2}\right)\right],
\end{equation}
that is, by
\begin{equation}\label{eq:rot.12}
 \tan\left(\varphi + \alpha \right) = \varepsilon \tan\frac{\bar \theta}{2}.
\end{equation}
In this case,
\begin{equation}
\begin{split}
C^{\,\prime}_{3n} &= 0, 
\qquad\qquad\qquad C^{\,\prime}_{4n} = -\bar{m} \, r(\bar \theta),
\\
D^{\,\prime}_{3n} &= \varepsilon \, \bar{m} \, r^{-1}(\bar \theta),
\qquad
D^{\,\prime}_{4n} = m_{\star} \, \sin \bar\theta \, r^{-1}(\bar \theta),
\end{split}
\label{eq:rot.12prime}
\end{equation}
just as it results from Eq. \eqref{eq:QCDt.8f}.

Equation \eqref{eq:rot.12}
is the counterpart of Eq. \eqref{eq:QCDt.13}, which was found 
by imposing the vacuum alignment condition at the level of the QCD Lagrangian. 
What the field redefinitions \eqref{eq:rot.1} and \eqref{eq:rot.2} do is 
to realize in the EFT a chiral rotation that, composed with the rotation  
in Eq. \eqref{eq:QCDt.8}, changes the angle 
$\alpha \rightarrow \alpha + \varphi$. 
After the field redefinition, there are no leading-order tadpoles;
we have effectively resummed in one go all terms generated by the third
term in Eq. \eqref{eq:rot.20} and by all other $S_3$s.

\subsection{Alignment in the presence of electromagnetism}
\label{nosalami}

We now show 
that the transformations \eqref{eq:rot.1} and \eqref{eq:rot.2} do not change 
the realization of isospin-breaking operators generated by the 
electromagnetic interaction of the quarks.

In the presence of electromagnetism, the covariant derivatives change
according to Eq. 
\eqref{eq:minimal.1}. 
This does not change the results \eqref{eq:rot.6} and \eqref{eq:rot.7}, 
\begin{equation}\label{eq:minimal.2}
 D_{\mu, \, \rm{em}} \pi_i  = 
\sum_j O^{\prime}_{ij} D^{\prime}_{\mu,\, \rm{em}} \pi_j^{\prime} 
\end{equation}
and
\begin{equation}\label{eq:minimal.2prime}
 \mathcal D_{\mu, \, \rm{em}} N  = 
U^{\prime} \mathcal D^{\prime}_{\mu, \, \rm{em}} N^{\prime}.
\end{equation}
As a consequence, chiral-invariant operators constructed with the minimally 
coupled pion and nucleon covariant derivatives are unchanged by the 
field redefinitions \eqref{eq:rot.1} and \eqref{eq:rot.2}.

Following the method of App. \ref{appB}, the chiral-variant operators
involving electromagnetism can be constructed from the components
of $SO(4)$ antisymmetric tensors:
the 4-$i$ component, $T_{4 i}[0,N]$, which is an isovector, 
and under the field redefinitions \eqref{eq:rot.1} and \eqref{eq:rot.2} 
transforms like Eq. \eqref{eq:rot.9}; and the $i$-$j$ component,
$T_{i j}[0,N]$, which transforms as
\begin{equation}
 T_{i j}[0,N] = 
\sum_{l, m} O^{\prime}_{i l} O^{\prime}_{j m} T^{\prime}_{lm}[0,N^{\prime}].
\end{equation}
Since under the redefinitions \eqref{eq:rot.1} and \eqref{eq:rot.2}
\begin{equation}\label{eq:rot.19}
\left[\frac{1}{D}\left(1-\frac{\boldpi^2}{F^2_{\pi}}\right)\delta_{3 i}
+\frac{2\pi_3\pi_{i}}{F_{\pi}^2 D} \right] T_{4 i}[0,N] = 
\left[\frac{1}{D}
\left(1-\frac{\boldpi^{\prime 2}}{F^2_{\pi}}\right)\delta_{3 i} 
+\frac{2 \pi^{\prime}_3 \pi^{\prime}_{i}}{F^2_{\pi} D^\prime } \right] 
T^{\prime}_{4 i}[0,N^{\prime}]
\end{equation}
and
\begin{equation}\label{eq:rot.19prime}
\frac{2}{F_{\pi} D}\left(\delta_{3 j}\pi_i -\pi_j\delta_{3 i}\right) 
T_{i j}[0, N] =
\frac{2}{F_{\pi} D^{\prime}} 
\left(\delta_{3 j}\pi^{\prime}_i -\pi^{\prime}_j \delta_{3 i}\right) 
T^{\prime}_{i j}[0, N^{\prime}],
\end{equation}
the tensor is also invariant.

\section{Interactions from the QCD $\bar{\theta}$ Term}
\label{thetainteractions}

In the rest of this paper we work with the Lagrangian \eqref{eq:QCDt.8f},
or equivalently, with the Lagrangian \eqref{eq:QCDt.8} followed by the 
rotation \eqref{eq:rot.1} and \eqref{eq:rot.2}  with angle given
by Eq. \eqref{eq:rot.12}.
In this section we construct the most important chiral-variant operators
in the low-energy EFT,
neglecting electromagnetic interactions.
(We correct this defect in Sec. \ref{EMinteractions}.)

The first class of interactions originates entirely from the
first term in Eq. \eqref{eq:QCDt.8f}. 
These interactions break $SO(4)$ explicitly down to the $SO(3)$
of isospin. 
They are well known, and examples are given in App. \ref{appB}.
The most important effect is an $S_4$ that gives rise to 
the pion mass,
\begin{equation}
{\mathcal L}^{(0)}_{\slashchi, f=0}= 
\frac{m^2_{\pi}F_\pi^2}{4} -\frac{m^2_{\pi}}{2D} \boldpi^2,
\label{eq:pimass}
\end{equation}
where $m^2_{\pi} = {\mathcal O}(r(\bar \theta)\bar{m}M_{QCD})$.
Also relevant for what follows is a similar $S_4$ but containing two
nucleon fields, the nucleon sigma term
\begin{equation}
 \mathcal L^{(1)}_{\slashchi, \, f = 2}  
= \Delta m_N \bar N N  \left(1
- \frac{2\boldpi^2 }{F^2_{\pi}D}
\right),
\label{sigma}
\end{equation}
where the nucleon mass correction
$\Delta m_N= \mathcal O\left(r(\bar \theta)\bar{m}\right)
=\mathcal O\left( m^2_{\pi}/M_{QCD} \right)$.
There is, of course, an infinite number of other $S_4$s,
all of which will bring in interactions $\propto \bar{m}$. In addition,
there are interactions from
tensor products of $S_4$s proportional to higher powers of $\bar{m}$.
For example, $S_4\otimes S_4$ with the same $S_4$ that generates
Eq. \eqref{eq:pimass} produces
\begin{equation}
{\mathcal L}^{(2)}_{\slashchi, f=0}= 
\frac{F_\pi^2 \, \Delta^{(2)} m^2_{\pi} }{8}
-\frac{\Delta^{(2)} m^2_{\pi}}{2D^2} \boldpi^2,
\label{eq:pimasscorr}
\end{equation}
where $\Delta^{(2)} m^2_{\pi}$ is an ${\cal O}(m_\pi^4/M_{QCD}^2)$
contribution to the pion mass.
All such chiral-variant interactions
have strengths proportional to powers of 
$m^2_{\pi}$ times appropriate powers of $M_{QCD}$.
Since by all evidence $r(\bar \theta)$ is not small,
the dimensionless factors are expected to be of ${\mathcal O}(1)$.
When we are interested in processes with typical momenta
$Q\sim m_\pi$, the power counting of $f\le 2$ interactions,
Eq. \eqref{Delta}, can be straightforwardly generalized 
by defining $d$ to count powers of $m_\pi$ as well. 

More interesting are the low-energy interactions stemming from the other
two terms in  Eq. \eqref{eq:QCDt.8f}.
The second breaks the $SO(3)$ of isospin down to $SO(2)$ of
rotations in the 1-2 plane in $\boldzeta$ space. 
In particular, it is also charge-symmetry breaking (CSB) 
---charge symmetry is a discrete
isospin rotation of $\pi$ around the 2 axis that exchanges
(up to a phase) the $u$ and $d$ quarks \cite{jerry,CSBth}.
The third term is $P$- and $T$-violating but also
breaks $SO(4)$.
The crucial point is that these two terms are linked because
they break chiral symmetry through components of the {\it same}
chiral four-vector. 
Therefore, $T$ violation from the $\bar \theta$ term is intrinsically
linked to CSB because of chiral symmetry: for each  $T$-violating 
hadronic interaction 
with an odd (even) number of pions from a $P_4$, there is a
CSB interaction with an even (odd) number of pions from the associated $P_3$.
The ratio between the coefficients of the $P_4$ and $P_3$ components 
is fixed by the ratio in Eq. \eqref{eq:QCDt.8f},
\begin{equation}\label{eq:lagr.2}
\frac{\textrm{$T$ violation}}{\textrm{isospin violation}} 
= \frac{m_{\star}}{\varepsilon \bar m} \sin \bar{\theta}
= \frac{1-\varepsilon^2}{2\varepsilon} \sin \bar{\theta}
\equiv \rho(\bar{\theta}, \epsilon).
\end{equation}
This ratio $\rho$ is small 
when $\sin \bar\theta \simeq \bar\theta$ for $|\bar\theta|\ll 1$ and 
$\sin \bar\theta \simeq \pi- \bar{\theta}$ for $|\pi- \bar{\theta}| \ll 1$.

Unfortunately this link becomes ineffective when sufficiently
complicated tensor products have to be included.
We show in App. \ref{appE} that in the pion-nucleon sector of the purely 
hadronic Lagrangian this problem only appears when considering operators 
suppressed by $m^4_{\pi}/M^4_{QCD}$ relative to
the leading $T$-violating interaction. 
As we will see in Sec. \ref{EMinteractions}, the electromagnetic interaction 
makes this problem more acute, so that Eq. \eqref{eq:lagr.2} is ineffective 
already for the leading short-distance contributions to the nucleon EDM.  

We now proceed to build the low-energy interactions
that arise from the isospin- and $T$-violating terms in Eq. \eqref{eq:QCDt.8f}.
They will generate terms proportional to powers of 
${\bar m} \varepsilon r^{-1}(\bar \theta)
=\mathcal{O}(\varepsilon m_{\pi}^2/M_{QCD}r^2(\bar \theta))$
and 
$m_\star \sin\bar\theta \, r^{-1}(\bar \theta)
=\mathcal{O}(\rho\varepsilon m_{\pi}^2/M_{QCD}r^2(\bar \theta))$.
As for the chiral-variant but isospin- and $T$-symmetric terms,
we will for simplicity take $r(\bar \theta)$ to be $\mathcal{O}(1)$ for
power-counting purposes.
Note that mixed operators that combine symmetry breaking from various
sources have to be included.
Since the chiral-symmetry-breaking operators
involving only $S_4$s do not directly affect the link \eqref{eq:lagr.2},
we do not list them.

We consider here only the lower chiral-index $\Delta$ operators,
classified according to the number $f$ of nucleon fields.
As $d$ and $f$ increase, interactions decrease in importance 
\cite{ulfreview,paulo},
but obviously the procedure can be continued {\it ad nauseum}.
As we will show, non-aligned operators, like pion tadpoles, 
appear in power-suppressed terms in the Lagrangian, but they can be dealt
with in perturbation theory.

\subsection{Pion sector}
\label{masspion}

Here we construct the leading interactions that violate isospin
and $T$, which involve only pion fields, that is, with $f=0$.
As it turns out, one cannot construct any terms that transform as $P_3$ and 
$P_4$. Higher-order terms have the same transformation properties as the 
third and fourth components of $SO(4)$ tensors that correspond to tensor 
products of the different symmetry-breaking sources in the QCD Lagrangian.

The chiral-symmetry-breaking Lagrangian with $\Delta = 2 $ 
could receive contributions from the tensor products $S_4 \otimes P_b$ 
and $P_a \otimes P_b$. 
No purely-pionic operator of the first type can be constructed,
while the tensor $P_a \otimes P_b$  can be reduced to a chiral invariant and 
a two-index symmetric tensor, $P_a \otimes P_b = \delta_{ab}\, I + S_{ab}$.
In the pion sector the invariant is a constant, and can be discarded.  
The 3-3 component of the symmetric tensor yields an isospin-breaking correction
to the pion mass, the 4-4 component a chiral-breaking but isospin-conserving 
correction to the mass, while the 3-4 component breaks isospin, parity and 
time reversal:
\begin{equation}\label{eq:lagrsuppr.1}
 \mathcal L^{(2)}_{\slashchi, f = 0} 
=  \frac{\rho^2 F_\pi^2 \, \delta^{(2)} m^2_{\pi}}{8}
+\frac{\delta^{(2)} m^2_{\pi}}{2D^2} 
\left[\pi_3^2 -\rho^2 \boldpi^2
+\rho  F_{\pi}\left(1- \frac{\boldpi^2}{F^2_{\pi}}\right)\pi_3 \right],
\end{equation}
where we introduced the coefficient 
\begin{equation}
\delta^{(2)} m^2_{\pi} =  
\mathcal O\left(\frac{\varepsilon^2 m^4_{\pi}}{r^{4}(\bar\theta )M^2_{QCD}} 
\right),
\end{equation}
which is the largest quark-mass contribution to the
pion-mass splitting \cite{vanKolck}.
Since the latter receives a much larger electromagnetic contribution
(see App. \ref{appB})
$\delta m^2_{\pi,{\rm em}}=\mathcal O\left(\alpha_{\rm em}M^2_{QCD}/\pi\right)$
\cite{vanKolck,isoviolphen}, it is unlikely that this term is 
of any phenomenological use in itself.

However,
Eq. \eqref{eq:lagrsuppr.1} presents a simple illustration of the link
between isospin and $T$ violation.
It also has the 
interesting feature that, even after we chose to align the vacuum
linearly in the chiral-symmetry breaking parameters, non-aligned 
operators appear in the power-suppressed Lagrangian. 
We discuss the role of such tadpoles in Sec. \ref{distad}.

\subsection{Pion-nucleon sector}

Interactions with $f=2$ are potentially the most
important for $T$-violation
phenomenology, because they appear at the lowest chiral index.
Already at $\Delta=1$ we can find a $P_a$ vector,
whose  third and fourth components give the operators 
\begin{equation}\label{eq:lagr.3}
\mathcal L^{(1)}_{\slashchi, f = 2} = 
\delta m_N \left\{\bar N t_3 N 
            -\frac{2\pi_3}{F^2_{\pi} D} \bar N \boldt \cdot \boldpi N 
-   \frac{2\rho }{F_{\pi} D} \bar N \boldt \cdot \boldpi N\right\}.
\end{equation}
These operators provide the leading isospin-breaking \cite{vanKolck} and 
$T$-violating \cite{CDVW79} 
pion-nucleon interactions, respectively.
The low-energy constant
\begin{equation}
\delta m_N=
\mathcal O \left(\frac{\varepsilon m^2_{\pi}}{r^{2}(\bar\theta)M_{QCD}} 
                 \right)
\end{equation}
is the main quark-mass contribution to the 
nucleon mass splitting \cite{vanKolck,isoviolphen}.
Equation \eqref{eq:lagr.3} links the leading $T$-violating 
pion-nucleon coupling 
to the strong nucleon mass splitting $\delta m_N$ via the characteristic 
factor $\rho$, Eq. \eqref{eq:lagr.2}.
We return to this issue in Sec. \ref{FF}.

Considering $\Delta = 2 $, we can construct operators that contain one 
covariant derivative,
\begin{equation}
\mathcal{L}^{(2)}_{\slashchi, f=2}=  
\frac{\beta_1}{F_{\pi}} \left\{D_{\mu} \pi_3 
- \frac{2 \pi_3 }{F^2_{\pi} D} \boldpi \cdot D_{\mu} \boldpi 
- \frac{2\rho}{F_{\pi} D}\boldpi\cdot D_{\mu}\boldpi 
\right\} \bar{N}S^{\mu}N,
\label{TviolPiN2}
\end{equation}
with 
\begin{equation}
\beta_1 =
\mathcal O\left(\frac{\varepsilon m^2_{\pi}}{r^{2}(\bar\theta)M^2_{QCD}} 
                         \right).
\end{equation}
The subleading $T$-violating interaction
thus consists of a seagull vertex
(and its chiral partners) \cite{HvK}, which is related to  
isospin violation in the pion-nucleon coupling constant
\cite{vanKolck,isoviolphen,isoviolOPE}.

Increasing the index by one, we find terms with two covariant derivatives
and two powers of symmetry-breaking parameters.
We write
\begin{equation}
\mathcal L^{(3)}_{\slashchi, f=2} =
\mathcal L^{(3)}_{\slashchi^1, f=2}+\mathcal L^{(3)}_{\slashchi^2, f=2}.
\end{equation}
With two covariant derivatives we find
\begin{eqnarray}\label{eq:2deriv.1}
\mathcal L^{(3)}_{\slashchi^1, f=2} &=&   
\left\{
\frac{\zeta_1}{F_{\pi}} \left( D_{\nu} \boldpi \right) \cdot 
\bar N  \left[S^{\mu},S^{\nu}\right]    
\boldt \left( \mathcal D_{\mu} - \mathcal D^{\dagger}_{\mu}\right) N +
\frac{\zeta_2}{F_{\pi}}\, (v\cdot\mathcal D \, v\cdot D \,\boldpi) 
\cdot \bar N \boldt N  
\right. \nonumber \\ 
&& \left. 
+ \frac{\zeta_3}{F_{\pi}} 
\left( \mathcal D_{\mu} D^{\mu}\,  \boldpi \right) \cdot \bar N  \boldt N  
+\frac{\zeta_4}{F^2_{\pi}} \left(D_{\mu} \boldpi \times D_{\nu} \boldpi \right)
\cdot  \bar N \left(S^{\mu} v^{\nu} - S^{\nu}v^{\mu}\right)\boldt N  
\right\} \nonumber \\ & &
\frac{1}{D}\left[\frac{2\pi_3}{F_{\pi}}
+\rho\left(1-\frac{\boldpi^2}{F^2_{\pi}}\right)\right]
\nonumber \\ 
&& 
+\left\{-\frac{\zeta_{5}}{4}\bar N t_i 
\left(\mathcal D_{||} -\mathcal D_{||}^{\dagger}\right)^2 N  
- \frac{\zeta_6}{4} 
  \bar N  t_i \left(\mathcal D_{\perp} -\mathcal D_{\perp}^{\dagger}\right)^2 
N  
\right. \nonumber \\ 
&& \left. 
+ \frac{\zeta_7}{F_{\pi}}\bar N S^{\mu} \left(\boldt\times 
(v \cdot \mathcal D D_{\mu} \boldpi) \right)_i  N
+ \frac{i\zeta_8}{2F_{\pi}} \, (v\cdot D \pi_i)\, 
\bar N S^{\mu}\left(\mathcal D_{\mu} - \mathcal D_{\mu}^{\dagger} \right) N 
\right. \nonumber \\ 
& & \left.
+ \frac{i\zeta_{9}}{F^2_{\pi}}
\left(D_{\mu}\boldpi \times D_{\nu} \boldpi \right)_i 
\bar N [S^{\mu},S^{\nu}] N 
+ \frac{\zeta_{10}}{F^2_{\pi}} \, (D_{\mu} \pi_i) \, (D^{\mu}\boldpi) \cdot 
\bar N \boldt N \right. \nonumber \\  
&& \left. 
+ \frac{\zeta_{11}}{F^2_{\pi}} \, (v \cdot D \pi_i) \, (v \cdot D\boldpi) \cdot
\bar N  \boldt N  
+ \frac{\zeta_{12}}{F^2_{\pi}} \, (D_{\mu} \boldpi) \cdot (D^{\mu} \boldpi ) 
 \bar N  t_i  N  
+ \frac{\zeta_{13}}{F^2_{\pi}} (v\cdot D \boldpi)^2 \, \bar N  t_i    N 
\right\}
\nonumber \\  
&& 
\left( \delta_{i3} - \frac{2\pi_3\, \pi_i}{F^2_{\pi}D}
- \frac{2 \rho \pi_i}{F_{\pi} D} \right),
\end{eqnarray}
where the coefficients 
\begin{equation}
\zeta_i=
\mathcal O\left(\frac{\varepsilon m^2_{\pi}}{r^{2}(\bar\theta)M^3_{QCD}}
                        \right).
\end{equation} 
Again, we see that the $T$-violating terms have an extra factor of $\rho$ 
compared to their isospin-breaking partners. 
In Eq. \eqref{eq:2deriv.1}, the subscript $_{||}$ ($_\perp$) on the 
nucleon covariant derivatives signifies that we are considering only the 
component parallel (orthogonal) to the velocity, 
\begin{eqnarray}
{\mathcal D}_{|| \lambda} &=& v_{\lambda} v \cdot {\mathcal D},
\\
\mathcal D_{\perp \lambda}&=& \mathcal D_{\lambda} -\mathcal D_{|| \lambda},
\end{eqnarray}
and 
$\bar N \mathcal D_\mu^{\dagger}$
stands for $\overline{\mathcal D_\mu N}$.
In addition, 
for the operators proportional to $\zeta_1$, $\zeta_5$, and $\zeta_6$ 
we used the short-hand notation
\begin{eqnarray}
t_i \left(\mathcal D_{\mu} - \mathcal D^{\dagger}_{\mu}\right) &=& 
t_i \mathcal D_{\mu} - \mathcal D^{\dagger}_{\mu} t_i,
\label{eq:covprime}
\\
t_i \left(\mathcal D - \mathcal D^{\dagger}\right)^2 &=& 
t_i \, \mathcal D_{\mu} \mathcal D^{\mu} 
+ \mathcal D_{\mu}^{\dagger} \mathcal D^{\mu \dagger} \, t_i 
- 2 \mathcal D_{\mu}^{\dagger} \, t_i \, \mathcal D^{\mu}.
\label{eq:cov}
\end{eqnarray}
Note that the $\zeta_2$ and $\zeta_5$ terms vanish for on-shell nucleons. 
Lorentz invariance relates the coefficients of some of the operators in 
Eq. \eqref{eq:2deriv.1}   
to $\delta m_N$ and $\beta_1$. 
We discuss such relations in App. \ref{rel}, where we find 
\begin{equation}
\label{eq:rpi}
\begin{split}
\zeta_1 = 
\zeta_6 = \frac{\delta m_N}{2 m^2_N}, 
\qquad \zeta_8 = \frac{g_A \delta m_N}{m^2_N} - \frac{\beta_1}{m_N}. 
\end{split}
\end{equation}
The relation between $\zeta_6$ and $\delta m_N$ is in agreement with the one 
found in Ref. \cite{isoviolphen}.
Equation \eqref{eq:rpi}  reproduces the relations in 
Ref. \cite{Fettes:2000gb}, once a field redefinition is used to eliminate 
time derivatives acting on the nucleon field from the subleading chiral 
Lagrangian \cite{jordy2}.

Contributions to $\mathcal L^{(3)}_{\slashchi, f = 2}$ that do not contain 
derivatives come from consideration of the tensor products
$P_a \otimes P_b$ and $S_4 \otimes P_b$.
As noticed earlier, the representation of $P_a \otimes P_b$ contains a 
chiral invariant and a symmetric tensor.
In the pion-nucleon sector, the chiral-invariant operator gives an 
inconsequential
correction to the nucleon mass,
while 
the symmetric tensor yields a $P$- and $T$-conserving isospin-breaking term 
from its 3-3 component,
a $P$- and $T$-violating isospin-breaking term ($\propto \rho$) 
from its 3-4 component,
and a $P$- and $T$-conserving chiral-symmetry breaking but isospin-conserving 
term ($\propto \rho^2$) from its 4-4 component.
The tensor product $S_4 \otimes P_b$, in turn,
contributes the 3-4 (which is isospin-breaking)
and 4-4 (which is $P$- and $T$-violating and down by a factor
of $\rho$) components of a symmetric tensor, 
and the 3-4 component (which is isospin-breaking) of an antisymmetric tensor.
We thus find the additional $\Delta=3$ terms,
\begin{eqnarray}\label{eq:lagrsuppr.2}
\mathcal L^{(3)}_{\slashchi^2, f = 2} 
&=&  c^{(3)}_1 \left[\frac{4 \pi_3^2}{F_{\pi}^2D^2} 
+ \rho^2 \left(1 - \frac{4\boldpi^2}{F^2_{\pi}D^2}\right)
+ \frac{4 \rho \pi_3}{F_{\pi}D^2}
  \left(1- \frac{\boldpi^2}{F^2_{\pi}}\right)\right] \bar N N 
\nonumber\\
&&+  c^{(3)}_{2} 
\left\{ \bar N \left[ t_3 +
\frac{2}{F^2_{\pi}D}
\left(6 \pi_3\left(1 - \frac{4\boldpi^2}{3F^2_{\pi}D}\right) 
\boldpi \cdot \boldt-\boldpi^2t_3 \right) 
 \right] N  
\right.\nonumber\\
&&\left.\qquad \quad 
+ \frac{4\rho}{F_{\pi} D^2} \left(1 - \frac{\boldpi^2}{F^2_{\pi}} \right) 
\bar N \boldpi \cdot \boldt N \right\} 
\nonumber\\
&&+ c^{(3)}_3 \bar N \left[
 t_3  + \frac{2}{F^2_{\pi}D} \left(\pi_3\boldpi \cdot \boldt - \boldpi^2t_3  
\right) 
\right] N,
\end{eqnarray}
where
we can estimate the coefficients,
\begin{equation}
c^{(3)}_1 = \mathcal O \left(
\frac{\varepsilon^2 m^4_{\pi}}{r^{4}(\bar \theta)M_{QCD}^3} \frac{}{} 
\right), 
\qquad 
c^{(3)}_{2,3}=\mathcal O \left(
\frac{\varepsilon m^4_{\pi}}{r^{2}(\bar\theta)M_{QCD}^3}  
\right). 
\end{equation}
The $T$-violating interaction associated to $c^{(3)}_{2}$ is similar 
to the leading $T$-violating pion-nucleon interaction in Eq. \eqref{eq:lagr.3},
and it is also linked to a contribution to the nucleon mass splitting,
but it is suppressed by an extra $m_\pi^2/M_{QCD}^2$.
More interesting is the $T$-violating interaction associated to $c^{(3)}_{1}$,
since it involves only the neutral pion. Because of its
isospin character, it contributes differently to observables
than the leading $T$-violating pion-nucleon interaction. 
As one can see, 
it is suppressed with respect to the latter by a factor 
$\varepsilon m_{\pi}^2/M_{QCD}^2$,
and it is linked to an isospin-breaking two-neutral-pion-nucleon 
seagull interaction.
Note that  
there is no 
$T$-violating operator directly associated to $c^{(3)}_{3}$.
This term has exactly the same form as the main electromagnetic
contribution to the nucleon mass difference, Eq. \eqref{emnucleonmass},
and can only be distinguished by the dependence of
its coefficient on $m_\pi^2$; in our power counting,
it is suppressed by one power of $m_\pi/M_{QCD}$.

With the interactions constructed so far we can draw a few conclusions
about the $T$-violating interactions of pions and nucleons.
For example, in Sec. \ref{FF} we examine the pion-nucleon form factor
to one-loop level.
One can of course continue the procedure to higher orders.
It is hard to imagine they would have much phenomenological use,
but they are not entirely devoid of structural interest.
For example, it is from tensor products of three vectors (at
$\Delta=5$) that the first non-electromagnetic $\pi_3 \bar N t_3 N$
interaction appears. 
Also, at this point the connection between isospin- and $T$-violation
ceases to be useful.
These tensor products are discussed in App. \ref{appE}.

\section{Electromagnetic Interactions}
\label{EMinteractions}

In this section we are interested in studying how the combined effects of the 
electromagnetic interaction of quarks and of the QCD $\bar \theta$ term 
manifest themselves in the low-energy Lagrangian. 
As we have mentioned, we have to consider interactions of two types,
with and without soft photons. The former provide short-range contributions
to EDMs. 
The latter involve the exchange of at least 
one hard photon, 
which cannot be 
resolved in the low-energy EFT and is therefore integrated out 
---these interactions are purely hadronic and sometimes called indirect
electromagnetic effects.
Such indirect effects include
pion-nucleon $T$-violating vertices,
which result from a $T$-violating interaction accompanied by a hard-photon 
exchange.

The simplest operators are linear in the chiral-breaking 
parameters $\bar m$ and $e$, and thus necessarily involve a soft photon.
Under $SO(4)$, these operators have the transformation properties of tensor 
products of the chiral-symmetry-breaking terms in $\mathcal L_{m}$,
Eq. \eqref{eq:QCDt.8f},
and $\mathcal L_e$, Eq. \eqref{eq:rot.16},
\begin{equation}\label{eq:em.1}
 \left[\bar m r(\bar\theta) S_4 
- \bar m \varepsilon r^{-1}(\bar\theta) \left( P_3 + \rho P_4 \right) \right]
\otimes \, e A_{\mu} \left( \frac{I^{\mu}}{6} + T^{\mu}_{34}\right).
\end{equation}
We have therefore to construct operators that transform as components of 
$SO(4)$ vectors, $S_4$ and $P_a$,
or components of tensor products, $S_4 \otimes T_{34}$ and 
$P_a \otimes T_{34}$, with $a = 3,4$. 
The tensor product of the antisymmetric tensor $T_{a b}$ and the vector 
$P_c$ gives rise to a vector $(\boldV,V_4)$ and a three-index tensor 
$Z_{ab, c}$, antisymmetric in the first two indices. 
As far as parity and time reversal are concerned, the vector 
$(\boldV, V_4)$ has the same properties as $(\boldS,S_4)$:
$V_4$ is $P$ and $T$ even while  $\boldV$ is $P$ and $T$ odd.
On the other hand, the tensor product $T_{ab} \otimes S_c$ generates 
a vector with the same properties as $P$ and a three-index tensor.
For soft-photon interactions, our index $\Delta$
counts also the number of photon fields and their derivatives.

Another possibility is to construct operators that 
have higher powers of
the chiral-symmetry-breaking parameters $\bar m$ and $e$. 
Those with odd (even) powers of $e$ generate operators with odd (even)
number of external photons.
The simplest 
of the indirect electromagnetic effects come from
operators that
under the group $SO(4)$ 
have the transformation properties of tensor product of 
\begin{equation}\label{eq:em.3}
 \left[\bar m r(\bar\theta) S_4 
- \bar m \varepsilon r^{-1}(\bar\theta) \left( P_3 + \rho P_4 \right) \right]
\otimes \, e A_{\mu} \left( \frac{I^{\mu}}{6}  + T^{\mu}_{34}\right) 
\otimes \, e A_{\nu} \left( \frac{I^{\nu}}{6}  + T^{\nu}_{34}\right),
\end{equation}
in which the photon is integrated out.
In this case we need components of 
$SO(4)$ vectors, $S_4$ and $P_a$,
and of tensor products, $S_4 \otimes T_{34}$,
$P_a \otimes T_{34}$,  $S_4 \otimes T_{34}\otimes T_{34}$,
and $P_a \otimes T_{34}\otimes T_{34}$.
These contributions are proportional to the electromagnetic
fine-structure constant $\alpha_{\rm em}$. Typically,
there is also an extra inverse factor of $\pi$,
so for power counting purposes we assign them a factor of 
$\alpha_{\rm em}/\pi$.
Recall that in this paper we enlarge the chiral index to count also powers of 
$\alpha_{\rm{em}}/\pi$, with the assumption 
$\alpha_{\rm{em}}/\pi \sim \varepsilon m^3_{\pi}/M^3_{QCD}$. 

Clearly, more complicated operators can be constructed, which involve
either more external photons and/or more powers of 
$\bar m$ and $\alpha_{\rm em}/\pi$.
Operators with two photons have been discussed in Ref. \cite{compton}
in connection with nucleon Compton scattering;
since 
they give small contributions 
even to atomic EDMs
\cite{texas}, we do not list here operators with more than a 
single soft photon.
Higher-order terms in the Lagrangian can also be realized 
by building
operators 
with the transformation properties above
that contain covariant derivatives of the nucleon and pion fields 
or higher-dimension gauge-invariant operators.

In the following we catalog the most important $T$-violating interactions,
classifying them by the number of nucleon fields $f$ and the
number of external photons.
These interactions of course arise from $P_4$ and are
always linked to operators from $P_3$, just like in Eq. \eqref{eq:lagr.2}.
Below we list these partners together.
Unlike the operators in Sec. \ref{thetainteractions}, however, here the $S_4$s
play an important role: although the interactions they generate
are of course $T$ conserving, when combined with $T_{34}$s
they lead to isospin-breaking interactions that spoil
the link \eqref{eq:lagr.2} between $T$-violating and $T$-conserving 
interactions already for the leading electromagnetic terms.
These $T$-conserving interactions are given in 
App. \ref{TconsEM}.
We assess the impact on the link \eqref{eq:lagr.2} in Sec. \ref{discon}.

\subsection{Pion sector}
\label{pionalpha}

The tensor products that generate operators $\propto \alpha_{\rm em}/\pi$
in the pion sector are $S_4$, $S_4 \otimes T_{34} \otimes T_{34}$, and 
$P_a \otimes T_{34}$. 
The last two tensors do not belong to irreducible representations of $SO(4)$ 
and they both contain a vector  
with the same transformation properties as $S$. 
$T$ violation is found only in $P_a \otimes T_{34}$:
\begin{equation}\label{eq:lagrsuppr.9}
 \mathcal L^{(3)}_{\slashchi, f = 0, \rm{em}} = 
\frac{\delta^{(3)}_{3, \textrm{em}}  m^2_{\pi} \; F_\pi^2}{4}
- \frac{\delta^{(3)}_{3, \textrm{em}}  m^2_{\pi}}{2D} 
\left( \boldpi^2 +\rho F_{\pi}\, \pi_3 \right),
\end{equation}
the two terms corresponding to $P_4 \otimes T_{34} $ and 
$P_3 \otimes T_{34}$.
Here the coefficient
\begin{equation}
\delta^{(3)}_{3, \textrm{em}} m^2_{\pi} =
\mathcal O\left(\frac{\alpha_{\textrm{em}}}{\pi} 
\frac{\varepsilon m^2_{\pi}}{r^{2}(\bar \theta)} \right).
\end{equation}

Equation \eqref{eq:lagrsuppr.9} is exactly of the form of 
Eq. \eqref{eq:rot.20}.
The only $T$-violating operator is the pion tadpole
(and its associated interactions). 
Since $\delta^{(3)}_{3, \rm{em}} m^2_{\pi} \ll m^2_{\pi}$, 
it does not signal vacuum instability and can be treated in 
perturbation theory.
Because these are already small terms, we do not bother
to consider higher orders.

\subsection{Pion-nucleon sector}
\label{piNalpha}

In the $f=2$ sector, 
interactions start only at $\Delta=4$. They involve no derivatives
but are linear in the quark masses and the fine-structure constant.
The $T$-violating  interactions originate from
$P_a \otimes \, e A_{\mu} \left(I^{\mu}/6 + T^{\mu}_{34}\right) 
\otimes \, e A_{\nu} \left(I^{\nu}/6 + T^{\nu}_{34}\right) $,
and are given by
\begin{eqnarray}\label{eq:lagrsuppr.10}
\mathcal L^{(4)}_{\slashchi,\, f = 2,\, \textrm{em}} & = &
\left[ 
c^{(4)}_{1, \rm{em}} 
+ \frac{4  c^{(4)}_{2, \rm{em}} }{F^2_{\pi} D^2}
\left(\boldpi^2 -\pi_3^2 \right)
\right]
\bar N \left[t_3 - \frac{2 \pi_3}{F^2_{\pi} D} \boldt \cdot \boldpi
- \frac{2\rho}{F_{\pi} D} \boldt \cdot \boldpi \right] N 
\nonumber\\ 
&& +
\left(1 - \frac{2\boldpi^2}{F^2_{\pi}D}-\frac{2\rho\pi_3}{F_{\pi}D} \right)
\left\{
c^{(4)}_{3, \rm{em}}\bar N N 
+c^{(4)}_{4, \rm{em}}
\bar N \left[t_3
+ \frac{2}{F^2_{\pi}D}\left(\pi_3 \boldpi \cdot \boldt- \boldpi^2t_3 \right) 
\right] N 
\right\}
\nonumber\\ 
& & +
\frac{c^{(4)}_{5, \rm{em}}}{F_{\pi} D^2} 
\left[
\frac{2\pi_3}{F_{\pi}}
+\rho \left(1 - \frac{\boldpi^2}{F^2_{\pi}}\right)  
\right]
\bar N \left[
\left(1+\frac{2}{F^2_{\pi} D} \left(\pi_3^2-\boldpi^2\right) \right) 
\boldpi \cdot \boldt - \pi_3 t_3 \right] N.
\end{eqnarray}
The 
operators with coefficient $c^{(4)}_{1, \rm{em}}$
transform like the third and fourth component of a chiral vector, 
coming from the realization of $P_a$ and of the vector in 
$P_a \otimes T_{34} \otimes T_{34}$. 
The operators with coefficient $c^{(4)}_{3, \rm{em}}$
transform as the fourth and third component of a vector with the 
same properties as $S$;
they are generated by the vector in the tensor product $P_a \otimes T_{34}$. 
The operators with coefficient $c^{(4)}_{2, \rm{em}}$
correspond to a three-index tensor in the representation of 
$P_a \otimes T_{34}$, 
while those 
with coefficients $c^{(4)}_{4, \rm{em}}$ and $c^{(4)}_{5, \rm{em}}$
are the realization of $P_a \otimes T_{34} \otimes T_{34}$.
The coefficients scale as
\begin{equation}
c^{(4)}_{1-5, \rm{em}} = \mathcal O\left(\frac{\alpha_{\rm{em}}}{\pi}
\frac{\varepsilon m^2_{\pi}}{ r^{2}(\bar\theta) M_{QCD}} \right).
\end{equation}

Equation \eqref{eq:lagrsuppr.10} 
shows that electromagnetic corrections contribute at the same order
to all of the possible 
$T$-violating, non-derivative pion-nucleon interactions \cite{TVpiN}:
$ \bar N \boldpi \cdot \boldt N$, 
$ \pi_3  \bar N  N$, and $\pi_3  \bar N  t_3 N$.
Their coefficients are linked by chiral symmetry
to $T$-conserving operators.
However,  there are other operators (see  App. \ref{TconsEM}) 
at this order that 
destroy this link 
with $T$-conserving counterparts.

\subsection{Photon-nucleon sector}
\label{photonnucleon}

Interactions with soft photons can be obtained using the $U(1)$-gauge
covariant derivatives \eqref{eq:minimal.1} in existing operators.
More interesting are the interactions that arise through
the field strength $F_{\mu\nu}$, which we describe here.
Since the pion has spin 0, we cannot construct an EDM operator
in the $f = 0$ sector.
In contrast, there are plenty of 
$T$-violating interactions
in the $f=2$ sector. 

The leading terms come at $\Delta=3$. $T$ violation appears in
\begin{eqnarray}\label{eq:lagr.5}
 \mathcal L^{(3)}_{\slashchi, f=2, \rm{em}} & =&  
c^{(3)}_{1, \rm{em}} \left[ \frac{2 \pi_3}{F_{\pi} D}  
+ \rho  \left(1 - \frac{2\boldpi^2}{F^2_{\pi}D}\right) \right]
\bar N \left( S^{\mu}v^{\nu} - S^{\nu} v^{\mu} \right) N\, e F_{\mu \nu} 
\nonumber\\
&& +c^{(3)}_{2, \rm{em}}   
\bar N \left(  t_{3} 
- \frac{2 \pi_3}{F^2_{\pi} D}  \boldpi \cdot \boldt  
-  \frac{2 \rho  }{F_{\pi} D} \boldpi \cdot \boldt \right)\, 
i \left[ S^{\mu},  S^{\nu} \right] N\, e F_{\mu \nu}  
\nonumber\\
&& -c^{(3)}_{3, \rm{em}}  
\bar{N}\left[\frac{2 }{F_{\pi} D} \boldpi \cdot \boldt 
+\rho \left(t_3 - \frac{2\pi_3}{F^2_{\pi}D} \boldpi \cdot \boldt \right)\right]
\left(S^{\mu} v^{\nu} - S^{\nu} v^{\mu}\right) N\, e F_{\mu \nu}
\nonumber\\  
&& + c^{(3)}_{4, \rm{em}}
\left( 1-\frac{2\boldpi^2}{F^2_{\pi}D}  
-  \frac{2 \rho \pi_3}{F_{\pi}D} \right)
\bar{N} i \left[S^{\mu} , S^{\nu} \right] N\, e F_{\mu \nu}
\nonumber\\
&& +  c^{(3)}_{5, \rm{em}}
\left[ \frac{2 \pi_3}{F_{\pi}D} 
+ \rho \left(1-\frac{2\boldpi^2}{F^2_{\pi}D}\right) \right]
\nonumber\\
&& \quad
\bar{N} \left[ \left(  1 - \frac{2\boldpi^2}{F^2_{\pi}D}\right) t_3 
+ \frac{2 \pi_3}{F^2_{\pi}D} \boldpi \cdot \boldt  \right]  
\left(S^{\mu} v^{\nu} - S^{\nu} v^{\mu}\right) N\, e F_{\mu \nu}.
\end{eqnarray}
Here the first two sets of interactions have
the transformation properties of $P_3$ and $P_4$,
while the other three 
represent the tensor product $T_{34} \otimes P_a$.
The $T$-violating operators with coefficient 
$c^{(3)}_{1, \rm{em}} $ 
contribute to the isoscalar nucleon EDM,
while  $c^{(3)}_{3,\rm{em}}$
and $c^{(3)}_{5,\rm{em}}$
are isovector contributions. 
With coefficients $c^{(3)}_{4,\rm{em}}$ and $c^{(3)}_{2, \rm{em}}$,
we find 
$T$-violating interactions that contribute to pion photoproduction,
and are associated with isoscalar and isovector 
contributions (suppressed by $m^2_{\pi}/M^2_{QCD}$ with respect 
to Eq. \eqref{eq:magnetic.1})
to the nucleon magnetic dipole moment, respectively.
The coefficients 
in Eq. \eqref{eq:lagr.5} scale as
\begin{equation}
c^{(3)}_{1-5, \rm{em}} = 
\mathcal O \left(\frac{\varepsilon m^2_{\pi}}{ r^{2}(\bar\theta)M ^3_{QCD}} 
\right). 
\end{equation}

Just like in the previous subsection, here too we find
chiral-symmetry breaking not associated with $T$ violation,
shown in App. \ref{TconsEM}, which destroys our ability to extract
information about the $T$-violating operators from their
$T$-conserving partners.
This, unfortunately, is true in particular for the coefficients 
of the short-distance contributions to the nucleon EDM.

Increasing the index by one, we construct operators that contain one covariant 
derivative of the nucleon or of the pion, or one derivative of the 
electromagnetic field 
strength\footnote{The $T$-violating terms in the $\Delta = 4$ electromagnetic 
Lagrangian in Eqs. \eqref{eq:oneder.1} and \eqref{eq:oneder.2} were 
independently constructed by J. de Vries, 
who first obtained the relations \eqref{eq:rpiEDM}.
We thank him for many discussions on the subject.}. 
We can write the $\Delta = 4$ electromagnetic Lagrangian as
\begin{equation}
\mathcal L^{(4)}_{\slashchi, f=2, \rm{em}} = 
\mathcal L^{(4)}_{\slashchi^V, f=2, \rm{em}} 
+ \mathcal L^{(4)}_{\slashchi^T, f=2, \rm{em}}. 
\end{equation}
In $\mathcal L^{(4)}_{\slashchi^V, f=2, \rm{em}}$ we include the operators 
that realize the components $P_3$ and $P_4$ and the vector components in 
$P_a \otimes T_{34}$. We find
\begin{eqnarray}\label{eq:oneder.1}
 \mathcal L^{(4)}_{\slashchi^V, f=2, \rm{em}} &=& 
\left[ \frac{2 \pi_3}{F_{\pi}D} 
+ \rho\left(1 - \frac{2\boldpi^2}{F^2_{\pi}D}\right) \right] 
e F_{\mu \nu}
\nonumber\\
&& 
\left\{
c^{(4)}_{6, \rm{em}} \bar N  i S^{\mu} \left(  \mathcal D^{\nu}_{\perp}
-   \mathcal D^{\nu \, \dagger}_{\perp} \right) N  
+ \frac{c^{(4)}_{7, \textrm{em}}}{F_{\pi}} D^{\mu} \boldpi \cdot 
\bar N v^{\nu} \boldt N \right\}
\nonumber \\
&&+  \left[\delta_{i3} - \frac{2\pi_3 \pi_i}{F^2_{\pi} D}
- \frac{2\rho \pi_i}{F_{\pi} D}\right] 
e F_{\mu \nu}  
\nonumber \\
&&\left\{ c^{(4)}_{8, \textrm{em}}\mathcal D^{\mu}_{\perp} 
\left(\bar N t_i v^{\nu} N \right) 
+ i c^{(4)}_{9, \textrm{em}}\varepsilon^{\lambda \alpha \mu \nu} 
\bar N t_i \, S_{\alpha} 
\left(\mathcal D_{\perp \lambda} - \mathcal D^{\dagger}_{\perp\lambda}\right) N
\right. 
\nonumber \\
&&\left.  
+ \frac{c^{(4)}_{10, \textrm{em}}}{F_{\pi}} 
\bar N \left(D^{\mu} \boldpi \times \boldt \right)_i S^{\nu} N 
+ \frac{c^{(4)}_{11, \textrm{em}}}{F_{\pi}}\varepsilon^{\lambda \alpha \mu \nu}
v_{\alpha} D_{\lambda} \pi_i \bar N N \right\} 
\nonumber \\
&&+ 
\left[\frac{2\pi_i}{F_{\pi} D} 
+ \rho \left(\delta_{i3} - \frac{2\pi_3 \pi_i}{F^2_{\pi} D}\right)\right] 
eF_{\mu \nu}
\left\{
c^{(4)}_{12, \textrm{em}} \bar N t_i\, i S^{\mu} \left(\mathcal D^{\nu}_{\perp}
-   \mathcal D^{\nu \, \dagger}_{\perp} \right) N 
\right. 
\nonumber \\ 
&&\left. 
+\frac{c^{(4)}_{13, \textrm{em}}}{F_{\pi}} D^{\mu} \pi_i \, 
\bar N v^{\nu} N 
+ \frac{c^{(4)}_{14, \textrm{em}}}{F_{\pi}} 
\varepsilon^{\lambda \alpha \mu \nu} 
\bar N \left(D_{\lambda} \boldpi \times \boldt\right)_i S_{\alpha} N 
\right\}
\nonumber\\
&&+ 
\left[1- \frac{2\boldpi^2}{F^2_{\pi}D} - \frac{2\rho \pi_3}{F_{\pi} D}\right] 
e F_{\mu \nu}
\left\{ c^{(4)}_{15, \textrm{em}}
\mathcal D^{\mu}_{\perp} \left(\bar N v^{\nu} N \right) 
\right. 
\nonumber \\ 
&&\left.
+  c^{(4)}_{16, \textrm{em}} \varepsilon^{\lambda \alpha \mu \nu} 
\bar N i S_{\alpha} \left(\mathcal D_{\perp \lambda} 
- \mathcal D_{\perp \lambda}^{\dagger} \right) N 
+\frac{c^{(4)}_{17, \textrm{em}}}{F_{\pi}} 
\varepsilon^{\lambda \alpha \mu \nu} v_{\alpha} D_{\lambda} \boldpi \cdot 
\bar N \boldt N  \right\} ,
\end{eqnarray}
where the operators associated with $c^{(4)}_{6-11, \textrm{em}}$ 
transform as the component 
$P_4$ accompanied by an isospin-breaking operator that transforms as 
the third component of the same vector, 
while the operators associated with $c^{(4)}_{12-17, \textrm{em}}$ 
are induced by the tensor $P_a \otimes T_{34} $ and 
have the same transformation as the components 4 and 3 of the vector $S$.
Instead, $\mathcal L^{(4)}_{\slashchi^T, f=2, \rm{em}}$ contains the purely 
tensor components of $P_a \otimes T_{34}$, and it is given by
\begin{eqnarray}\label{eq:oneder.2}
 \mathcal L^{(4)}_{\slashchi^T, f=2, \rm{em}} &=& 
\left[ \frac{2 \pi_3}{F_{\pi}D} 
+ \rho\left(1 - \frac{2\boldpi^2}{F^2_{\pi}D}\right) \right] 
\left[ \left(1 - \frac{2\boldpi^2}{F_{\pi}^2 D}\right) \delta_{i3} 
+ \frac{2\pi_3 \pi_i}{F^2_{\pi}D}\right]e F_{\mu \nu} 
\nonumber \\ 
&&  
\left\{
 c^{(4)}_{18, \rm{em}} \bar N  t_i\, i S^{\mu} \left(  \mathcal D^{\nu}_{\perp}
-   \mathcal D^{\nu \, \dagger}_{\perp} \right) N  
+\frac{c^{(4)}_{19, \textrm{em}}}{F_{\pi}} 
\varepsilon^{\lambda \alpha \mu \nu} 
\bar N S_{\alpha} \left( D_{\lambda} \boldpi \times \boldt \right)_i N 
\right. 
\nonumber \\ 
&&\left. 
+ \frac{c^{(4)}_{20, \textrm{em}}}{F_{\pi}} D^{\mu} \pi_i  
\bar N v^{\nu} N  \right\} 
\nonumber \\
&&+ \left[\frac{2 \pi_3}{F_{\pi}D} 
+ \rho\left(1 - \frac{2\boldpi^2}{F^2_{\pi}D}\right) \right] 
e F_{\mu \nu} 
\left\{ \frac{c^{(4)}_{21, \textrm{em}}}{F_{\pi}} i \varepsilon^{3 i j} 
\pi_i \mathcal D_{\perp \lambda} 
\left(\bar N t_j \left[S^{\lambda}, S^{\nu}\right] v^{\mu} N\right) 
\right. 
\nonumber \\ 
&&\left. 
+\frac{c^{(4)}_{22, \textrm{em}}}{F^2_{\pi}} 
\left(D^{\mu} \boldpi \times \boldpi \right)_3 \bar N S^{\nu} N 
+ \frac{c^{(4)}_{23, \textrm{em}}}{F^2_{\pi}} 
\left(\boldpi \cdot D_{\lambda} \boldpi \, \delta_{i3} 
- \pi_i D_{\lambda} \pi_3 \right) \varepsilon^{\lambda \alpha \mu \nu} 
v_{\alpha} \bar N t_i N \right\} 
\nonumber \\
&&+ \left[\delta_{3i} - \frac{2\pi_3 \pi_i}{F^2_{\pi}D}
- \frac{2\rho \pi_i}{F_{\pi} D}\right]  
e F_{\mu \nu}\,\varepsilon^{\lambda \alpha \mu \nu}
\nonumber \\ 
&& 
\left\{ \frac{c^{(4)}_{24, \textrm{em}}}{F^2_{\pi}D} D_{\lambda} \pi_i 
\bar N \left(\boldpi \times \boldt \right)_3 S_{\alpha} N 
+\frac{c^{(4)}_{25, \textrm{em}}}{F^2_{\pi} D}
\left(\boldpi \times D_{\lambda} \boldpi\right)_3 
\bar N t_i S_{\alpha} N\right\}
\nonumber\\
&&+ \frac{c^{(4)}_{26, \textrm{em}}}{F_{\pi}} D_{\lambda} \pi_i 
\left\{ 
\left(1- \frac{2\boldpi^2}{F^2_{\pi} D}\right)\delta_{i3} \delta_{j3} 
+ \frac{2\boldpi^2 \pi_3}{F_{\pi}^4 D^2} (\pi_j \delta_{i3}+ \pi_i \delta_{j3})
- \frac{4\pi_3^2\pi_i \pi_j}{F^4_{\pi} D^2} \right.
\nonumber \\ 
&&\left. 
-\frac{\rho }{F_{\pi}D} 
\left[\left(1 - \frac{2\boldpi^2}{F^2_{\pi} D}\right) 
\left(\pi_j \delta_{i3}+ \pi_i \delta_{j3}\right) 
+ \frac{4\pi_3\pi_i \pi_j}{F^2_{\pi}D}\right]\right\} 
e F_{\mu \nu}\, \varepsilon^{\lambda \alpha \mu \nu}
\bar N v_{\alpha} t_j N .
\end{eqnarray}
In Eqs. \eqref{eq:oneder.1} and \eqref{eq:oneder.2} 
we used the same conventions as in Eq. \eqref{eq:2deriv.1}.
The coefficients $c^{(4)}_{i, \rm{em}}$ are of order 
\begin{equation}
c^{(4)}_{i, \rm{em}} = 
\mathcal O\left(\frac{\varepsilon m^2_{\pi}}{r^{2}(\bar\theta)M^4_{QCD}}
\right).
\end{equation}
Some of the coefficients in Eqs. \eqref{eq:oneder.1} and \eqref{eq:oneder.2}
are related by Lorentz invariance to coefficients in 
$\mathcal L^{(3)}_{\slashchi, f = 2, \rm{em}}$; we find
(see App. \ref{rel})
\begin{equation}\label{eq:rpiEDM}
c^{(4)}_{6, \rm{em}} = \frac{c^{(3)}_{1, \rm{em}}}{m_N}, 
\qquad  
c^{(4)}_{9, \rm{em}} = \frac{c^{(3)}_{2, \rm{em}}}{2 m_N}, 
\qquad
c^{(4)}_{12, \rm{em}} = \frac{c^{(3)}_{3, \rm{em}}}{m_N}, 
\qquad  
c^{(4)}_{16, \rm{em}} = \frac{c^{(3)}_{4, \rm{em}}}{2 m_N}, 
\qquad 
c^{(4)}_{18, \rm{em}} = \frac{c^{(3)}_{5, \rm{em}}}{m_N}.
\end{equation}

Equations \eqref{eq:oneder.1} and \eqref{eq:oneder.2} include interactions
that are simply recoil ($\propto 1/m_N$) corrections to the short-range
nucleon EDM, and interactions of
the nucleon simultaneously with a pion and a photon.
Gory as they are, they might be of some interest in nuclear EDM calculations.
Note that
the $T$-conserving operators 
in these equations
receive contributions also from the realization of  
$S_4 \otimes I$ and $S_4 \otimes T_{34}$,
which have no $T$-violating partners. 
Therefore, as it happens in the case of 
$\mathcal L^{(3)}_{\slashchi, f =2, \rm{em}}$, the relation  
\eqref{eq:lagr.2} is not sufficient to constrain the coefficients 
of the short-distance contributions to the electric dipole moment.
This is, of course, consistent with the constraint
\eqref{eq:rpiEDM} from Lorentz invariance. 

One can also
have higher powers of
the chiral-symmetry-breaking parameters $\bar m$ and $e$. 
As far as operators that contain one photon are concerned, the tensorial 
structure of higher-order operators quickly becomes complicated.
However, if one neglects multiple pions, the basic structure of short-range 
contributions to the EDM is
\begin{equation}\label{eq:em.2}
 \bar N \left( \tilde d_0  + \tilde d_3  t_3 \right) 
\left(S^{\mu} v^{\nu} - S^{\nu} v^{\mu}\right) N e F_{\mu \nu}.
\end{equation}
Operators proportional to higher powers of $\bar m$ and $e$ provide 
$\mathcal O(m^2_{\pi}/M^2_{QCD})$, $\mathcal O(m^4_{\pi}/M^4_{QCD})$, 
$\mathcal O(\alpha_{\rm{em}}/\pi)$, $\ldots$ corrections to the coefficients 
$\tilde d_0$ and $\tilde d_3$.

\section{Discussion}
\label{Discussion}

In Secs. \ref{thetainteractions} and \ref{EMinteractions}
we constructed the 
$T$-violating chiral Lagrangian involving pions and 
up to one nucleon in lowest orders,
including in particular corrections of 
${\cal O}(m^2_{\pi}/M^2_{QCD})$ and ${\cal O}(\alpha_{\rm{em}}/\pi)$ 
with respect to the leading 
$T$-violating pion-nucleon interaction, and short-distance contributions to 
the nucleon EDM up to relative ${\cal O}(m_{\pi}/M_{QCD})$. 
Table \ref{tabI} summarizes the sizes of the various terms considered
explicitly in the text, with factors of $\rho$ (Eq. \eqref{eq:lagr.2}) 
omitted. 
Some of the pion-nucleon terms of
${\cal O}(m^6_{\pi}/M^5_{QCD})$ are discussed in App. \ref{appE}.
Our emphasis has been on interactions that could impact calculations
of hadronic and nuclear EDMs. 
Interactions with delta isobars and more nucleon fields 
can be constructed similarly, and do not affect the qualitative
discussion of this section.

\begin{table}[b]
\begin{tabular}{|c||c|c|c|}
\hline 
Correction  & pion   &  pion-nucleon           & photon-nucleon \\
\hline  
mass        & none   & $m^2_{\pi}/M_{QCD}$     & $eQ m^2_{\pi}/M^3_{QCD}$   \\
derivative  & none   & $Q m^2_{\pi}/M^2_{QCD}$ & $eQ^2 m^2_{\pi}/M^4_{QCD}$ \\
mass$^2$    & $m^4_{\pi}/M^2_{QCD}$ & $m^4_{\pi}/M^3_{QCD}$           & --- \\
derivative$^2$ & none & $Q^2 m^2_{\pi}/M^3_{QCD}$ & --- \\
indirect E\&M & $\alpha_{\rm{em}} m^2_{\pi}/\pi$ 
              & $\alpha_{\rm{em}}m^2_{\pi}/\pi M_{QCD}$ & --- \\
\hline	
\end{tabular}
\caption{Size of the terms in the Lagrangian constructed in the text.
$T$-violating terms have an extra factor of $\rho$, Eq. \eqref{eq:lagr.2}.}
\label{tabI}
\end{table}

The first noticeable aspect of Table \ref{tabI} is that all interactions
are proportional to negative powers of the large scale $M_{QCD}$,
or to $\alpha_{em}/\pi$.
This is a simple consequence of two facts: {\it i)} the $\bar\theta$ term 
can be traded for a mass term, which then brings at least one power of
$m_\pi^2$ in the EFT; {\it ii)} no $P$ vector can be constructed out
of pion fields alone. Time reversal is an accidental symmetry,
in the sense that it only appears in the subleading effective Lagrangian,
even though it is (for $\bar\theta\ne 0$) a leading interaction 
(that is, represented by
a dimension-four operator) in the underlying theory.
$T$ violation would thus be somewhat suppressed at low energies, even if 
$\bar\theta$ had natural size.
The same is true of isospin violation and $\varepsilon$ \cite{vanKolck}.

\subsection{Connection to isospin violation}
\label{discon}

As we have shown, in the purely hadronic sector of the theory, chiral symmetry 
links the coefficients of the leading
$T$-violating operators to those of isospin-breaking 
operators.
Therefore, 
the measurement of $T$-conserving but isospin-breaking 
observables can determine the contribution of the QCD dynamics to
$T$-violating coupling constants.
Since isospin violation that is linear in the quark masses always breaks 
charge symmetry, while this is not necessarily true of 
indirect electromagnetic
interactions, it is in CSB observables
that we have the best chance of making inferences about $T$ violation
from the $\bar \theta$ term.
If we consider the latter as the only source of $T$ violation, 
this link would 
leave $\bar \theta$ as the only parameter to be determined in the direct 
observation of $T$ violation. 

The most important example of the link to isospin violation is in the 
lowest-order terms \cite{CDVW79,Hockings,Savage}:
Eq. \eqref{eq:lagr.3} links the leading $T$-violating interaction
\begin{eqnarray}\label{eq:piNagain}
\mathcal L^{(1)}_{\slashT, \pi N} =
-\frac{2\bar g_0 }{F_{\pi} D} 
\bar N \boldt \cdot \boldpi N
\end{eqnarray}
to the quark-mass contribution to the nucleon mass splitting $\delta m_N$:
\begin{equation}\label{eq:lagr.3.bis}
\bar g_0  =  \rho \, \delta m_N = 
\mathcal O\left(\frac{\rho\varepsilon m^2_{\pi}}{r^{2}(\bar\theta)M_{QCD}} 
                \right).
\end{equation}
It is well-known that this interaction produces the dominant
long-range contribution to the nucleon EDM and form factor 
\cite{CDVW79,su3,scott,HvK}.
With $\delta m_N$ known, a determination of $\bar g_0$ 
would allow one to obtain the value of $\bar\theta$
via Eq. \eqref{eq:lagr.3.bis}.

Now, $\delta m_N$ cannot be determined solely from the observed mass splitting,
since the latter also receives an indirect electromagnetic contribution of 
similar size (see App. \ref{appB}),
$\delta m_{N, \rm{em}}=\mathcal O(\alpha_{\rm em}M_{QCD}/\pi)$ 
\cite{vanKolck,isoviolphen}. (Note that this contribution
is larger by a power of $(\varepsilon m_\pi^2/M^2_{QCD})^{-1}$
than the indirect electromagnetic contributions to the nucleon-mass
splitting that are linked to $T$ violation, which appear in 
Eq. \eqref{eq:lagrsuppr.10}.)
One can use models for higher-energy physics in order
to extract $\delta m_{N, \rm{em}}$ from the Cottingham sum rule,
$\delta m_{N, \rm{em}}= -(0.76 \pm 0.30)$ MeV \cite{Cott}, 
thus inferring $\delta m_N$.
There is also a lattice calculation,
$\delta m_N=2.26 \pm 0.57 \pm 0.42 \pm 0.10$ MeV \cite{latticedeltamN}.
Alternatively, one would like to determine $\delta m_N$ directly
from low-energy data. This is in principle possible \cite{vanKolck}
because $\delta m_{N, \rm{em}}$ originates from a chiral tensor, and thus
generates different interactions between the nucleon and an even
number of pions than does Eq. \eqref{eq:lagr.3}. 
In fact, CSB observables in
pion production reactions such as $pn\to d \pi^0$ 
\cite{CSBd} and $dd\to \alpha \pi^0$ \cite{CSBa} are 
quite sensitive to $\delta m_N$.
Unfortunately they are also sensitive to other CSB
parameters and the calculation of the strong interactions themselves
are not easy, so that at present 
there is room for improvement in the extraction of $\delta m_N$ from 
data \cite{CSBexp}. 
This should, nevertheless, be possible
as we hone our theoretical and experimental tools \cite{CSBth}.
The link with $T$ violation should serve as an extra motivation
for this program.

The connection with CSB is in no way limited to leading 
order. The first correction in the pion-nucleon sector \cite{HvK},
\begin{equation}
\mathcal{L}^{(2)}_{\slashT, 2\pi N}=  
- \frac{2\bar h^{(2)}_1}{F_{\pi}^2 D}\boldpi\cdot D_{\mu}\boldpi 
\bar{N}S^{\mu}N,
\label{TviolPiN2prime}
\end{equation}
has a coefficient related by Eq. \eqref{TviolPiN2}
to the quark-mass contribution 
to isospin breaking in the pion-nucleon coupling constant,
\begin{equation}
\bar h^{(2)}_1 = \rho\beta_1
=\mathcal O\left(\frac{\rho\varepsilon m^2_{\pi}}{r^{2}(\bar\theta)M_{QCD}^2} 
                \right).
\end{equation}
At present there are only bounds on $\beta_1$. For example,
from a phase-shift analysis of two-nucleon data \cite{isoviolphen,isoviolOPE},
$\beta_1=0(9) \cdot 10^{-3}$, which
is comparable to estimates of $\beta_1$ from $\pi$-$\eta$ mixing.

Note that when we face interactions that are no longer
linear in $\varepsilon$, the connection is not necessarily
to CSB; it might be merely to more general isospin violation.
For example, the new $T$-violating structure $\pi_3 \bar N N$ in 
Eq. \eqref{eq:lagrsuppr.2},
\begin{eqnarray}\label{eq:piNred}
\mathcal L^{(3)}_{\slashT, \pi N} &=& 
\frac{4\bar h^{(3)}_{1} }{F_{\pi} D} 
\left(1 - \frac{2\boldpi^2}{F^2_{\pi}D}\right)
\pi_3 \bar N N +\ldots, 
\end{eqnarray}
with
\begin{equation}
\bar h^{(3)}_1 = \rho c_1^{(3)}
=\mathcal O\left(\frac{\rho\varepsilon^2 m^4_{\pi}}{r^{4}(\bar\theta)M_{QCD}^3}
                 \right),
\end{equation}
has a
partner that is isospin-breaking but does respect 
charge symmetry. 
The parameter $c_1^{(3)}$ 
can in principle be extracted from isospin violation
in pion-nucleon scattering, but it is not easy:
even the very sophisticated, state-of-the-art analysis of 
Ref. \cite{panzer} stops one order shy of it, at which level
many other poorly determined parameters already appear.

We can thus obtain information about some strong-interaction matrix
elements that appear in $T$ violation from an analysis of isospin violation.
However, at higher orders in the strong-interaction sector
this connection disappears.
The consideration of subleading $T$-violating interactions requires the 
construction of operators that transform as tensor products of the 
chiral-breaking terms in the QCD Lagrangian. 
In this case, the relation \eqref{eq:lagr.2} applies to the ratio of the 
coefficients of the $T$-violating and $T$-conserving components of the tensors.
In general, however, the tensors thus obtained belong to some reducible 
representation of $SO(4)$, and they have to be decomposed as the sum of 
independent operators belonging to irreducible representations of the group. 
High-order tensor products may generate operators that have the same 
chiral properties as the vectors $S$ and $P$ in the QCD Lagrangian.
$T$ violation is still in $P_4$s and $S_3$s proportional
to $\rho$, but their number might no longer match those
of $T$-conserving $S_4$s and $P_3$s. 

One example, given in App. \ref{appE}, is that
of a $T$-violating operator with the same transformation properties as 
$S_3$, which is linked by Eq. \eqref{eq:lagr.2} to 
a chiral-breaking operator that transforms as $S_4$.
This $S_4$ is merely a subleading correction to another $S_4$,
the nucleon sigma term, which does not have
a $T$-violating partner.
The correction cannot be separated experimentally from the lower-order
term; it could potentially be separated
theoretically via lattice simulations with varying quark masses
(although if it is necessary to appeal to lattice calculations
then one could calculate the strong-interaction
coefficient of the $T$-violating operator directly).
Worse still, another $S_4$ without $T$-violating
partner can appear at the same order as the $S_4$ we are interested
in, which is in fact the case of the example in App. \ref{appE}.
In this case,
the connection with a $T$-conserving observable is completely lost.

Similarly, in another App. \ref{appE} example, only
part of a $P_4$ is linked to a $P_3$, which is a correction
to the nucleon mass splitting. As a consequence, the connection 
between the coefficient of a $\bar N\boldpi\cdot\boldt N$ 
interaction and $\delta m_N$, shown in leading order
in Eq. \eqref{eq:lagr.3.bis}, no longer holds four orders down
in the $m_\pi/M_{QCD}$ expansion.

As soon as the electromagnetic interaction is turned on, 
the combined 
isospin-breaking effects of the electromagnetic coupling 
and of the quark-mass difference destroy the validity of 
the relation \eqref{eq:lagr.2}.
An example is provided by corrections to 
the $\bar N \boldpi\cdot\boldt N$ coupling.
First, there is a correction ($\bar h_2^{(3)}=\rho c_2^{(3)}$) 
in Eq. \eqref{eq:lagrsuppr.2}
that is linked
to a contribution to the nucleon mass splitting ($c_2^{(3)}$).
The $c_3^{(3)}$ terms in Eq. \eqref{eq:lagrsuppr.2}
also contribute to the nucleon mass splitting without
a $T$-violating partner, and threaten to spoil our ability
to determine $c_2^{(3)}$.
In the absence of electromagnetism,
$c_2^{(3)}$ and $c_3^{(3)}$ could be separated by their different 
pion seagulls.
Once electromagnetism is considered, however,
it provides a contribution $\delta m_{N, \rm{em}}$
to the nucleon mass difference, whose full operator
has exactly the same form as the $c_3^{(3)}$ term in 
Eq. \eqref{eq:lagrsuppr.2}. 
Therefore $c_3^{(3)}$ cannot be separated
from $\delta m_{N, \rm{em}}$ experimentally, and thus 
the ${\cal O}(m_\pi^2/M_{QCD}^2)$ correction 
to the $T$-violating pion-nucleon coupling constant $\bar g_0$,
$\bar h_2^{(3)}$, cannot be determined.
At next order, 
$\bar h^{(4)}_{1,\rm{em}} = \rho c^{(4)}_{1,\rm{em}}$, 
the Eq. \eqref{eq:lagrsuppr.10} correction of 
${\cal O}(\alpha_{\rm{em}}/\pi)$ 
to $\bar g_0$, 
cannot be inferred by measuring isospin violation, either,
due to the presence at the same order of an unpaired
contribution to the nucleon mass difference, the 
$c^{(4)}_{28,\rm{em}}$ term in Eq. \eqref{eq:lagrsuppr.11}.

More important is the case of the short-distance 
contributions to the nucleon EDM,
extracted from Eqs. \eqref{eq:lagr.5}, \eqref{eq:oneder.1},
\eqref{eq:oneder.2}, and \eqref{eq:rpiEDM}:
\begin{eqnarray}\label{eq:sd}
 \mathcal L^{(3+4)}_{\slashT, N\gamma}  & = &   
e F_{\mu \nu}
\bar{N} 
\left[\bar h^{(3)}_{1, \rm{em}}
+\left(\bar h^{(3)}_{5, \rm{em}}- \bar h^{(3)}_{3, \rm{em}}\right) t_3 
+\frac{2}{F^2_{\pi}}
\left(\frac{\bar h^{(3)}_{5, \rm{em}}}{D} 
      + \frac{\bar h^{(3)}_{3, \rm{em}}}{2-D} \right)
\left(\pi_3\boldpi \cdot \boldt -\boldpi^2t_3\right) \right] 
\nonumber\\
&&
\qquad\quad
\left[ 
S^{\mu} \left(v^{\nu}+\frac{i \mathcal D^{\nu}_{\perp} }{m_N} \right)
- S^{\nu} \left( v^{\mu} +\frac{i\mathcal D^{\mu}_{\perp}}{m_N}\right)^\dagger
\right] N
\; \left(1-\frac{2\boldpi^2}{F^2_{\pi}D}\right),
\end{eqnarray}
with coefficients 
\begin{equation}
\bar h^{(3)}_{i,\rm{em}} = \rho c^{(3)}_{i,\rm{em}}
=\mathcal O\left(\frac{\rho\varepsilon m^2_{\pi}}{r^{2}(\bar\theta)M_{QCD}^3} 
                \right).
\end{equation}
(In Eq. \eqref{eq:sd} $\mathcal D^{\mu\, \dagger}_{\perp}$ is to be understood
as acting on $\bar N$ only.)
We see that 
$\tilde d_0=\bar h^{(3)}_{1,\rm{em}}$
and 
$\tilde d_3=\bar h^{(3)}_{5,\rm{em}}- \bar h^{(3)}_{3,\rm{em}}$
are, respectively,
isoscalar and isovector short-range contributions to the EDM.
Since by power counting 
they should be comparable 
to the EDM generated by 
a pion loop \cite{CDVW79,su3,HvK},
the 
nucleon EDM up to next-to-leading order
depends on three $T$-violating parameters:
$\bar g_0$, $\tilde d_0$, and $\tilde d_3$.

These operators
are linked by Eq. \eqref{eq:lagr.5} 
to operators that contribute to pion photoproduction on the nucleon.
(We do not see 
here a direct link to anomalous magnetic moments \cite{susan}.)
However, the coefficients cannot be extracted 
from the measurement of isospin violation in pion photoproduction
due to the existence in Eq. \eqref{eq:lagr.6}
of operators with the same chiral properties 
as $c^{(3)}_{1,\rm{em}}$ and $c^{(3)}_{3,\rm{em}}$  
that are not linked to $T$-violating operators.
Even if one assumes $T$ violation to arise solely from the $\bar \theta$ term,
the measurement of the neutron and proton EDMs alone 
would not be sufficient to fix $\bar g_0$ 
(and thus extract the value of the angle $\bar \theta$
using $\delta m_N$),
unless the short-distance operators are calculated in lattice QCD.

These conclusions are obtained by considering the chiral group 
$SU(2)_{L} \times SU(2)_R$. 
If one assumes the strange quark mass to provide a suitable
expansion parameter, one works with $SU(3)_L \times SU(3)_R$ instead.
This larger symmetry increases our ability to extract strong-interaction
matrix elements needed for an analysis of $T$ violation from
$T$-conserving measurements \cite{CDVW79,su3}.
The limitation of this approach comes, of course, from the poorer convergence 
of
the chiral expansion in the $SU(3)_L \times SU(3)_R$ case.

If on the one hand chiral symmetry is not sufficient to fully constrain the
QCD dynamics that enters the $T$-violating couplings, on the other hand it is 
a powerful tool to organize the $T$-violating Lagrangian as a series of terms 
suppressed by more and more powers of $m_{\pi}/M_{QCD}$ and
$Q/M_{QCD}$.
It can be used to extrapolate lattice calculations,
in this case the nucleon EDM \cite{lattice}, to realistic
values of $m_\pi$,
and to take one-nucleon information, experimental or numerical,
into nuclear systems \cite{paulo}.

In Sec. \ref{FF}, we examine in more detail the implications
of this organizational principle to the interaction of a single pion with 
a nucleon. A new ingredient beyond leading order is the pion tadpole,
which we discuss first.

\subsection{Role of tadpoles}
\label{distad}

We imposed vacuum alignment in first order in
the symmetry-breaking parameters
by choosing $\varphi$ according to Eq. \eqref{eq:rot.12}
so that no 
$S_3$ was present in the leading Lagrangian.
Yet, non-aligned terms, 
tadpoles in particular, germinate in second order.
At $\Delta =2$, Eq. \eqref{eq:lagrsuppr.1} brings in a tadpole,
\begin{equation}\label{eq:lagrtad.1}
\mathcal L^{(2)}_{\rm{tadpole}} = 
\frac{\delta^{(2)} m^2_{\pi}}{2 D^2} \rho F_{\pi} 
\left(1- \frac{\boldpi^2}{F^2_{\pi}}\right) \pi_3,
\end{equation}
with 
\begin{equation}
\delta^{(2)} m^2_{\pi}  = 
\mathcal O\left(\frac{\varepsilon^2 m^4_{\pi}}{r^{4}(\bar\theta)M^2_{QCD}}
                \right).
\end{equation} 

These suppressed non-aligned operators can be dealt with in perturbation theory
and do not constitute a problem.
This can be seen, for example, in the case of the two-point pion 
Green's function, already discussed in Sec. \ref{need}.
{}From Eqs. \eqref{eq:pimass}, \eqref{eq:pimasscorr}, 
and \eqref{eq:lagrsuppr.1},
the non-derivative terms in the pion Lagrangian up to $\Delta = 2$ read
\begin{eqnarray}
\mathcal L^{(\Delta\le 2)}_{\slashchi, f = 0}  &=&  
\frac{m^2_{\pi}F_\pi^2}{4}
\left(1+\frac{\Delta^{(2)}m_\pi^2+\rho^2 \delta^{(2)}m^2_{\pi}}{2m^2_{\pi}}
\right)
-\frac{m^2_{\pi}}{2D}
\left(1+\frac{\Delta^{(2)}m_\pi^2+\rho^2 \delta^{(2)}m^2_{\pi}}{m^2_{\pi}D}
\right) \boldpi^2
\nonumber\\
&&+
\frac{\delta^{(2)}m^2_{\pi}}{2D^2} \pi_3^2
+
\rho F_\pi \frac{\delta^{(2)}m^2_{\pi}}{2D} 
\left(1- \frac{2\boldpi^2}{F^2_{\pi}D}\right) \pi_3.
\label{eq:rot.22}
\end{eqnarray}
The Lagrangian \eqref{eq:rot.22} generates Feynman diagrams similar to those 
in Fig. \ref{Fig.2a}. 
If we limit ourselves to power counting and neglect the details of the 
diagrams, the tadpole term generates interactions analogous to those in 
Eq. \eqref{eq:rot.20}, with the replacement 
\begin{equation}
g \sim \rho \frac{\delta^{(2)} m^2_{\pi}}{m^2_{\pi}} =
\mathcal O\left(\frac{\rho\varepsilon^2 m^2_{\pi}}{r^4(\bar\theta )M^2_{QCD}} 
                \right).
\end{equation}
However, in the present case $g$ is small, suppressed  by two powers of 
$m_{\pi}/M_{QCD}$. 
Therefore, the contribution of the diagrams in 
Fig. \ref{Fig.2a} is suppressed by at least 
$\rho^2 \varepsilon^4 m^4_{\pi}/M^4_{QCD}$,
and it is negligible for most purposes.
Differently from the case discussed in Sec. \ref{need}, the tadpole in 
Eq. \eqref{eq:lagrtad.1}
does not cause vacuum instability, that is, the choice of vacuum done in the 
construction of the chiral Lagrangian is still viable and the explicit 
symmetry-breaking terms can be handled in $\chi$PT.
A toy model that illustrates this fact can be found in App. \ref{appC}.

However, generally pion tadpoles 
need to be 
considered when calculating any observable.
Due to the smallness of their 
coefficients, only a 
manageable number of them contribute to the calculation 
of an 
observable at a given accuracy in the expansion in powers of $Q/M_{QCD}$. 
A concrete example where tadpoles play a role is the $T$-violating 
pion-nucleon form factor at relative
${\cal O}(Q^2/M_{QCD}^2)$, which will be discussed in Sec. \ref{FF}
and App. \ref{piNFFwtad}.  

Still, one can rotate the tadpoles away using
Eqs. \eqref{eq:rot.1} and \eqref{eq:rot.2}.
To kill the tadpole at $\Delta=2$, the angle $\varphi$ is 
${\cal O}(\rho \varepsilon^2 m_\pi^2/M_{QCD}^2)$,
such that
\begin{equation}
\sin \varphi = -  \rho \frac{\delta^{(2)} m^2_{\pi}}{m^2_{\pi}}. 
\label{subrot}
\end{equation}
This rotation induces changes in all chiral-breaking terms,
but the new interactions it generates are always two orders higher in the
$m_\pi/M_{QCD}$ expansion.
In terms of the new fields (but dropping the primes from the rotated fields),
\begin{eqnarray}
\mathcal L^{(\Delta\le 2)}_{\slashchi, f = 0}  &=&  
\frac{m^2_{\pi}F_\pi^2}{4}
\left(1+\frac{\Delta^{(2)}m_\pi^2+\rho^2 \delta^{(2)}m^2_{\pi}}{2m^2_{\pi}}
\right)
-\frac{m^2_{\pi}}{2D}
\left(1+\frac{\Delta^{(2)}m_\pi^2+\rho^2 \delta^{(2)}m^2_{\pi}}{m^2_{\pi}D}
\right) \boldpi^2
\nonumber\\
&&
+\frac{\delta^{(2)}m^2_{\pi}}{2D^2} \pi_3^2
-\rho \frac{\delta^{(2)}m^2_{\pi}}{F_\pi D^2} \pi_3 \boldpi^2.
\label{eq:rot.22rotated}
\end{eqnarray}
To this order, then, the only change in the pion sector is the elimination
of the tadpole. Note that, because the operator \eqref{eq:lagrsuppr.1}
originates in a tensor structure more complicated than an $S_3$,  
residual $T$-violating interactions
involving an odd number of pions
are left behind. In many processes they will only contribute
at loop level and, consequently, at high order.

This rotation affects the other sectors of the theory as well.
In the nucleon sector,
the $\Delta=1$ chiral-breaking terms from 
Eqs. \eqref{sigma} and \eqref{eq:lagr.3},
will generate changes in the $\Delta=3$ Lagrangian \eqref{eq:lagrsuppr.2}:
\begin{eqnarray}
 \mathcal L^{(\Delta\le 3)}_{\slashchi, \, f = 2} 
&=& 
\Delta m_N \bar N N  \left(1 - \frac{2\boldpi^2 }{F^2_{\pi}D} \right)
+\delta m_N \left\{\bar N t_3 N 
            -\frac{2\pi_3}{F^2_{\pi} D} \bar N \boldt \cdot \boldpi N 
-   \frac{2\rho }{F_{\pi} D} \bar N \boldt \cdot \boldpi N\right\}
\nonumber\\
&&
+c^{(3)}_1 \left[\frac{4 \pi_3^2}{F_{\pi}^2D^2} 
+ \rho^2 \left(1 - \frac{4\boldpi^2}{F^2_{\pi}D^2}\right)
+ \frac{4 \rho \pi_3}{F_{\pi}D^2}
  \left(1- \frac{\boldpi^2}{F^2_{\pi}}\right)\right] \bar N N 
+\ldots
\label{sigmaetal}
\end{eqnarray}
goes into
\begin{eqnarray}
 \mathcal L^{(\Delta\le 3)}_{\slashchi, \, f = 2} 
&=& 
\Delta m_N \bar N N  \left(1 - \frac{2\boldpi^2 }{F^2_{\pi}D} \right)
+\delta m_N \left(1-\rho^2  \frac{\delta^{(2)} m^2_{\pi}}{m^2_{\pi}}\right) 
\left\{\bar N t_3 N 
       -\frac{2\pi_3}{F^2_{\pi} D} \bar N \boldt \cdot \boldpi N \right\}
\nonumber\\
&&
-   \frac{2\rho }{F_{\pi} D} 
\delta m_N\left(1+\frac{\delta^{(2)} m^2_{\pi}}{m^2_{\pi}}\right)
\bar N \boldt \cdot \boldpi N
+c^{(3)}_1 \left[\frac{4 \pi_3^2}{F_{\pi}^2D^2} 
+ \rho^2 \left(1 - \frac{4\boldpi^2}{F^2_{\pi}D^2}\right)\right] \bar N N 
\nonumber\\
&&
+ \frac{4 \rho }{F_{\pi}D}
  \left[c^{(3)}_1 -\frac{\Delta m_N}{2}\frac{\delta^{(2)} m^2_{\pi}}{m^2_{\pi}}
        - \frac{2c^{(3)}_1\boldpi^2}{F^2_{\pi}D}        
\right]\pi_3 \bar N N 
+\ldots
\label{sigmaetalprime}
\end{eqnarray}
This amounts to a very small shift in the nucleon-mass-splitting term,
an ${\cal O}(\varepsilon^2 m_\pi^2/M_{QCD}^2)$ shift in $\bar g_0$,
and a shift in the $\pi_3 \bar N N $ coupling.
Recall that the $c^{(3)}_1$ term
originated in the symmetric tensor contained in
a $P_a\otimes P_b$ structure just like the tadpole, as
it is obvious from the form of the corresponding terms.
However, since there is no {\it a priori} relation between
$c^{(3)}_1/\Delta m_N$ and $\delta^{(2)} m^2_{\pi}/2m^2_{\pi}$,
the $\pi_3 \bar N N $ coupling is not eliminated
when we rotate the tadpole away.

In 
Eq. \eqref{eq:lagrsuppr.9} we saw that another tadpole
arises from an $S_3$ at ${\cal O}(\alpha_{\rm em} \varepsilon m_\pi^2/\pi)$.
This is despite the fact that we eliminated the $S_3$ from
the QCD Lagrangian.
Much of what we just said about the hadronic tadpole can be adapted
to this electromagnetic look-alike, 
as well as others that appear at higher orders.
The suppression of their effects is even greater,
so they only show up in high orders.
One can also carry out a further rotation to eliminate each of them.
For the electromagnetic one, for example,
the angle is of the form \eqref{subrot} with
$\delta^{(2)} m^2_{\pi} \to -\delta^{(3)}_{3,\, \rm{em}} m^2_{\pi}$.
In this case there are no residual odd-pion vertices as in 
Eq. \eqref{eq:rot.22rotated}: this $S_3$ is completely rotated away.
One might think that then other $S_3$s ---such as the
$c^{(4)}_{3,\rm{em}}$ term in Eq. \eqref{eq:lagrsuppr.10}---
would be eliminated by the same rotation.
Alas, just like in Eq. \eqref{sigmaetalprime},
the lack of an {\it a priori} relation between parameters
---such as $c^{(4)}_{3,\rm{em}}/\Delta m_N$ and 
$\delta^{(3)}_{3,\rm{em}} m^2_{\pi }/m^2_{\pi}$--- 
means that the other $S_3$s survive. 

Since in EFT a field redefinition does not change the result 
for any observable, it is our choice whether to keep
or eliminate tadpoles. 
We give an example of this flexibility in the next section.

\section{Pion-Nucleon Coupling and Form Factor}
\label{FF}

The most important element in the evaluation of
hadronic and nuclear EDMs is the $T$-violating pion-nucleon coupling.
As an example of application of our framework, 
we study in this section this coupling 
and the associated form factor, which has
recently been considered in Ref. \cite{TVpiNNFF}.
We carry out the calculation
to one-loop level, that is, up to a suppression by two
powers of $Q/M_{QCD}$.
For simplicity we employ the field redefinition of Sec. \ref{distad};
the calculation with tadpoles is a bit subtler and is discussed
in App. \ref{piNFFwtad}.

Traditionally \cite{nuclear}, implications of $T$ violation in nuclear physics
have been drawn from the possible isospin structures of non-derivative 
pion-nucleon interactions, without prejudice about their relative sizes.
In  Tab. \ref{Tab:hierarchy} we list the non-derivative
$T$-violating pion-nucleon couplings found in 
Secs. \ref{thetainteractions} and \ref{EMinteractions},
along with their chiral transformation properties
and estimated sizes.
The realization of explicit symmetry breaking in $\chi$PT
implies that the leading pion-nucleon vertex has the form of
Eq. \eqref{eq:piNagain};
it is nothing but the pion-nucleon interaction
of Ref. \cite{CDVW79} with its chiral partners, in stereographic
coordinates. 
This coupling receives both hadronic corrections of 
${\cal O}(m^2_{\pi}/M^2_{QCD})$ and electromagnetic corrections of
${\cal O}(\alpha_{\rm{em}}/\pi)$. 
The $T$-violating coupling $\pi_3 \bar N N$ is suppressed by two powers of 
$m_{\pi}/M_{QCD}$ or by one power of $\alpha_{\rm{em}}/\pi$.
Numerically $\alpha_{\rm{em}}/\pi \sim \varepsilon m^3_{\pi}/M^3_{QCD}$
(using $M_{QCD}\sim m_\rho$, the mass of the rho-meson),
so the most important contribution is presumably the hadronic one. 
Finally, the most relevant contribution to the third possible vertex, 
$\pi_3 \bar N t_3 N$,  has electromagnetic origin and is suppressed by 
$\alpha_{\rm{em}}/\pi$ with respect to $\bar g_0$. 
Hadronic contributions to $\pi_3 \bar N t_3 N$ are suppressed by 
$m^4_{\pi}/M^4_{QCD}$, as shown in App. \ref{appE}.

\begin{table}[b]
\centering
 \begin{tabular}{|c|c|c|c|}
\hline
$T$-violating $\pi N$ vertex 
& coefficient 
$\times\left(\rho\varepsilon m^2_{\pi}/M_{QCD}\right)^{-1}$
&  $SO(4)$ properties & Equation\\ 
\hline
$\bar N \boldt \cdot \boldpi N$    & $1$    & $P_4$ 
& \eqref{eq:lagr.3}\\
& $m_{\pi}^2/M_{QCD}^2$  & $P_4 \otimes S_4$ 
& \eqref{eq:lagrsuppr.2}\\
& $\alpha_{\rm{em}}/\pi$ & $T_{34} \otimes T_{34} \otimes P_4$ 
& \eqref{eq:lagrsuppr.10}  \\
\hline 
$\pi_3 \bar N N$  & $\varepsilon m_{\pi}^2/M_{QCD}^2$ & $P_{3} \otimes P_4 $  
& \eqref{eq:lagrsuppr.2} \\
& $\alpha_{\rm{em}}/\pi$ & $S_3$  
& \eqref{eq:lagrsuppr.10}\\
\hline
$\pi_3 \bar N t_3 N$ & $\alpha_{\rm{em}}/\pi$ 
& $T_{34}\otimes T_{34}\otimes P_4$  
& \eqref{eq:lagrsuppr.10}\\
\hline
\end{tabular}
\caption{List of possible non-derivative $T$-violating pion-nucleon vertices, 
up to 
${\cal O}(\rho\varepsilon m^4_{\pi}/M^3_{QCD})$
and ${\cal O}(\rho\alpha_{\rm{em}}m^2_{\pi}/\pi M_{QCD})$. 
We give the form of the one-pion interaction, 
the size of the contributions to the 
interaction strengths in units 
of $\rho \varepsilon m^2_{\pi}/M_{QCD}$, 
the $SO(4)$ tensor properties of the full operator,
and the equation where it can be found.
For simplicity we assumed $r(\bar\theta)={\cal O}(1)$;
otherwise, $\varepsilon \to \varepsilon/r^2$ above.
}
\label{Tab:hierarchy}
\end{table}

Thus, as far as the $\bar\theta$ term is concerned, the 
isospin-breaking $T$-violating couplings are much smaller than $\bar g_0$.
As a consequence, using the estimates in Ref. \cite{nuclear}, 
we would expect the $^3$He EDM to be somewhat larger than
the neutron EDM, which in turn would be somewhat
larger than the deuteron EDM.
Note, however, that the calculations in Ref. \cite{nuclear} do not 
incorporate the systematic power counting discussed here.
It would be interesting to repeat these calculations 
within our framework.

As a step in this direction, 
note that other pion-nucleon couplings
are expected to be as important as the isospin-breaking non-derivative
couplings. 
{}From Secs. \ref{thetainteractions} and \ref{EMinteractions}
we see that the $T$-violating  pion-nucleon interaction 
receives corrections two orders down.
After the field redefinition of Sec. \ref{distad}, and 
expanding in the number of pions,
\begin{eqnarray}\label{eq:piN}
\mathcal L^{(3)}_{\slashT, \pi N} &=& 
\frac{2}{F_{\pi}} 
\left[\left(2\bar h^{(3)}_{2} 
-\bar g_0 \frac{\delta^{(2)} m^2_{\pi}}{m^2_{\pi}}\right)\boldpi
+\frac{\bar\eta_2}{2} (v\cdot\partial)^2 \boldpi
+\frac{\bar\eta_3}{2}  \partial^2  \boldpi \right] \cdot \bar N \boldt  N
\nonumber\\
&& + \frac{1}{2F_{\pi}} 
\boldpi\cdot 
\bar N \boldt 
\left[\bar\eta_{5} \left(\partial_{||} - \partial_{||}^{\dagger}\right)^2 
+ \frac{\bar g_0}{2 m_N^2} 
\left(\partial_\perp - \partial_\perp^{\dagger}\right)^2 \right] N  
\nonumber\\
&&
+\frac{\bar g_0}{2 m_N^2 F_{\pi}} \left( \partial_{\nu} \boldpi \right)\cdot
\bar N\left[S^{\mu},S^{\nu}\right]  \, \boldt 
\left(\partial_{\mu} - \partial_{\mu}^{ \dagger}\right) N
\nonumber\\
&&
+\frac{2}{F_{\pi}} 
\left(2\bar h^{(3)}_{1} 
-\rho \Delta m_N\frac{\delta^{(2)}m_\pi^2}{m_\pi^2}\right) 
\pi_3 \bar N N +\ldots ,
\end{eqnarray}
where
\begin{equation}
\bar h^{(3)}_i = \rho c_i^{(3)} = 
\mathcal O\left(\frac{\rho\varepsilon m^4_{\pi}}{r^{2}(\bar\theta)M_{QCD}^3}
          \right),
\qquad
\bar\eta_i = \rho \, \zeta_i =
\mathcal O\left(\frac{\rho\varepsilon m^2_{\pi}}{r^{2}(\bar\theta)M_{QCD}^3} 
          \right).
\end{equation}
In addition to a correction $\bar h^{(3)}_{2}$ 
to $\bar g_0$
and to the isospin-breaking non-derivative $\bar h^{(3)}_{1}$ coupling,
plus the two terms from the field redefinition,
the remaining terms all involve derivatives, either of the pion
or the nucleon.

In fact, some of these other couplings
are necessary 
to renormalize processes involving the coupling of pions and nucleons,
while others lead to momentum dependence.
To make this point evident, let us consider the
three-point Green's function for an incoming (outgoing) nucleon
of momentum $p^\mu$ ($p^{\prime\, \mu }$) 
and a pion of momentum $q^{\, \mu} = p^{\,\mu} - p^{\prime\, \mu }$ 
and isospin $a$.
It can be written as
\begin{equation}
V_a(q, K) = \frac{2i}{F_{\pi}} 
\left[F_1(q, K)  t_a + F_2(q, K) \delta_{a 3} 
+ F_{3}(q,K) \delta_{a3} t_3 \right],
\label{FFform}
\end{equation}
in terms of the functions $F_{1,2,3}$ of $q^{\, \mu}$ and
$K^{\, \mu} = (p^{\, \mu} + p^{ \prime\, \mu })/2$.
We will work up to relative ${\cal O}(Q^2/M_{QCD}^2)$, when
the form factors $F_{1,2,3}(q,K)$ receive contributions
from the $T$-violating pion-nucleon vertex 
\eqref{eq:piNagain} at  tree and one-loop levels,
and from the $T$-violating pion-nucleon vertices \eqref{eq:piN}
at tree level.

\begin{figure}
\centering
\includegraphics[width=2.5cm]{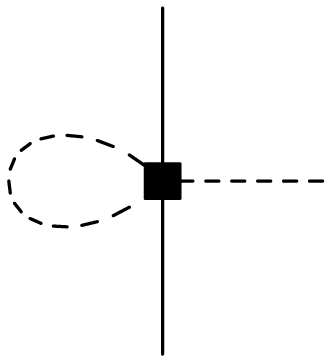} 
\includegraphics[width=2.5cm]{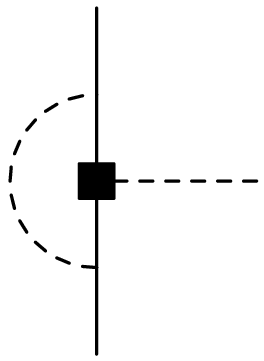} 
\includegraphics[width=2.5cm]{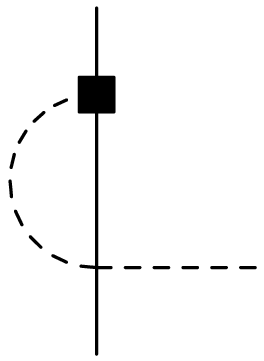} 
\includegraphics[width=2.5cm]{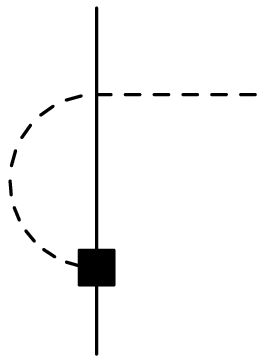} 
\includegraphics[width=2.5cm]{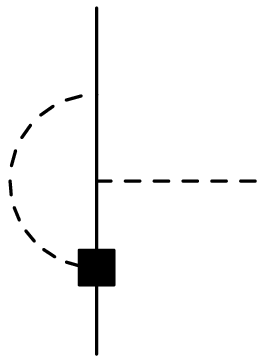} 
\includegraphics[width=2.5cm]{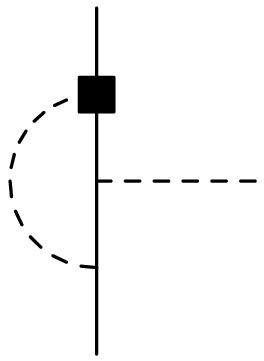} 
\caption{One-loop contributions of 
$\mathcal O(\bar g_0 m_\pi^2/(2\pi F_\pi)^2)$
to the pion-nucleon form factor $F_1(q, K)$ .
A nucleon (pion) is represented by a solid (dashed) line;
the $T$-violating vertex 
\eqref{eq:piNagain}
is indicated by
a square, while other vertices come from Eq. \eqref{eq:QCD.16}.
} 
\label{Fig.6}
\end{figure}

The loops, shown in Fig. \ref{Fig.6},
only contribute to $F_1(q, K)$.
(Note that we do not include wavefunction renormalization here; this can
be easily done if needed.)
The leading $T$-violating interaction \eqref{eq:piNagain} is dressed by
$T$-conserving interactions from the $\Delta=0$ 
Lagrangian \eqref{eq:QCD.16}.
The one-loop diagrams are of course 
divergent; we use dimensional regularization
in $d$ spacetime dimensions, which introduces the renormalized scale $\mu$
and 
\begin{equation}\label{L}
L =  \frac{1}{d-4} 
+ \frac{1}{2} \left( \gamma_E -1 -\ln4\pi\right),
\end{equation}
with $\gamma_E=0.55721\ldots$.
We define the renormalized parameters
\begin{eqnarray}
\bar{\bar h}^{(3)}_2 
& =& \bar h^{(3)}_2 + \frac{\bar g_0}{4} \frac{m^2_{\pi}}{(2\pi F_{\pi})^2} 
\left[  \left(1 + 3 g^2_A\right)
\left( L +\ln \frac{m_{\pi}}{\mu}\right) + 3 g^2_A  \right], 
\label{eq:rencoeff} \\ 
\bar{\bar \eta}_2  
& =& \bar\eta_2 + \frac{ \bar g_0}{(2\pi F_{\pi})^2} 
\left(6 - \frac{g^2_A}{2} \right) 
\left( L + \ln \frac{m_{\pi}}{\mu} - \frac{1}{2} \right),
\label{renorm1} \\
\bar{\bar \eta}_5  & =&  
\bar\eta_5  + \frac{\bar g_0}{(2\pi F_{\pi})^2} 
\left(3 g_A^2 - 4\right) 
\left(L + \ln \frac{m_{\pi}}{\mu} - \frac{1}{2} \right).
\label{renorm2}
\end{eqnarray}

At one loop, the form factors are found to be 
\begin{eqnarray}
 F_1(q, K) & =&  - \bar g_0 
\left[ 1 + \frac{\delta^{(2)} m^2_{\pi}}{m^2_{\pi}}
+ \frac{m^2_{\pi}}{(2\pi F_{\pi})^2} 
f\left(\frac{v\cdot q}{2m_\pi}, \frac{v\cdot K}{m_\pi}\right)\right] 
+ 2 \bar{\bar{h}}^{(3)}_2 
- \frac{1}{2} \left(\bar{\bar{\eta}}_2 
  + \bar \eta_3\right) \left(v \cdot q \right)^2    
\nonumber\\
&&  - \bar{\bar{\eta}}_5 \left(v \cdot K\right)^2
 + \frac{\bar \eta_3}{2}  \vec q^{\;2} + \frac{\bar g_0}{2 m_N^2} \vec K^2
+ i \frac{\bar g_0}{2 m_N^2} \vec S \cdot \left(\vec K \times \vec q\right),
\label{formfactor}\\
F_2(q,K) &= & 2 \bar h^{(3)}_{1}-\rho \Delta m_N
\frac{\delta^{(2)}m_\pi^2}{m_\pi^2},
\label{formfactor2} \\
F_3(q,K) &= & 0, 
\label{formfactor3}
\end{eqnarray}
with 
\begin{eqnarray}\label{ff2}
  f(x, y) &=&
g^2_A
- \sqrt{1- \left(y + x\right)^2} 
\left\{2y + 6 x 
+ \frac{g^2_A }{2x }
  \left[1 -\left(y+x\right)^2\right] \right\} 
\arccos\left(- y - x\right)
\nonumber\\ 
&&- \sqrt{1 - \left(y - x\right)^2} 
\left\{2 y -6 x 
-  \frac{g^2_A }{2x} 
  \left[1 - \left(y - x \right)^2\right]
\right\}
\arccos\left(- y + x\right).
\end{eqnarray}

The result greatly simplifies if we let the nucleons go on shell,
which we write in a short-hand notation in terms of the nucleon isospin 
as
\begin{eqnarray}
v\cdot q &=& \frac{\vec K \cdot \vec q}{m_N} 
+ i\delta m_N \varepsilon^{3 a b} t_b +\ldots, 
\label{onshell1}\\
v\cdot K &=& \frac{1}{2m_N}\left(\vec K^2 + \frac{\vec q^{\, 2}}{4}\right) 
-\Delta m_N - \delta m_N \delta_{a3} t_3 +\ldots
\label{onshell2}
\end{eqnarray}
In this limit, 
$f(v\cdot q/2m_\pi, v\cdot K/m_\pi)$
is subleading, 
\begin{equation} 
f\left(\frac{v\cdot q}{2m_\pi}, \frac{v\cdot K}{m_\pi}\right)
= 0 + \mathcal O\left( \frac{m_\pi^3}{M^3_{QCD}}\right),
\end{equation}
and, at the accuracy to which we are working, it can be neglected. 
The form factors for on-shell nucleons become
\begin{eqnarray}
F_1(q, K) &=&  - \bar g_0  
\left(1+ \frac{\delta^{(2)} m^2_{\pi}}{m^2_{\pi}}\right)
+ 2 \bar{\bar{h}}^{(3)}_2  
+ \frac{\bar \eta_3}{2}   \vec q^{\;2} + \frac{\bar g_0}{2 m_N^2} \vec K^2
+ i \frac{\bar g_0}{2 m_N^2} \vec S  \cdot  \left(\vec K \times \vec q\right),
\label{formfactorshell}\\ 
F_2(q,K) &=& 2 \bar h^{(3)}_{1}-\rho \Delta m_N
\frac{\delta^{(2)}m_\pi^2}{m_\pi^2}  ,
\label{formfactorshell2}\\
F_3(q,K) &=& 0. 
\label{formfactorshell3}
\end{eqnarray}
They are in the form of a local expansion in momenta.
The coupling 
$-2\bar{\bar{h}}^{(3)}_2$ and the tadpole factor 
$\bar g_0  \delta^{(2)} m^2_{\pi}/m^2_{\pi}$
are chiral corrections of
${\cal O}(m_\pi^2/M_{QCD}^2)$ to the leading coupling $\bar g_0$.
The vertex $2\bar{h}^{(3)}_1$ 
and the tadpole correction $-\rho \Delta m_N\delta^{(2)}m_\pi^2/m_\pi^2$ 
are the leading contributions to $F_2$.
As far as the pion-nucleon form factor goes, 
one could as well absorb the tadpole terms in 
$\bar{\bar{h}}^{(3)}_2$ and $\bar{h}^{(3)}_1$.
Note, however, that the tadpole contributions and the
vertices have different tensorial properties
and could, in principle, be separated in other reactions.
In App. \ref{piNFFwtad} we show how the tadpole terms are generated
from tree-level diagrams when we do not do the field redefinition
of Sec. \ref{distad}.
The $\bar \eta_3$ term gives the $F_1$ form-factor radius,
while the remaining two terms in Eq. \eqref{formfactorshell}
are relativistic corrections.
Note that $F_3$ does not receive any contribution up to this order.

The $T$-violating pion-nucleon form factor has recently been studied in 
Ref. \cite{TVpiNNFF}
using a model relativistic Lagrangian 
for the interactions of nucleons, pions, $\rho$, $\omega$ and $\eta$ mesons. 
The $T$-violating sector of the Lagrangian 
in Ref. \cite{TVpiNNFF} contains all the possible 
non-derivative one-pion/two-nucleon interactions ---in particular an 
isoscalar coupling with coupling constant $c_{\pi}$--- and 
interactions of the $\rho$, $\omega$ and $\eta$ mesons with the nucleon. 
Similarly, the $T$-conserving sector includes 
a pseudo-vector pion-nucleon coupling with coupling constant $g_{\pi}$, 
the coupling of the $\rho$ meson to the nucleon and to two pions
with constants $g_{\rho}$ and $g_{\rho \pi}$ respectively, 
and the couplings of the $\eta$ and $\omega$ mesons to the nucleon. 
The model Lagrangian in Ref. \cite{TVpiNNFF} does not include multi-pion terms 
and, therefore, it is not fully consistent with the chiral symmetry of the QCD 
Lagrangian. 

On the other hand, our framework is limited to
momentum transfer $Q\sim m_{\pi}\ll m_\rho$.
It is instructive to make contact between Ref. \cite{TVpiNNFF} 
and the formalism presented here. 
For that, 
we have first of all to integrate out the contributions of the $\rho$, 
$\omega$ and $\eta$ mesons. 
At energies much smaller than the $\eta$ and $\omega$ masses, loops 
containing the $\omega$ and $\eta$ mesons appear as short-distance 
contributions, absorbed in the coupling $\bar g_0$. 
At energies much smaller than $m_{\rho}$, 
the $T$-conserving processes in which 
the nucleon emits a $\rho$ meson that subsequently decays into two pions 
appear like a local seagull
vertex, the Weinberg-Tomozawa term. We can thus establish the relation  
$g_{\rho} g_{\rho \pi}/m^2_{\rho} = -1/F^2_{\pi}$. 
Analogously, the emission of a $\rho$ meson via a $T$-violating interaction, 
followed by its decay into two pions, matches onto a $T$-violating seagull 
interaction with the form of the operator $\zeta_4$ in 
Eq. \eqref{eq:2deriv.1}. 
Loops containing such a vertex are subleading in the power counting. 
Terms cubic in the $T$-violating couplings are similarly of higher orders.  
Expanding the result 
of Ref. \cite{TVpiNNFF} in powers of $m^2_{\pi}/m^2_N$ and 
$m^2_{\pi}/m^2_{\rho}$, we find that
the sum of the last five diagrams in Fig. \ref{Fig.6} reproduces the 
infrared behavior of the fully relativistic calculation,
that is, the factors of $\ln m_{\pi}/\mu$ exactly match in the two 
calculations,
provided that we use 
$c_{\pi} = -\bar g_0/F_{\pi}$
and the Goldberger-Treiman relation $g_{\pi} = 2 m_N/F_{\pi} g_A$.
However, the first diagram of Fig. \ref{Fig.6}
does not have a counterpart in the 
calculation of Ref. \cite{TVpiNNFF}, whose model Lagrangian does not 
include multi-pion terms. These multi-pion terms follow from the 
chiral properties of the $T$-violating operators, 
which are tied to their roots in 
the $\bar \theta$ term in the QCD Lagrangian.  

The framework presented here thus affords a method to carry out
hadronic calculations where the QCD symmetries are included properly.
It also allows a systematic ordering of the infinite number 
of contributions allowed by the symmetries.
The results \eqref{formfactorshell}, \eqref{formfactorshell2},
and \eqref{formfactorshell3} can be used as input, for example, 
in nuclear calculations.
If more accuracy is needed, one can compute the form factor
in higher orders.
For example,
as we saw in Table \ref{Tab:hierarchy}, $F_3$ first appears at 
relative
${\cal O}(\alpha_{\rm{em}}/\pi)$,
which, in the way we count powers of $\alpha_{\rm{em}}/\pi$,
is the next order in the $Q/M_{QCD}$ expansion.
At this order one would have to include photon loops as well.

\section{Conclusion}
\label{Conclusion}

As is well known, $\bar{\theta}$ is unnaturally small.  
Given the current value for the neutron EDM \cite{currentbound}
and the chiral estimate from the non-analytic part \cite{CDVW79}
of chiral loops,
there is a tight upper bound $\bar{\theta}\lesssim 10^{-10}$.
Nevertheless, when a neutron EDM is measured, it could be assigned
to a $\bar{\theta}$ of just the correct size, if no other relevant
information is available. 

In this paper we discussed some of the information that could be used 
to challenge such an interpretation.
The basic idea is that the $\bar{\theta}$ term breaks not only
$P$ and $T$, but also, for two quark flavors,
chiral $SU(2)_{L} \times SU(2)_R$ symmetry.
The way chiral symmetry is explicitly
broken determines the form of hadronic interactions
associated with $\bar{\theta}$.
As a consequence, the $\bar{\theta}$ term should give rise
to a particular pattern of nuclear $T$-violating observables.

Care must be taken because a $P$- and $T$-violating interaction
can lead to a misalignment between 
spontaneous and explicit chiral-symmetry breaking,
in the form of
pion tadpoles. We have found a hadronic field
redefinition that enforces vacuum alignment directly in 
the hadronic theory. In first order in
the quark masses, it is just the hadronic counterpart
of Baluni's rotation at the quark level \cite{Baluni}.
Our hadronic field
redefinition provides a convenient way to extend alignment
to any desired order in an expansion
in $m_\pi/M_{QCD}$.

Starting from a vacuum aligned in first order,
we have constructed the leading interactions from the $\bar{\theta}$ term
in the low-energy EFT involving pions and nucleons, $\chi$PT.
(The extension to delta isobars is straightforward.)
We have observed that pion tadpoles reappear in higher orders,
and can either be eliminated with our field redefinition,
or kept and treated in perturbation theory.
We have also shown that chiral symmetry provides a
relation, via a parameter $\rho\simeq (m_u+m_d) \sin(\bar\theta) /2(m_d-m_u)$, 
between $T$ violation 
and charge-symmetry breaking, which could be used 
to constrain $\bar{\theta}$. 
However, 
this link is lost at higher orders,
and also in the presence of
electromagnetic interactions.

Because the link to $T$-conserving interactions is limited,
more than just the neutron and proton EDMs need to be measured
for testing a $\bar{\theta}$ origin of a positive signal.
One thus is led to look elsewhere.
The implications for the sources of $T$ violation 
in deuteron and $^3$He EDMs, for example, have been 
studied with particular models \cite{nuclear}. 
An important ingredient in these calculations is $T$ violation
in the pion-nucleon interaction, the isospin structure of which is crucial.
Most calculations assume a non-derivative coupling \cite{TVpiN}.
We have shown that,
because of the way chiral symmetry is broken,
different isospin structures first appear at different orders:
$\boldpi \cdot \bar N \boldt N$ at 
${\cal O}(\rho\varepsilon m^2_{\pi}/M_{QCD})$,
$\pi_3 \bar N \boldt N$ at 
${\cal O}(\rho\varepsilon^2 m^4_{\pi}/M_{QCD}^3)$,
and $\pi_3 \bar Nt_3 N$ at 
${\cal O}(\alpha_{\rm em}\rho\varepsilon m^2_{\pi}/\pi M_{QCD})$.
However, already at ${\cal O}(\rho\varepsilon Q^2 m^2_{\pi}/M_{QCD}^3)$
derivative interactions appear,
which endow the $T$-violating pion-nucleon interaction
with a momentum dependence not usually taken into account \cite{nuclear}.
We have calculated the corresponding form factor 
two orders beyond leading, which has a very simple
structure when the nucleons are on-shell.

Although limited to low energies, the 
EFT formulated here includes the approximate symmetries 
of QCD correctly.
The chiral Lagrangian we constructed provides a basis
for a systematic improvement in studies of 
nuclear $T$ violation.

\acknowledgments
We have benefited from discussions with Dani\"el Boer, Jordy de Vries, 
Jim Friar, Claudio Maekawa, Mike Ramsey-Musolf, Rob Timmermans, and, 
long ago, Scott Thomas. 
We thank Sid Coon for comments on the manuscript.
We acknowledge useful correspondence with 
Sergiy Kondratyuk about the pion-nucleon form factor.
UvK thanks the hospitality extended to him
at the Kernfysisch Versneller Instituut of the Rijksuniversiteit Groningen
and at 
the National Institute for Nuclear Theory at the University of Washington,
where portions of this work were carried out. 
This research was supported by the US Department of Energy
under grants DE-FG02-04ER41338 (EM, WH, UvK) and DE-FG02-06ER41449 (EM).

\appendix
\section{Spontaneous Chiral Symmetry Breaking}
\label{appA}

Here we discuss spontaneous chiral symmetry breaking 
for two quark flavors combined in an isospin doublet,
Eq. (\ref{eq:q=ud}), following Refs. \cite{Wbook,vanKolck}. 
The QCD Lagrangian (\ref{eq:QCDt.1})
for massless and chargeless quarks
is invariant under a chiral transformation (\ref{chiraltrans}).
The $(G^A)=(\boldt, \boldx)$ have the commutation relations
\begin{equation}
\begin{split}
 \left[t^i, t^j \right] &= i \varepsilon^{ijk}\, t^k,\\
 \left[x^i, x^j \right] &= i \varepsilon^{ijk}\, t^k, \\
 \left[t^i, x^j \right] &= i \varepsilon^{ijk}\, x^k,
\end{split}
\label{eq:QCD.4}
\end{equation}
and generate the chiral group $SU_{L}(2) \times SU_R(2)$,
which is isomorphic to the group $SO(4)$ of rotations in 
four-dimensional Euclidean space.
Acting on four-dimensional vectors, the generators are written as
\begin{eqnarray}
& \left(\mathcal T^{a}\right)_{b c} =  - i \varepsilon^{abc}, 
& \qquad  \left(\mathcal T^{a}\right)_{b4} = \left(\mathcal T^{a}\right)_{4b} 
  = \left(\mathcal T^{a}\right)_{44} = 0,  \\
& \left(\mathcal X^{a}\right)_{b 4} =  - \left(\mathcal X^{a}\right)_{4 b} 
  = - i \delta_{ab}, 
& \qquad  \left(\mathcal X^{a}\right)_{bc} 
  = \left(\mathcal X^{a}\right)_{44} = 0 .
\label{defgen}
\end{eqnarray}

The chiral symmetry of the QCD Lagrangian is spontaneously broken to its
vector (isospin) subgroup $SU_V(2)$, isomorphic to $SO(3)$: 
$SU_L(2)\times SU_R(2) \rightarrow SU_V(2)$ 
or, equivalently, $SO(4)\to SO(3)$. 
Goldstone's theorem requires that for each broken symmetry a 
massless particle exists, with spin 0 and the same parity and 
internal quantum numbers as the current associated to the broken generators.
Here there are three broken generators, $\boldx$, 
and thus three massless, spin-0 Goldstone bosons with negative parity,
identified with the pions.
The Goldstone bosons live in the coset space 
$SU_L(2)\times SU_R(2)/SU_V(2)\sim SO(4)/SO(3)\sim S^3$,
the ``chiral circle''.
We can parametrize this space with 
stereographic coordinates $\boldzeta(x)=\boldpi(x)/F_\pi$,
where $\boldpi(x)$ is the canonically normalized pion field
and $F_\pi\simeq 186$ MeV (the ``pion decay constant'') is the diameter
of the chiral circle.
The point on the chiral circle labeled by $\boldzeta$ is obtained
by a rotation $R_{\alpha\beta}[\boldzeta]$,
\begin{equation}
\sum_{\gamma=1}^4 R_{\alpha \gamma} R_{\beta \gamma} = \delta_{\alpha \beta},
\label{RR=1}
\end{equation}
from the north pole $({\mathbf 0} \;\, 1)^T$, given by
\begin{equation}
R_{\alpha\beta}[\boldzeta]=
\left(
\begin{array}{cc}
\delta_{ij}-\frac{2}{D}\zeta_{i}\zeta_{j}& \frac{2}{D}\zeta_{i}\\
  -\frac{2}{D}\zeta_{j}& \frac{1}{D}(1-\boldzeta^2)
\end{array}
\right),
\label{RotMatZeta}
\end{equation}
where
\begin{equation}
D=1+\boldzeta^2.
\end{equation}

Under an infinitesimal
isospin transformation, the Goldstone-boson 
field $\boldzeta$ transforms like an isovector 
\begin{equation}
\delta \boldzeta = \boldtheta_V \times \boldzeta,
\label{eq:QCD.13}
\end{equation}
while an
infinitesimal axial transformation is non-linear in the field,
\begin{equation}
\delta \boldzeta = (1 - \boldzeta^2) \boldtheta_A 
                  + 2 \left(\boldtheta_A \cdot \boldzeta\right) \boldzeta .
\label{eq:QCD.14}
\end{equation}
It is convenient to introduce the covariant derivative of the pion field,
\begin{equation}
 D_{\mu} \boldzeta = \frac{\partial_{\mu} \boldzeta}{D},
\label{eq:QCD.12}
\end{equation}
which has simpler transformation properties:
it is an isovector,
\begin{equation}
\delta D_{\mu} \boldzeta = \boldtheta_V \times D_{\mu} \boldzeta,
\label{eq:QCD.13prime}
\end{equation}
that under an axial transformation transforms in the same way, but
with a field-dependent angle $\boldzeta \times \boldtheta_A$,
\begin{equation}
\delta D_{\mu}\boldzeta = 2 (\boldzeta \times \boldtheta_A) 
                           \times D_{\mu} \boldzeta .
\label{eq:QCD.15}
\end{equation}
One can also construct the covariant derivative of this covariant 
derivative,
\begin{equation}\label{eq:pioncov}
\mathcal D_{\nu} D_{\mu} \boldzeta = \partial_{\nu}D_{\mu} \boldzeta 
-  2 \left(\boldzeta \cdot D_{\mu} \boldzeta \right)  D_{\nu} \boldzeta 
+ 2 \left( D_{\nu} \boldzeta \cdot D_{\mu} \boldzeta \right) \boldzeta , 
\end{equation}
and so on.
These covariant objects make it simpler to construct
interactions with the desired transformation properties. 

Being light, pions are important degrees of freedom at low energies.
In addition, the lightest baryons, the proton ($p$) and the neutron ($n$),
are present in the ground state of strong-interacting systems.
We then have to include a field $N(x)=(p \;\, n)^T$ and its interactions 
with pions. 
We choose $N$ to transform non-linearly in the same way as $D_\mu\boldzeta$.
Being an isospin doublet, under an $SU(2)_V$ transformation,
\begin{equation}
\delta N = i \boldt \cdot \boldtheta_V N,
\end{equation}
and under an infinitesimal axial transformation,
\begin{equation}
\delta N  = 2 i \boldt \cdot \left(\boldzeta \times \boldtheta_A\right)N. 
\label{eq:QCD.19}
\end{equation}
It is straightforward to show that the chiral-covariant derivative
of this nucleon field is
\begin{equation}
\mathcal D_{\mu} N = 
\left(\partial_{\mu} +2i \boldt\cdot\boldzeta\times D_\mu \boldzeta \right) N.
\label{eq:QCD.18}
\end{equation}
As before, we can also define higher covariant derivatives. 
Notice that the covariant derivative of $D_{\mu} \boldzeta$ 
in Eq. \eqref{eq:pioncov} is nothing but Eq. \eqref{eq:QCD.18} 
in the adjoint representation, $(t^j)_{ik} = i \varepsilon^{i j k }$.

At the energies we are working at, a nucleon is non-relativistic since
its typical momentum is much smaller than its mass, $Q\sim m_\pi \ll m_N$. 
The nucleon momentum can be written as
\begin{equation}
 p^{\mu} = m_N v^{\mu} + k^{\mu},
\label{parametrization}
\end{equation}
where the nucleon velocity satisfies $v^2 = 1$ and in the nucleon rest frame 
$v^{\mu} = (1, \vec{0})$, and
the residual momentum $k\sim Q$.
Pion-nucleon interactions do not modify the nucleon velocity but only 
the residual momentum. 
In this regime, the nucleon mass is not a dynamical scale and it can 
be eliminated from the theory by defining a velocity-dependent nucleon field
\cite{heavybaryon}
\begin{equation}
N_{v} = \exp \left( i m_N \slashchar v v \cdot x \right) N.
\end{equation}
Derivatives of $N_v$ are proportional to the residual momentum.
The field $N_v$ satisfies
\begin{equation}
 \frac{1+ \slashchar v}{2} N_{v} = N_{v},
\end{equation}
which allows us to reduce the possible Dirac matrices to be used in the 
construction of operators bilinear in the heavy nucleon field to 
$\Gamma = \{ 1, S^{\mu} \}$.
 Here $S^{\mu}$ is the spin operator, satisfying
\begin{equation}
 v \cdot S = 0, \qquad S^2 N_v = - \frac{3}{4} N_v, \qquad 
[S^{\lambda}, S^{\sigma}] 
= i \varepsilon^{\lambda \sigma \alpha \beta} v_{\alpha} S_{\beta}, \qquad 
\{S^{\lambda}, S^{\sigma} \} 
= \frac{1}{2} \left( v^{\lambda} v^{\sigma} - g^{\lambda \sigma}\right).
\end{equation}
In the nucleon rest frame, $S^\mu=(0, \vec{\sigma}/2)$.
In the rest of this paper we drop the label $v$ from the nucleon field.

The same procedure can be followed for other baryons.
Since its mass difference to the nucleon is only a factor 2 larger
than the pion mass, $m_\Delta -m_N \simeq 300$ MeV, the delta isobar is 
the most
important of these resonances. 
For simplicity we neglect the delta in this paper.
The method can be easily generalized for any baryon.

Chiral symmetry strongly constrains the form of the 
interactions among Goldstone bosons and other particles in the theory. 
The most general Lagrangian containing nucleons and pions,
invariant under chiral symmetry, can be constructed 
by including all the operators that are invariant under isospin
and contains 
the covariant derivative of the pion field $D_{\mu}\boldpi$,
the nucleon field $N$, and their covariant derivatives.
Equations \eqref{eq:QCD.16prime} and \eqref{eq:QCD.16}
are the most important examples.

\section{Explicit Chiral Symmetry Breaking}
\label{appB}

Explicit symmetry breaking terms can be included in the effective Lagrangian
by mimicking the breaking in the QCD Lagrangian \cite{Wbook,vanKolck}. 
Consider the generic case in which the symmetry is explicitly broken by 
a linear combination of the components $\mathcal O_A$ of some representation 
$D$ of the group:
\begin{equation}
 \Delta \mathcal L = \sum_A c_A \mathcal O_A
\label{eq:ex.4}
\end{equation}
with 
\begin{equation}
           \mathcal O_A \rightarrow \sum_B D_{AB}[g]\mathcal O_B
\label{eq:ex.5}
\end{equation}
under a transformation $g$ belonging to the symmetry group. 

In a non-linear realization of the symmetry, two statements can be proved.
First, there exists an element of the group $\gamma(\zeta)$ such that
\begin{equation}
\mathcal O_A[\boldzeta, \psi]
=\sum_B D[\gamma(\boldzeta)]_{AB} \mathcal O_B[0,\psi],
\label{eq:ex.61}
\end{equation}
where $\psi$ is a shorthand notation for the possible 
chiral-covariant fields 
in the theory, including nucleons, nucleon covariant derivatives, 
pion covariant derivatives, {\it etc}.
Thus, operators with explicit Goldstone bosons, $\mathcal O[\boldzeta, \psi]$,
can be found if 
their representations and the form 
of the operators without Goldstone bosons, $\mathcal O[0,\psi]$, are known.
Second,
\begin{equation}
\mathcal O_A[0, h \psi]    = \sum_B D[h]_{AB} \mathcal O_B[0,\psi],
\label{eq:ex.62}
\end{equation}
where $h$ belongs to the unbroken subgroup $SO(3)$. 
That is, the operators without Goldstone bosons $\mathcal O[0, \psi]$ 
transform linearly under the unbroken subgroup, according to one 
of the representations of the subgroup that can be found in $D_{AB}$. 

In the simplest example, the $SO(4)$ representation is the defining 
(vector) representation, so  
$D[\gamma(\zeta)]= \gamma(\zeta)$ and $\gamma(\zeta)$ 
has the form of Eq. \eqref{RotMatZeta}.
Thus, for $V=S,P$,
\begin{equation}
\begin{split}
V_4[\boldpi,N]& = \sum_{\alpha = 1}^4 R_{4\alpha}[\boldpi] V_\alpha[0, N] 
 = \frac{1}{D} \left(1-\frac{\boldpi^2}{F_{\pi}^2}\right) V_4[0,N] 
 - \frac{2 \boldpi}{D F_{\pi}} \cdot \boldV [0,N], \\
V_i[\boldpi,N]& = \sum_{\alpha = 1}^4 R_{i\alpha}[\boldpi] V_\alpha[0,N] 
= \frac{2 \pi_i}{D F_{\pi}} V_4[0,N] 
 + \sum_{j = 1}^3\left(\delta_{ij}-\frac{2\pi_i \pi_j}{D F^2_{\pi}}
   \right) V_j[0,N].
\end{split}
\label{eq:ex.7}
\end{equation}
Moreover, $S_4[0,N]$ ($P_4[0,N]$) is isoscalar, parity-even (parity-odd)
and time-reversal-even (time-reversal-odd), 
while 
$\boldS [0,N]$ ($\boldP [0,N]$) is isovector, parity-odd (parity-even)
and time-reversal-odd (time-reversal-even). 

The simplest vector, containing no nucleon fields 
nor pion covariant derivatives, is
\begin{equation}
 S[0, 0] = \left(\begin{array}{c} 
           \mathbf{0}\\ 
           v_0 
           \end{array}
           \right),
 \end{equation}
with $v_0$ a real number determined by the details 
of the dynamics of spontaneous chiral symmetry breaking.
{}From Eq. \eqref{eq:ex.7},
\begin{equation}
 S[\boldzeta, 0] =  \frac{v_0}{D} 
            \left(\begin{array}{c} 
            2\boldzeta\\ 
            1-\boldzeta^2
            \end{array}
            \right).
\label{Spimass}
 \end{equation}
As a second example,
\begin{equation}
 S[0, N] = \left(\begin{array}{c} 
           \mathbf{0}\\ 
            v_1 \bar N N 
           \end{array}
           \right),
 \end{equation}
with $v_1$ another real number, yields 
\begin{equation}
 S[\boldzeta, 0] =  \frac{v_1}{D} 
            \left(\begin{array}{c} 
            2\boldzeta\bar N N\\ 
            \left(1-\boldzeta^2\right)\bar N N
           \end{array}
           \right).
\label{Ssigma}
 \end{equation}

This method can be used to construct the chiral-variant terms in the
effective Lagrangian.
Consider ${\mathcal L}_m$ from Eq. \eqref{eq:QCDt.8f}
when $\varepsilon =0$ and $\bar \theta =0$. 
In this case, the fourth component of Eq. \eqref{Spimass}
generates, apart from a constant,
a pion mass term in the Lagrangian, Eq. (\ref{eq:pimass}),
where 
we introduce
the pion mass $m^2_{\pi} = 4v_0 \bar m/F^2_{\pi}= 
\mathcal O\left( \bar m M_{QCD} \right)$.
Similarly,
Eq. \eqref{Ssigma} gives rise to the so-called sigma term,
Eq. \eqref{sigma}, where we introduce the nucleon mass correction 
$\Delta m_N=v_1 \bar m  = \mathcal O\left( m^2_{\pi}/M_{QCD} \right)$.
Chiral symmetry relates this mass correction to a pion-nucleon 
seagull interaction. 
This procedure can be repeated {\it ad infinitum}.

In analogous fashion, one can construct the operators originating
from the other mass terms of the QCD Lagrangian, Eq. \eqref{eq:QCDt.8f}, 
as we explicitly do in Sec. \ref{thetainteractions}

Finally, we realize the chiral-symmetry-breaking operators due to the 
electromagnetic interaction of the quarks \cite{vanKolck}. 
An obvious class of electromagnetic operators consists of operators that 
contain soft photons. 
These are obtained by minimally coupling the charged pions and the proton to 
the photon, using the covariant derivatives defined in 
Eq. \eqref{eq:minimal.1}, and by constructing the most general gauge-invariant
operators involving $F_{\mu\nu}$.
{}From Eq. (\ref{eq:rot.16}), these 
operators are either chiral invariant 
or transform as the 3-4 component of an antisymmetric chiral tensor. 
For such a tensor,
\begin{equation}
T_{i4}[\boldpi, N] = - \frac{1}{D}
   \left[ \delta_{ik} \left(1-\frac{\boldpi^2}{F_\pi^2}\right)
         + \frac{2 \pi_i \pi_k}{F_\pi^2} \right] T_{4 k}[0,N] 
+ \frac{2}{D F_\pi} 
  \left(\pi_j \delta_{i k} - \delta_{i j} \pi_k \right) T_{j k}[0, N],
\label{eq:rot.18.b}
\end{equation}
where $T_{4 i}[0,N]$ is an isovector 
and $T_{i j}[0,N]$, an antisymmetric tensor.

In the nucleon sector, the simplest objects with two Lorentz-tensor indices
are
\begin{equation}
I^{\mu\nu}[0,N] = c^{(1)}_{s} \bar N i \left[S^{\mu}, S^{\nu}\right] N,
\end{equation}
and 
\begin{equation}
T^{\mu\nu}[0,N] = c^{(1)}_{v} \left(
\begin{array}{cc}
\mathbf{0} & \bar N i\left[S^{\mu}, S^{\nu}\right]t_i N \\
-\bar N i\left[S^{\mu}, S^{\nu}\right]t^i N & 0\\
\end{array}
\right),
\label{eq:b.13}
\end{equation}
which lead to the lowest-order contribution of this type: 
\begin{equation}\label{eq:magnetic.1}
\mathcal L^{(1)}_{f=2, \rm{em}} = c^{(1)}_{s}
\bar N i \left[S^{\mu}, S^{\nu}\right] N e F_{\mu \nu}
+ c^{(1)}_{v}
\bar N \left[t_3 +\frac{2}{F^2_{\pi}D}
\left(\pi_3 \boldpi \cdot \boldt -\boldpi^2t_3 \right) \right] 
i \left[S^{\mu}, S^{\nu}\right] N e F_{\mu \nu},
\end{equation} 
where the coefficients scale as
$c^{(1)}_{s,v}= \mathcal O (1/M_{QCD})$.
The two operators in Eq. \eqref{eq:magnetic.1} are leading contributions to
the isosinglet and  isovector magnetic dipole moments of the nucleon. 
Other such ``direct'' electromagnetic interactions can be derived similarly.

There is, however, another type of electromagnetic contribution.
As discussed in Sec. \ref{EMinteractions}, exchanges of hard photons 
between quarks cannot be resolved in the effective theory and 
generate purely hadronic operators. 
At lowest order in $\alpha_{\rm{em}}$ these operators involve the exchange of 
one hard photon and, as consequence of Eq. (\ref{eq:rot.16}), 
they have the $SO(4)$ transformation properties of 
the tensor product 
$\left(I^{\mu}/6 + T^{\mu}_{34}\right) 
\otimes \left(I_{\mu}/6  + T_{34 \, \mu}\right)$.
The resulting chiral-invariant operators simply represent
${\cal O}(\alpha_{\rm{em}})$ corrections to their strong-interaction
counterparts. The mixed terms transform as antisymmetric tensors,
Eq. (\ref{eq:rot.18.b}).
For the tensor product of two antisymmetric tensors,
\begin{eqnarray}\label{eq:b.14}
T_{i4 j4}[\boldpi, N] & = &  \frac{1}{D^2}
   \left[ \delta_{ik} \left(1-\frac{\boldpi^2}{F_\pi^2}\right)
         + \frac{2 \pi_i \pi_k}{F_\pi^2} \right]  \left[ \delta_{j l} 
           \left(1-\frac{\boldpi^2}{F_\pi^2}\right)
         + \frac{2 \pi_j \pi_l}{F_\pi^2} \right] T_{4 k 4 l}[0,N] 
		 \nonumber \\ 
&& - \frac{2}{ F_{\pi} D^2}
   \left[ \delta_{ik} \left(1-\frac{\boldpi^2}{F_\pi^2}\right)
         + \frac{2 \pi_i \pi_k}{F_\pi^2} \right]   
   \left(\pi_l \delta_{j m} - \delta_{j l} \pi_m \right) T_{4 k l m} [0, N]
		\nonumber\\
& & - \frac{2}{ F_{\pi} D^2}
      \left(\pi_l \delta_{i m} - \delta_{i l} \pi_m \right) \left[ \delta_{jk} 
      \left(1-\frac{\boldpi^2}{F_\pi^2}\right)
         + \frac{2 \pi_j \pi_k}{F_\pi^2} \right] T_{l m 4 k} [0, N]
\nonumber \\ 
& & + \frac{4}{D^2 F^2_\pi} 
  \left(\pi_l \delta_{i k} - \delta_{i l} \pi_k \right)  
  \left(\pi_m \delta_{j n} - \delta_{j m} \pi_n \right) T_{k l m n}[0, N].
\end{eqnarray}

In the 
mesonic sector, the first chiral-breaking operator induced by the 
electromagnetic interaction has the transformation properties of 
$T_{34} \otimes T_{34}$.
The choices 
\begin{equation}
\label{eq:b.15}
T_{4k 4l}[0,0] = v_{0, \rm{em}} \delta_{k l}, \qquad 
T_{kl mn}[0,0] =\frac{v_{0,\rm{em}}'}{4} 
            \left( \delta_{km} \delta_{ln} - \delta_{kn} \delta_{lm}\right),
\end{equation}
with real numbers $v_{0, \rm{em}}$ and $v_{0, \rm{em}}'$,
produce an isospin-breaking correction to the pion mass,
\begin{equation}\label{eq:b.16}
\mathcal L^{(1)}_{\slashchi,\, f=0, \, \rm{em}} = 
- \frac{\delta m^2_{\pi, {\rm em}} }{2D^2} 
\left( \boldpi^2 - \pi_3^2\right),
\end{equation}
where 
$\delta m^2_{\pi, {\rm em}} = 8 (v_{0,\rm{em}} -v_{0,\rm{em}}')/F^2_{\pi} 
= \mathcal O(\alpha_{\rm{em}} M^2_{QCD}/\pi)$ 
is the dominant contribution to the pion mass splitting.
Using $m_{\rho}$ for $M_{QCD}$, this estimate is very close to
the observed value, which corroborates our assignment of
a factor $\alpha_{\rm{em}}/\pi$ for the contribution of 
hard photons. 

In the pion-nucleon sector, the operators with the properties of 
$T_{34} \otimes T_{34}$ have a structure 
very similar to Eq. \eqref{eq:b.16},
\begin{equation}
\mathcal L^{(2)}_{\slashchi,\, f=2, \, \rm{em}} =  
\frac{\beta_{1, \textrm{em}}}{ F^2_{\pi} D^2} 
\left( \boldpi^2 - \pi_3^2\right) \bar N N,
\end{equation}
where $\beta_{1, \textrm{em}}=
\mathcal O(\alpha_{\rm{em}} M_{QCD}/\pi)$.
More interesting operators come from the realization of
the tensor product $T^{\mu}_{34} \otimes I_{\mu}/6$.
The simplest tensor has the structure of Eq. (\ref{eq:b.13}),
just without the commutator $i[S^{\mu}, S^{\nu}]$,
which induces the operator 
\begin{equation}
 \mathcal L^{(2)}_{\slashchi,\, f=2,\, \rm{em}} = 
\delta m_{N, \rm{em}} \left[ \bar N  t_3 N + \frac{2}{F^2_{\pi} D} 
\bar N \left( \pi_3 \boldpi \cdot \boldt - \boldpi^2 t_3\right) N\right],
\label{emnucleonmass}
\end{equation}
where 
$\delta m_{N, \rm{em}}
= \mathcal O(\alpha_{\rm{em}} M_{QCD}/\pi)$ 
is 
the leading electromagnetic contribution to the nucleon mass splitting.
Again, this estimate is within a factor of two of the observed value,
although in this case a quark-mass contribution of similar magnitude
has to be accounted for.

These and other ``indirect'' electromagnetic operators
have been discussed in more detail
in Refs. \cite{vanKolck,isoviolphen,isoviolOPE,isoviolNN,isoviolNNN,CSBd,CSBa}.

\section{
Linear Realization}
\label{appC}

Let us consider a linear realization $\Phi$ of the full group, here $SO(4)$.
Suppose the effective potential 
is
\begin{equation}
 V(\Phi) = V_0 (\Phi) + V_1 (\Phi),
\end{equation}
where $V_0$ is the effective potential generated by the symmetric part of 
the Lagrangian, while $V_1$ is the small correction due to 
explicit symmetry breaking. 
If the explicit breaking is small, the vacuum $\bar{\Phi}$ of the full theory 
will not be far from the vacuum $\bar{\Phi}_0$ calculated 
in the absence of explicit breaking: 
$\bar{\Phi} = \bar{\Phi}_0 + \bar{\Phi}_1$, with $\bar{\Phi}_1$ small.  
{}From the equilibrium condition for the vacuum,
\begin{equation}
\frac{\partial V(\Phi)}{\partial \Phi_\alpha}
\bigg|_{\Phi=\bar{\Phi}_0+\bar{\Phi}_1} 
= 0,
\end{equation}
using the invariance of the effective potential $V_0(\Phi)$, 
it can be shown \cite{Wbook}
that, if $V_1$ and $\bar{\Phi}_1$ are small, the following 
condition holds:
\begin{equation}
\sum_\alpha \left(G^A \bar{\Phi}_0 \right)_\alpha 
\frac{\partial V_1(\Phi)}{\partial \Phi_\alpha}\bigg|_{\Phi =\bar{\Phi}_0} = 0,
\label{eq:ex.1}
\end{equation}
where $G^A$ are the generators of the group, in our case $SO(4)$.
Equation \eqref{eq:ex.1} is called the ``vacuum alignment'' condition. 
If Eq. \eqref{eq:ex.1} does not hold, it means that the real vacuum is 
far from the unperturbed one, and the expansion around the vacuum 
$\bar{\Phi}_0$ is not perturbative.
Let us assume, for example, that the perturbation to the effective potential 
has the form 
\begin{equation}
V_{1}(\Phi) = \sum_\alpha u_\alpha \Phi_\alpha, 
\end{equation}
with $u_\alpha$ given parameters.
The vacuum alignment condition becomes
\begin{equation}
 \sum_\alpha u_\alpha (G^A \bar{\Phi}_0)_\alpha = 0,
\label{align}
\end{equation}
and, being the generators of $SO(4)$ antisymmetric, this condition is 
satisfied if the vectors $\bar{\Phi}_0$ and $u$ are parallel. 
This explains the name ``vacuum alignment''.

As a concrete example of chiral symmetry breaking 
we can consider a toy model ---the linear sigma model--- where
the Lagrangian is
\begin{equation}
\mathcal L_\sigma = \frac{1}{2} \partial_{\mu}\Phi \partial^{\mu}\Phi 
                   - V_0(\Phi)
=\frac{1}{2} \partial_{\mu} \Phi \partial^{\mu} \Phi - \frac{m^2}{2} \Phi^2 
 - \frac{\lambda}{4} \left(\Phi^2\right)^2,
\label{eq:toy.1}
\end{equation}
with two real parameters $m^2$ and $\lambda$.
When $m^2 < 0$,
the minimum of the potential $V_0(\Phi)$ is given by the condition
\begin{equation}
\bar{\Phi}_0^2 = -\frac{m^2}{\lambda} = v^2.
\end{equation}
We pick a vacuum in the fourth direction,
\begin{equation}
 \bar\Phi_0  = v \left(\begin{array}{c}
                       0 \\
                       0\\
                       0\\
                       1
                      \end{array}\right),
\end{equation}
a spontaneous breaking of $SO(4)$ symmetry.

Let us add a small explicit breaking potential, in the form
\begin{equation}
V_1(\Phi) =  g (\Phi_3 + \Phi_4), 
\label{eq:toy.2}
\end{equation}
with $g \ll m^2 v$. The vacuum we chose is not aligned with the 
symmetry-breaking potential
and the situation is analogous to the case we discussed 
in Sec. \ref{need} ---there are two explicit symmetry-breaking terms 
of the same order, one aligned with the vacuum ($\Phi_4$),
the other not ($\Phi_3$).
If we calculate the minimum of the potential $V_0 + V_1$, 
we find that it is no longer degenerate and it is
\begin{equation}
 \bar\Phi  = v \left(\begin{array}{c}
                       0 \\
                       0\\
                      \frac{1}{\sqrt{2}} + \frac{g}{2 m^2 v} \\
                      \frac{1}{\sqrt{2}} + \frac{g}{2 m^2 v} 
                      \end{array}\right) + \mathcal O(g^2).
\end{equation}
We see that even a small perturbation rotates the vacuum dramatically, 
the angle between the true and the old vacuum being approximately $\pi/4$. 

Consider now instead the explicit breaking 
\begin{equation}
V_1(\Phi) =  g \Phi_4  + \frac{g^2}{m^2 v} \Phi_3, 
\label{eq:toy.3}
\end{equation}
still with $g \ll m^2 v$. 
This situation resembles the second case we discussed in the text,
with a non-aligned perturbation much smaller than the aligned one 
---see Secs. \ref{thetainteractions} and \ref{distad}. 
We again can find a minimum,
\begin{equation}
\bar \Phi = v \left(\begin{array}{c}
                       0 \\
                       0\\
                       \frac{g}{m^2 v} \\
                       1 + \frac{g}{2 m^2 v} 
                      \end{array}\right) + \mathcal O(g^2).
\end{equation}
This time, the true vacuum is very close to the one we chose to 
expand the Lagrangian around. 

Once the vacuum is aligned with the dominant perturbation,
let us say, along the fourth direction,
we can perform an explicit field redefinition to exhibit 
the Goldstone modes \cite{Wbook}:
\begin{equation}
 \Phi_n = R_{n 4}(x) \sigma(x),
\label{eq:QCD.7}
\end{equation}
where $R$ is a rotation matrix that belongs to $SO(4)$,
that is, satisfies Eq. (\ref{RR=1}).
In the stereographic representation, we 
parameterize the rotation as in Eq. (\ref{RotMatZeta})
and define 
the fields
\begin{equation}
 \zeta_i = \frac{\Phi_i}{\Phi_4 + \sigma}, \qquad i = 1,2,3.
\label{eq:QCD.9}
\end{equation}
Under an infinitesimal isospin transformation with parameter $\boldtheta_V$,
\begin{eqnarray}
\delta \Phi_i &=& \sum_{jk}\varepsilon_{ijk} \theta_{Vj} \Phi_k,\\
\delta \sigma &=& 0,
\end{eqnarray}
it is easy to see that $\boldzeta$ is an isovector, 
Eq. (\ref{eq:QCD.13}).
Likewise, under an infinitesimal axial transformation $\boldtheta_A$,
\begin{eqnarray}
\delta \Phi_i &=& 2 \theta_{Ai} \Phi_4, \\
\delta \Phi_4 &=& - 2 \sum_{i} \theta_{Ai} \Phi_i,
\end{eqnarray}
the transformation of the Goldstone boson field is non-linear,
Eq. (\ref{eq:QCD.14}).

Defining the covariant derivative (\ref{eq:QCD.12}),
the Lagrangian \eqref{eq:toy.1} can be recast in the form
\begin{equation}
 \mathcal L =  \frac{1}{2} \partial_{\mu} \sigma \partial^{\mu} \sigma 
- \frac{1}{2} m^2 \sigma^2 - \frac{\lambda}{4} \sigma^4  
+\frac{1}{2} \sigma^2 D_{\mu} \boldzeta \cdot D^{\mu} \boldzeta .
\label{eq:QCD.11}
\end{equation}
It is easy to see that the Lagrangian 
is still invariant under $SO(4)$.
The aligned potential \eqref{eq:toy.3}, on the other hand,
when expressed in terms of the Goldstone boson fields, will depend
on $\boldzeta$ explicitly:
\begin{equation}
V_1(\Phi) =  g\sigma \frac{1-\boldzeta^2}{D} 
                    + \frac{g^2}{m^2 v} \sigma \frac{2\boldzeta_3}{D}.
\label{eq:toy.3prime}
\end{equation}

In the vacuum, $\bar{\sigma}=v +g/2m^2 +{\mathcal O}(g^2)$
and $\bar{\zeta}_i =\delta_{i3}g/2m^2 v +{\mathcal O}(g^2)$.
For processes at momenta $Q\ll m$, we can integrate out the 
fluctuations of the field $\sigma$,
obtaining a Lagrangian that for $g=0$
is a function of $D^{\mu} \boldzeta$ only.
For $g\ne 0$, one can recognize in Eq. (\ref{eq:toy.3prime}) the 
fourth and third components
of the vector $S[\boldzeta, 0]$ given by Eq. (\ref{Spimass}),
with coefficients in the ratio $g:g^2/m^2v$, just as
in the original perturbation \eqref{eq:toy.3}.

\section{Resummation of pion tadpoles}\label{appD}

In this appendix we show how to resum the tadpole diagrams generated by the 
Lagrangian \eqref{eq:rot.20}.
The method is general and can in principle be applied to other quantities,
but we illustrate it for the pion two-point Green's function at 
tree level. Some of the contributions from tadpoles in this case were
displayed in Fig. \ref{Fig.2a}.

\begin{figure}[t]
\includegraphics[width=2.5cm]{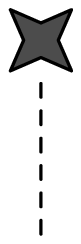} \, \raisebox{1.15cm}{=} \, 
\includegraphics[width=2.5cm]{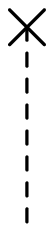} 
\, \raisebox{1.15cm}{+ $\frac{1}{2}$} \,\includegraphics[width=2.5cm]{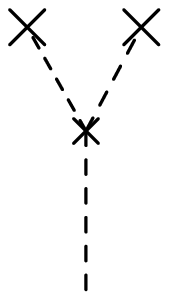} 
\, \raisebox{1.15cm}{+ $\frac{1}{3!}$}\,\includegraphics[width=2.5cm]{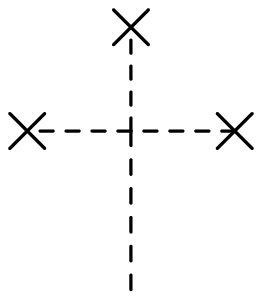} 

\vspace{0.5cm}
\, \raisebox{1.15cm}{$+\frac{1}{4!}$}\, \includegraphics[width=2.5cm]{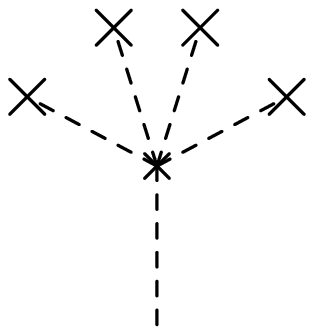} 
\, \raisebox{1.15cm}{$+\frac{1}{5!}$}\, \includegraphics[width=2.5cm]{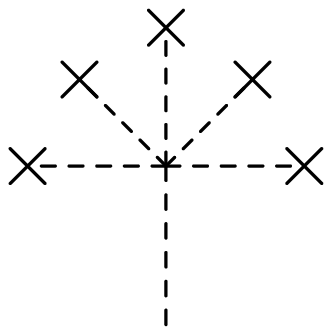} 
\, \raisebox{1.15cm}{$+\frac{1}{2}$}\, \includegraphics[width=2.5cm]{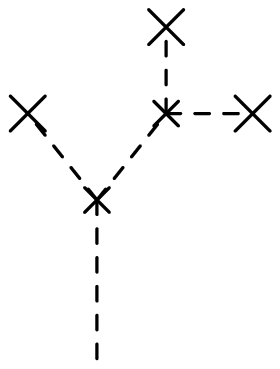} 

\vspace{0.5cm}
\, \raisebox{1.15cm}{$+\frac{1}{3!}$}\, \includegraphics[width=2.5cm]{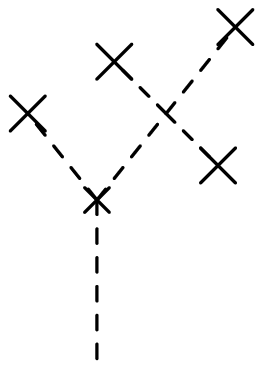} 
\, \raisebox{1.15cm}{$+\frac{1}{2!} \frac{1}{2}$} \,
\includegraphics[width=2.5cm]{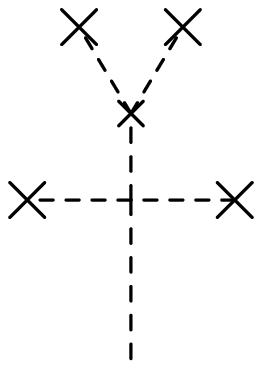}  
\, \raisebox{1.15cm}{+ $\frac{1}{3!} \frac{1}{2}$} \,
\includegraphics[width=2.5cm]{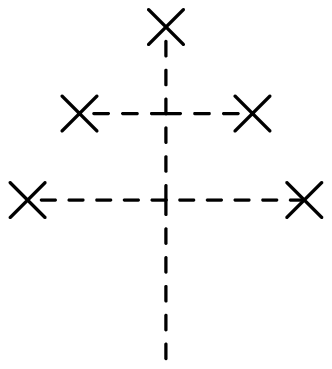} 
\, \raisebox{1.15cm}{+ $\ldots $}
\caption{Contributions to the pion one-point function $iT$
at tree level up to order $g^5$. 
Vertices are from the Lagrangian \eqref{eq:rot.20}.
For each diagram, the symmetry factor is explicitly indicated.} \label{Fig:5}
\end{figure}

\begin{figure}[t]
\includegraphics[width=2.5cm]{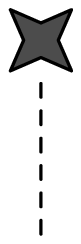} \, \raisebox{1.15cm}{=} \,
\includegraphics[width=2.5cm]{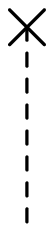} \, 
\raisebox{1.15cm}{+ $\frac{1}{2}$} \,\includegraphics[width=2.5cm]{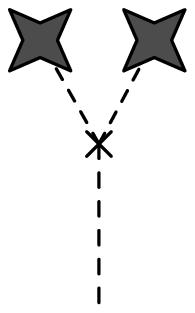} \, 
\raisebox{1.15cm}{+ $\frac{1}{3!}$}\,\includegraphics[width=2.5cm]{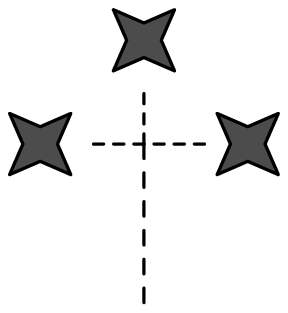} \, 

\vspace{0.5cm}
\raisebox{1.15cm}{+ $\frac{1}{4!}$}\,\includegraphics[width=2.5cm]{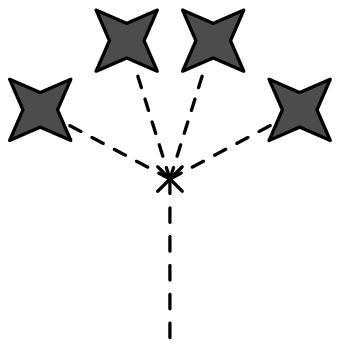} \, 
\raisebox{1.15cm}{+ $\frac{1}{5!}$}\,\includegraphics[width=2.5cm]{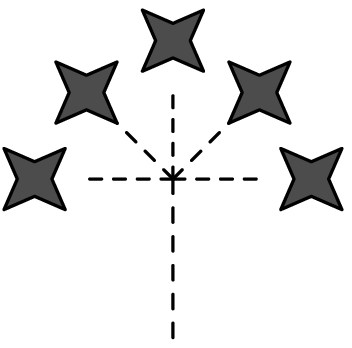} \,
\raisebox{1.15cm}{+ $\dots$}
\caption{Diagrammatic equation for the one-pion Green's function $i T$,
Eq. \eqref{eq:tad.9}.}\label{Fig:7}
\end{figure}

We start by defining the full one-pion Green's function 
\beq
iT_a= \frac{i}{2} g \tilde m^2_{\pi} F_{\pi} \tilde T \delta_{a3},
\label{Ta}
\eeq
where $a$ is the isospin index of the pion.
In lowest order in the chiral expansion we need to worry only
about tree-level diagrams.
In Fig. \ref{Fig:5} we display the corresponding 
diagrams contributing to $i T$ to order $g^5$ 
and, for convenience, we explicitly show the symmetry factor due to 
exchange of equivalent tadpoles.
In the diagrams in Fig. \ref{Fig:5}, the external neutral pion is connected 
to one of the basic vertices of the Lagrangian \eqref{eq:rot.20}, with three, 
four, $\ldots$, $n$-branches. 
Each branch then develops into a tadpole tree and ends up with the 
disappearance of an arbitrary number of $\pi_3$s into the vacuum. 
The diagrams in Fig. \ref{Fig:5} 
can be rearranged as in Fig. \ref{Fig:7}, 
and we can write the diagrammatic equation
\begin{equation}
\label{eq:tad.9}
\frac{i}{2} \tilde m^2_{\pi} g F_{\pi}\tilde T 
= \frac{i}{2} \tilde m^2_{\pi} g F_{\pi} 
+ \sum_{n =2}^{\infty} \frac{1}{n!} 
\left(-\frac{i}{\tilde m^2_{\pi}}\right)^n
\left( \frac{i}{2} \tilde m^2_{\pi} F_{\pi} g \tilde T\right)^n 
V^{\prime}_n (\tilde m^2_{\pi}), 
\end{equation}
where the factor $V^{\prime}_n (\tilde m^2_{\pi})$ can be obtained from 
the Lagrangian \eqref{eq:rot.20} and it is
\begin{equation}
 \label{eq:tad.10}
\begin{split}
V_{2m}^{\prime} (\tilde m^2_{\pi}) & 
= \frac{i}{2} (-1)^m (2m+1)! \, \frac{g \tilde m^2_{\pi}}{F_{\pi}^{2m-1}}, \\
V_{2m+1}^{\prime} (\tilde m^2_{\pi}) & 
=\frac{i}{2} (-1)^{m+1} (2m+2)!\, \frac{\tilde m^2_{\pi}}{F^{2m}_{\pi}}.
\end{split}
\end{equation}
Equation \eqref{eq:tad.9} can be rewritten as
\begin{equation}
 \label{eq:tad.11}
\sum_{m=0}^{\infty} (-1)^m (2m+1) \left(\frac{g^2}{4}\right)^m \tilde T^{2m} -
 \sum_{m=0}^{\infty} (-1)^m (m+1) \left(\frac{g^2}{4}\right)^m \tilde T^{2m+1} 
= 0.
 \end{equation}
The two series can be summed and we obtain
\begin{equation}\label{eq:tad.12}
 \frac{1}{\left(1 + \frac{g^2}{4} \tilde T^2\right)^2} 
\left[1 - \tilde T - \frac{g^2}{4}\tilde T^2 \right] = 0,
\end{equation}
which
admits two solutions,
\begin{equation}
\label{eq:tad.13}
\tilde T  = -\frac{2}{g^2} \left(1\pm \sqrt{1+g^2}\right).
\end{equation}
The non-analytic dependence on $g$ is a direct consequence of the
non-perturbative character of the problem.

\begin{figure}[t]
\includegraphics[width=2.5cm]{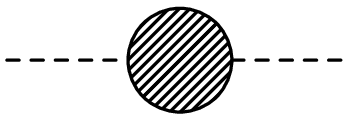}  \raisebox{1.15cm}{=}
\includegraphics[width=2.5cm]{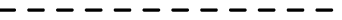}  \raisebox{1.15cm}{+}
\includegraphics[width=2.5cm]{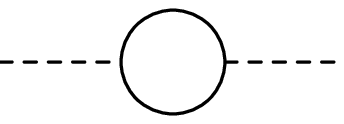}  \raisebox{1.15cm}{+}
\includegraphics[width=3.75cm]{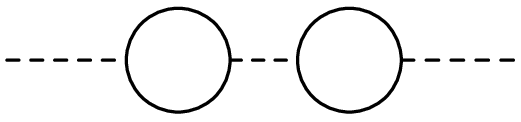}  \raisebox{1.15cm}{ + \ldots}
\caption{The full pion propagator $G$, denoted by a shaded blob,
as an iteration of the sum of 1PI diagrams $-i \Sigma$,
denoted by an empty circle.} \label{Fig:4}
\end{figure}

With the one-pion Green's function, Eqs. \eqref{Ta} and \eqref{eq:tad.13},
we can calculate the effect of pion tadpoles on quantities more directly
related to experiment.
Let us consider the two-point Green's function for a pion
of four-momentum $p$ and
isospin index $a$, $G_a(p^2, \tilde m^2_{\pi})$.
The diagrams that contribute to the two-point function up to order $g^4$ 
are shown in Fig. \ref{Fig.2a}.
Let us call, with abuse of language, ``one-particle irreducible'' (1PI), 
those diagrams that cannot be disconnected by cutting an internal line 
in which non-vanishing $p$ flows, 
and denote 
the sum of all the 1PI diagrams by 
$-i \Sigma_a(p^2, \tilde m^2_{\pi})$.
The full propagator can be expressed as the geometric sum of 1PI diagrams 
---see Fig.\ref{Fig:4}--- and
\begin{equation}\label{eq:tad.1}
\begin{split}
G_a(p^2, \tilde m^2_{\pi}) \delta_{ab} & 
= \frac{i\delta_{ab}}{p^2 - \tilde m^2_{\pi}+ i \varepsilon}
\left(1 
+\frac{\Sigma_a(p^2,\tilde m^2_{\pi})}{p^2 - \tilde m^2_{\pi}+ i \varepsilon} 
+\frac{\Sigma_a^2(p^2,\tilde m^2_{\pi})}
      {\left(p^2-\tilde m^2_{\pi}+i\varepsilon\right)^2} 
+ \ldots \right)  \\ 
& = \frac{i\delta_{ab}}
         {p^2-\tilde m^2_{\pi}-\Sigma_a(p^2,\tilde m^2_{\pi}) + i\varepsilon}.
\end{split}
\end{equation}

\begin{figure}[t]
\includegraphics[width=2.5cm]{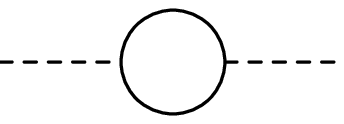} \, \raisebox{1.15cm}{=} \,
\includegraphics[width=2.5cm]{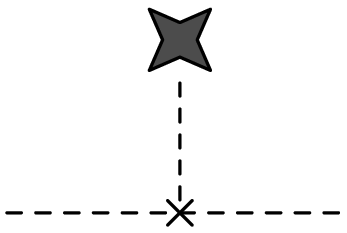} \, 
\raisebox{1.15cm}{$+\frac{1}{2}$} \,\includegraphics[width=2.5cm]{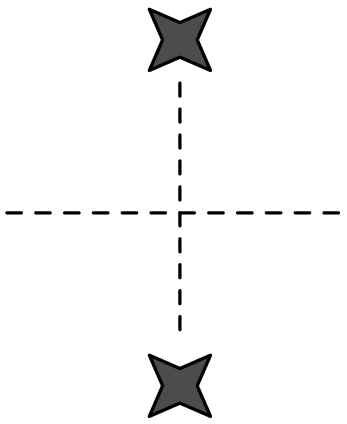}  

\vspace{0.5cm}
\, \raisebox{1.15cm}{$+\frac{1}{3!}$} \,\includegraphics[width=2.5cm]{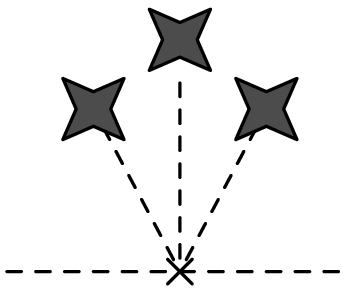}
\, \raisebox{1.15cm}{$+\frac{1}{4!}$}\,\includegraphics[width=2.5cm]{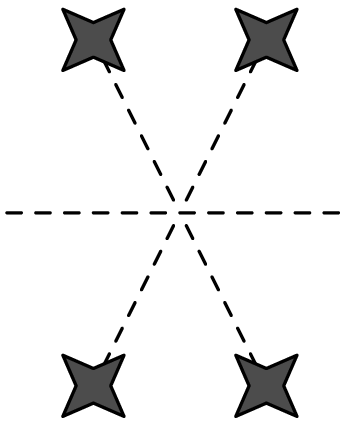}
\, \raisebox{1.15cm}{$+\ldots$} 
\caption{Diagrammatic equation for $- i \Sigma$ in terms of
the one-point Green's function $i T$, Eq. \eqref{eq:tad.2}.} \label{Fig:6}
\end{figure}

At tree level, contributions to the sum of 1PI diagrams 
$-i\Sigma_a(p^2,\tilde m^2_{\pi})$ have the following structure: 
the vertex connected to the two lines in which 
$p$ flows has a certain number of branches;
from each branch a tadpole tree sprouts, which ends with the disappearance of 
an arbitrary number of $\pi_3$s into the vacuum.  
Diagrammatically, the sum of 1PI diagrams can be expressed in terms of the  
pion one-point Green's function $iT$, as shown in Fig. \ref{Fig:6}:
\begin{equation}
\label{eq:tad.2}
-i\Sigma_a(p^2, \tilde m^2_{\pi}) = 
\sum_{n = 1}^{\infty} \frac{1}{n !}\left(\frac{- i}{\tilde m^2_{\pi}}\right)^n 
\left(\frac{i}{2}\tilde m^2_{\pi} F_{\pi} g \tilde T\right)^n 
V_{a;n}(p^2, \tilde m^2_{\pi}),
\end{equation}
where $V_n$ is a factor coming from the Feynman rules for the $(n+2)$-pion 
vertex. It can be derived from the Lagrangians 
\eqref{eq:QCD.16prime} and \eqref{eq:rot.20}, and it is
\begin{equation}
\label{eq:tad.3}
\begin{split}
V_{a;2m+1} (p^2, \tilde m^2_{\pi}) 
&= i (-1)^{m+1} \frac{g\tilde m^2_{\pi}}{F^{2m +1}_{\pi}} (m+1)(2m+1)!
 \left\{1+2(m+1)\delta_{a3}\right\}, \\
V_{a;2m+2} (p^2, \tilde m^2_{\pi}) 
&= i (-1)^{m+1}  \frac{g}{F^{2m +2}_{\pi}} (m+2)(2m+2)!
 \left\{p^2 -\left[1+2(m+1)\delta_{a3}\right]\tilde m^2_{\pi}\right\} .
\end{split}
\end{equation}
We can write
\begin{equation}\label{eq:tad.5}
\Sigma_a(p^2, \tilde m^2_{\pi})  = 
  \frac{g^2}{2} 
  \sum_{m =0}^{\infty} \left(-\frac{g^2}{4}\right)^m 
  \left[(p^2- \tilde m^2_{\pi})\frac{m+2}{2} \tilde T^{2m+2} 
       +\tilde m^2_{\pi}(m+1) \tilde T^{2m+1} \right]
  +\delta\Sigma_a(p^2, \tilde m^2_{\pi}),
\end{equation}
where
\begin{equation}\label{eq:tad.6}
\delta\Sigma_a(p^2, \tilde m^2_{\pi})  = 
\tilde m^2_{\pi} g^2 \sum_{m = 0}^{\infty} \left(-\frac{g^2}{4}\right)^m (m+1) 
\left[(m+1)  \tilde T^{2m+1} - \frac{m+2}{2} \tilde T^{2m+2} \right]
\delta_{a3}.
\end{equation}
Summing the series and using Eq. \eqref{eq:tad.12},
$\delta\Sigma_a(p^2, \tilde m^2_{\pi})$ vanishes and
the sum of 1PI diagrams 
becomes the same for charged and neutral pions:
\begin{equation}\label{eq:tad.7}
\Sigma_a(p^2, \tilde m^2_{\pi}) = 
\frac{g^2}{2\left(1+\frac{g^2}{4}\tilde T^2\right)^2} 
\left[\left( p^2 -\tilde m^2_{\pi} \right)\tilde T^2 
      \left( 1+ \frac{g^2}{8} \tilde T^2 \right) 
    +\tilde m^2_{\pi}\tilde T 
\right].
\end{equation}

The inverse of the propagator \eqref{eq:tad.1} is now
\begin{equation}
 \label{eq:tad.14}
G_a^{-1}(p^2, \tilde m^2_{\pi}) = 
\left(1+\frac{g^2}{4}\tilde T^2\right)^{-2}
\left[p^2 -\tilde m^2_{\pi} \left(1 + \frac{g^2}{2} \tilde T\right)\right],
\end{equation}
and it vanishes at the physical pion mass
\begin{equation}\label{eq:tad.15}
m^2_{\pi}= \tilde m^2_{\pi} \left(1 + \frac{g^2}{2} \tilde T\right)
= \pm \, \tilde m^2_{\pi} \sqrt{1 + g^2},
\end{equation}
where we used
the solutions \eqref{eq:tad.13}. 
Inserting the values of $\tilde m^2_{\pi}$ and $g$, 
Eqs. \eqref{eq:mass} and \eqref{eq:g},
\begin{equation}\label{eq:tad.17}
 m^2_{\pi} = \pm \frac{4 v_0}{F^2_{\pi}} \, \bar m \, r(\bar\theta).
\end{equation}
Equation \eqref{eq:tad.17} shows that, as it should be, the physical pion mass 
is independent of the arbitrary angle $\alpha$,
and its value is equal to the one we would get by working directly with the 
aligned Lagrangian \eqref{eq:QCDt.8f}.
Presumably the same can be shown 
for other observable quantities using this method.

\section{Higher-order interactions from the quark mass}
\label{appE}

We construct in this appendix operators that contribute to the 
chiral-breaking pion-nucleon Lagrangian 
$\mathcal L^{(5)}_{\slashchi, f = 2}$ 
and that do not contain covariant derivatives of the pion or of the nucleon 
field. 
Because they are strongly suppressed by $(m_{\pi}/M_{QCD})^4$ 
with respect to the 
leading pion-nucleon chiral-breaking vertices,
these operators are not relevant for 
any phenomenological application.
Nonetheless they are of some formal interest because, as we shall see, this is 
the first order in the purely hadronic sector of the chiral Lagrangian where 
the relation  \eqref{eq:lagr.2} between $T$-violating and isospin-breaking 
operators breaks down.
This is also the lowest order where a purely hadronic 
$T$-violating vertex $\pi_3 \bar N t_3 N$ appears.

The operators we consider here are obtained from the tensor products 
$P_a \otimes P_b \otimes S_4$, $P_{a} \otimes P_b \otimes P_c$,  
and $P_a \otimes S_4 \otimes S_4$, and we write the non-derivative
part of the Lagrangian as
\begin{equation}
\mathcal L^{(5)}_{\slashchi^3, f=2} =
\mathcal L^{(5)}_{\slashchi^3, f=2, PPS}
+\mathcal L^{(5)}_{\slashchi^3, f=2, PPP}
+\mathcal L^{(5)}_{\slashchi^3, f=2, PSS}.
\end{equation}

The tensor product $P_a \otimes P_b \otimes S_4$ can be decomposed into two 
$SO(4)$ vectors and a three-index tensor, symmetric in the first two indices:
$P_a \otimes P_b \otimes S_4 = \delta_{ab} V_4 + \left(\delta_{a4}\delta_{bd} 
+ \delta_{b4} \delta_{ad}\right) W_d + S_{ab,4}$.
The vectors $V$ and $W$ have the same properties under $P$ and $T$ as the 
vector $S$ introduced in Eq. \eqref{Sdef}.
In the realization of the third and fourth components of the tensor product, 
the third and fourth components of $W$ appear, but only the fourth component 
of $V$ does. 
In the $f=2$ sector, $P_a \otimes P_b \otimes S_4$ generates the operators
\begin{eqnarray}\label{eq:lagrsuppr.6}
\mathcal L^{(5)}_{\slashchi^3, f = 2, PPS} & =& 
\left[\rho^2 c^{(5)}_{1} +  \left(1+\rho^2\right) c^{(5)}_{2}   \right]
\left(1- \frac{2\boldpi^2}{F^2_{\pi}D}\right) \bar N N  
+ \frac{2 \rho c^{(5)}_{1 }}{F_{\pi} D} \pi_3 \bar N N 
\nonumber \\ 
&& + c^{(5)}_{3}\left(1-\frac{2\boldpi^2}{F^2_{\pi}D}\right)
\left[ \frac{4\pi_3^2}{F_{\pi}^2 D^2} 
+ \rho^2\left(1-\frac{4\boldpi^2}{F^2_{\pi}D^2}\right)
+ \frac{4\rho \pi_3}{F_{\pi} D^2} \left(1-\frac{\boldpi^2}{F^2_{\pi}}\right) 
\right] \bar N N.
\nonumber\\
\end{eqnarray}
Here the $c^{(5)}_{3}$ term has a similar structure to the $c^{(3)}_{1}$ term
in Eq. \eqref{eq:lagrsuppr.2}.
The $c^{(5)}_{2}$ term is nothing but a correction to the nucleon
sigma term, Eq. \eqref{sigma}.
More interestingly,
Eq. \eqref{eq:lagrsuppr.6} shows that even at the hadronic level the 
relation \eqref{eq:lagr.2} ceases to be valid at higher orders in 
the expansion in $m_{\pi}/M_{QCD}$. 
Indeed, it is not possible to disentangle the individual coefficients 
$c_1^{(5)}$ and $c^{(5)}_2$ by measuring a $T$-conserving observable, 
and, therefore, it is not possible to constrain the coefficient of the 
$T$-violating operator in Eq. \eqref{eq:lagrsuppr.6} with the properties 
of an $S_3$.

The tensor product $P_a \otimes P_{b} \otimes P_{c}$ yields symmetry-breaking 
terms that transform as components either
of a four-vector with the same properties 
as the vector $P$ defined in Eq. \eqref{Pdef}, 
or of a completely symmetric tensor. 
In the $f=2$ sector, the corresponding operators are
\begin{eqnarray}\label{eq:lagrsuppr.7}
 \mathcal L^{(5)}_{\slashchi^3, f = 2, PPP} & = & 
c^{(5)}_{4} \bar N \left[ t_3 - \frac{2\pi_3}{F^2_{\pi}D} \boldpi \cdot \boldt
 -\frac{ 2 \rho }{F_{\pi} D}  \boldpi \cdot \boldt\right] N
+ c^{(5)}_{5} \left(\frac{2\pi_3}{F_{\pi} D}\right)^2 \bar N 
  \left(t_3 - \frac{2\pi_3}{F_{\pi}D}\boldpi \cdot \boldt\right)N
\nonumber \\  
&& +  c^{(5)}_{5} \rho \frac{4 \pi_3}{F_{\pi} D} 
\bar N  \left[ 
t_3- \frac{2}{F^2_{\pi}D}
\left(
2\pi_3\left(1-\frac{3\boldpi^2}{2F^2_{\pi}}\right)\boldpi\cdot\boldt 
+ \boldpi^2  t_3 \right) \right]N
\nonumber \\ 
&& + 
c^{(5)}_5 \rho^2 \left(1-\frac{2\boldpi^2}{F^2_{\pi}D}\right)
\bar N \left[t_3 - \frac{10\pi_3 }{F^2_{\pi}D^2} 
\left(1+\frac{\boldpi^2}{5F^2_{\pi}}\right)
\boldpi \cdot \boldt\right]  N 
\nonumber \\ 
&&  -\frac{2c^{(5)}_5 }{F_{\pi}}
\rho^3 \left(1-\frac{4\boldpi^2}{F^2_{\pi}D^2}\right) 
\bar N \boldpi \cdot \boldt N.
\end{eqnarray}
The $c^{(5)}_{4}$ term
realizes the $SO(4)$ vector in the tensor product, and thus has a form
identical to Eq. \eqref{eq:lagr.3};
it is simply a correction to $\delta m_N$.
The other four operators correspond to the 3-3-3, 3-3-4, 3-4-4 and 4-4-4 
components of the symmetric tensor.
In them, a link between $T$ violation and isospin breaking survives.
The operator with coefficient $c^{(5)}_{5} \rho$ is the first purely hadronic 
contribution to the $T$-violating vertex $\pi_3 \bar N t_3 N$.

The representation of tensor products $P_a \otimes S_4 \otimes S_4$ contains 
two $SO(4)$ vectors with the same properties as $P$ and a three-index tensor. 
When we select the fourth component of the tensor product, the fourth component
of both vectors appears, while, if $a = 3$, we find only the third component 
of one of the two vectors.  This implies that, in the $f=2$ sector, 
the Lagrangian is
\begin{eqnarray}\label{eq:lagrsuppr.8}
\mathcal L^{(5)}_{\slashchi^3, f = 2, PSS} & =& 
c^{(5)}_7 \bar N\left(t_3-\frac{2\pi_3}{F^2_{\pi}D}\boldpi\cdot\boldt\right) N 
-\frac{2\rho}{F_{\pi}D}\left( c^{(5)}_{6}  + c^{(5)}_{7}\right) 
\bar N \boldpi \cdot \boldt N  
\nonumber \\ 
&& +c^{(5)}_{8} \left(1- \frac{4\boldpi^2}{F^2_{\pi}D^2}\right)
\bar N \left[ t_3 - \frac{2\pi_3}{F^2_{\pi}D} \boldpi \cdot \boldt  
- \frac{ 2\rho}{F_{\pi}D} \boldpi \cdot \boldt \right] N 
\nonumber \\ 
&& + \frac{2c^{(5)}_{9}}{F_{\pi} D^2} 
\left(1- \frac{2\boldpi^2}{F^2_{\pi}D}\right) 
\left[\frac{2\pi_3}{F_{\pi}}-\rho\left(1-\frac{\boldpi^2}{F^2_{\pi}}\right)
\right]   
\bar N  \boldpi \cdot \boldt  N. 
\end{eqnarray}
Here some links between $T$-violating interactions and isospin breaking 
survive. 
The $c^{(5)}_7$ term is identical in form to the $c^{(5)}_{4}$ term
in Eq. \eqref{eq:lagrsuppr.7}, so it also provides a 
correction to $\delta m_N$ in Eq. \eqref{eq:lagr.3}.
The $c^{(5)}_8$ term has a similar form.
The $c^{(5)}_9$ term links a $T$-violating interaction 
to an isospin-breaking seagull.
However, we see that 
the term with
coefficient $c^{(5)}_6 \rho$ does not have any 
$T$-conserving 
partner, and cannot be determined
from a $T$-conserving measurement.

The coefficients in Eqs. \eqref{eq:lagrsuppr.6}, \eqref{eq:lagrsuppr.7},
and \eqref{eq:lagrsuppr.8} 
scale as
\begin{equation}
 c^{(5)}_{1-3} = 
\mathcal O\left(\frac{\varepsilon^2 m^6_{\pi}}{r^4(\bar\theta)M^5_{QCD}} 
\right), 
\quad
 c^{(5)}_{4,5} = 
\mathcal O\left(\frac{\varepsilon^3 m^6_{\pi}}{r^6(\bar\theta)M^5_{QCD}} 
\right),
\quad
 c^{(5)}_{6-9} = 
\mathcal O\left(\frac{\varepsilon m^6_{\pi}}{r^2(\theta)M^5_{QCD}} 
\right) .
\end{equation}

\section{Lorentz-Invariance Constraints}
\label{rel}

In this appendix we derive the relations \eqref{eq:rpi} 
and \eqref{eq:rpiEDM}, which stem from Lorentz invariance.
In the heavy-baryon formalism, Lorentz invariance is implemented
order by order in a $Q/m_N$ expansion that goes hand-in-hand with
the $Q/M_{QCD}$ expansion of $\chi$PT. 
It relates the coefficients of operators at different orders.
There are many ways to derive such relations.
One method, intrinsic to the formalism and dubbed reparametrization invariance,
is to demand invariance under small
changes of the velocity $v^\mu$ in Eq. \eqref{parametrization}
\cite{reparam}. 
Going beyond $1/m_N$ corrections is complicated but can be done \cite{jordy2}.
Another method is to implement a Foldy-Wouthuysen transformation  
\cite{bw-nonrel}.
A third, more popular method \cite{Bernard:1992qa,Fettes:2000gb}
is to start from a relativistic 
Lagrangian and perform an integration over antinucleon
fields in the path integral.
Here we follow a variant of the latter,
where we match the non-relativistic Green's functions to their
relativistic counterparts
\footnote{Most of the results in this appendix were obtained independently
by J. de Vries using the method of Ref. \cite{jordy2}.
}.

The $T$-conserving dynamics of a relativistic nucleon are described by the
 Lagrangian
\begin{equation}\label{eq:RelLagr.0}
\mathcal L = \bar N \left(i \slashchar{\mathcal D} - m_N 
+ \frac{2 g_A}{F_{\pi}} \gamma^5  \boldt \cdot \slashchar{D} \boldpi \right) N 
 + \ldots,
\end{equation}
where ``$\ldots$'' denotes higher-dimension operators, with more nucleon or 
pion covariant derivatives and more
powers of chiral-symmetry breaking parameters.
The $T$-violating relativistic Lagrangian in the strong-interaction sector
with operators containing up to two derivatives with respect to the leading 
$T$-violating coupling $\bar g_0$ is  
\begin{eqnarray}
\mathcal L_{\slashT} & = & 
- \frac{2\bar g_0}{F_{\pi}D} \bar N \boldpi \cdot \boldt N 
- \frac{\bar h^{(2)}_1}{F_{\pi}^2 D} \boldpi \cdot D_{\mu} \boldpi 
  \bar N \gamma^{\mu} \gamma_5 N
+ \frac{\bar\eta_3}{F_{\pi}} \frac{1}{D} 
  \left(1 -\frac{\boldpi^2}{F_{\pi}^2} \right) \mathcal D_{\mu} D^{\mu} \boldpi
  \cdot \bar N \boldt N 
\nonumber \\ 
&& 
+ \frac{\bar\eta_4}{2F^2_{\pi}} \frac{1}{D} 
  \left(1 -\frac{\boldpi^2}{F_{\pi}^2} \right) 
  (D_{\mu} \boldpi \times D_{\nu} \boldpi) \cdot 
   \bar N \sigma^{\mu \nu} \gamma^5 \boldt N 
-  \frac{\bar\eta_9}{F^3_{\pi} D} \boldpi \cdot 
   \left(D_{\mu} \boldpi \times D_{\nu} \boldpi \right)  \, 
   \bar N i \sigma^{\mu \nu} N 
\nonumber \\ 
&&
-  \frac{2 \bar\eta_{10}}{F^3_{\pi} D} \boldpi \cdot D_{\mu} \boldpi  
   \, D^{\mu} \boldpi \cdot \bar N \boldt N 
-  \frac{2 \bar\eta_{12}}{F^3_{\pi} D} D_{\mu} \boldpi \cdot D^{\mu} \boldpi  
   \, \boldpi  \cdot \bar N \boldt N,
\label{eq:RelLagr.1}
\end{eqnarray}
where, with abuse of notation, we denote the 
relativistic coupling constants by the same symbols used in the
text for the non-relativistic constants.  

We find the 
$f=2$ $T$-violating heavy-baryon Lagrangian by equating 
(matching) the 
relativistic two-nucleon $n$-pion Green's functions, 
computed with the Lagrangians in Eqs. \eqref{eq:RelLagr.0} and
\eqref{eq:RelLagr.1}, 
to the non-relativistic Green's functions, obtained with the $T$-violating 
Lagrangians
\eqref{eq:lagr.3}, \eqref{TviolPiN2}, and \eqref{eq:2deriv.1} and 
the $T$-conserving chiral Lagrangians \eqref{eq:QCD.16} and 
\begin{eqnarray}
\mathcal L^{(1+2)}_{\chi, f = 2} & =& 
- \frac{1}{2 m_N}\bar N \mathcal D_{\perp}^2 N 
+ \frac{c^{(1)}}{F_{\pi} } \,(i v \cdot D \boldpi )\cdot 
\bar N \boldt S \cdot \left(\mathcal D - \mathcal D^{\dagger}\right)_{ \perp} N
\nonumber \\
&& 
+ \frac{c^{(2)}_1}{ F_{\pi} } D_{\mu} \boldpi \cdot 
\bar N \boldt S^{\mu} \left(\mathcal D_{\perp} - \mathcal D^{\dagger}_{\perp}
\right)^2 N
- \frac{c_2^{(2)}}{F_{\pi}} D_{\mu} \boldpi \cdot 
\bar N \boldt \left(\mathcal D^{\mu} - \mathcal D^{\mu \dagger}\right)_{\perp} 
S \cdot \left(\mathcal D- \mathcal D^{\dagger}\right)_{\perp}N 
\nonumber \\ 
& & 
+ \ldots
\label{strong}
\end{eqnarray}
The ``$\ldots$'' in Eq. \eqref{strong} denote multi-pion operators, 
which are not needed in the matching procedure below. 
Chiral-symmetry breaking operators induced by the quark mass $\bar m$ and 
by the quark mass difference $\bar m \varepsilon$ should be included in the 
relativistic and in the heavy-baryon Lagrangian, Eqs. \eqref{eq:RelLagr.0} 
and \eqref{strong}. However, it turns out that these terms do not affect the 
matching of the $T$-violating one-pion and two-pion Green's functions at 
the order we consider.

We set the external nucleon on shell, and expand the relativistic Green's 
function in powers of $1/m_N$, retaining terms up to order $1/m^2_N$.
We do the matching in the nucleon rest frame, $v = (1, \vec{0})$, 
where the spin operator is $S^{\mu} = (0, \vec{\sigma}/2)$.
In the relativistic part of the matching, the incoming and outgoing nucleons 
are represented by the Dirac spinors $u(\vec p)$ and 
$\bar u(\vec p^{\, \prime})$, whose explicit expressions are
\begin{equation}
u(\vec p) = \sqrt{\frac{E + m_N}{2 E}} \left(
\begin{array}{c}
\xi \\
\frac{\vec p \cdot \vec{\sigma}}{E + m_N} \xi
\end{array}
\right), \qquad
\bar u(\vec p^{\, \prime}) = \sqrt{\frac{E^{\prime} + m_N}{2 E^{\prime}}} 
\left(\xi^{\dagger}, 
- \xi^{\dagger} \frac{\vec p^{\, \prime} \cdot \vec{\sigma}}{E^{\prime} + m_N} 
\right),
\end{equation} 
where  $\xi$ is a two-component spinor, normalized to one, and the 
nucleon energy is $E = \sqrt{m_N^2 + \vec p^{\, 2}}$.
In the heavy-baryon part of the matching, the nucleons are represented by the 
spinor $\xi$. 

\begin{figure}[tb]
\includegraphics[width=2.5cm]{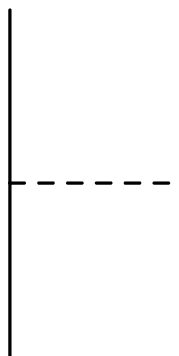} 
\hspace{0.5cm}\raisebox{1.13cm}{$\Longrightarrow$}
\includegraphics[width=2.5cm]{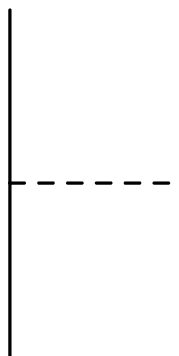} 
\includegraphics[width=2.5cm]{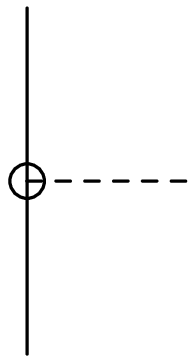}
\includegraphics[width=2.5cm]{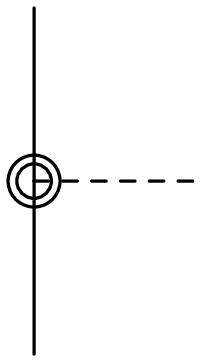}
\caption{Matching of the one-pion $T$-conserving Green's function. 
The l.h.s. represents the relativistic Lagrangian
\eqref{eq:RelLagr.0}. 
On the r.h.s., the unmarked vertex denotes the interaction in the leading-order
Lagrangian 
$\mathcal L^{(0)}_{\chi, f =2}$ \eqref{eq:QCD.16},  
while the vertices with one and two circles denote,
respectively, once- and twice-suppressed interactions in the 
Lagrangian $\mathcal L^{(1,2)}_{\chi, f =2}$ \eqref{strong}.
}
\label{Fig:Match.1a}
\end{figure}

\begin{figure}[tb]
\includegraphics[width=2.5cm]{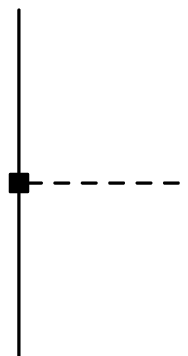} 
\hspace{0.5cm}\raisebox{1.13cm}{$\Longrightarrow$}
\includegraphics[width=2.5cm]{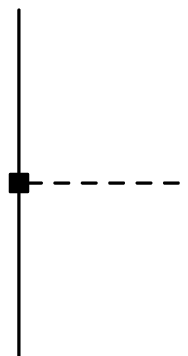}
\includegraphics[width=2.5cm]{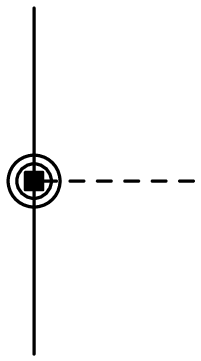}
\caption{Matching of the one-pion $T$-violating  Green's function. 
The l.h.s represents the relativistic Lagrangian
\eqref{eq:RelLagr.1}. 
On the r.h.s., the square denotes the $T$-violating 
vertex in the leading $T$-violating Lagrangian 
$\mathcal L^{(1)}_{\slashchi, f =2}$ \eqref{eq:lagr.3}, 
while the square with two circles the vertices in the power-suppressed 
Lagrangian $\mathcal L^{(3)}_{\slashchi^1, \, f =2}$ \eqref{eq:2deriv.1}. }
\label{Fig:Match.1b}
\end{figure}

The Feynman diagrams for the matching of the one-pion $T$-conserving 
and $T$-violating Green's function 
are depicted in Figs. \ref{Fig:Match.1a} and \ref{Fig:Match.1b}. 
On the relativistic side, the interactions are given by the 
Lagrangians \eqref{eq:RelLagr.0} and \eqref{eq:RelLagr.1}.
In Fig. \ref{Fig:Match.1a}, on the heavy-baryon side the diagram with an 
unmarked vertex denotes the leading pion-nucleon interaction in 
Eq. \eqref{eq:QCD.16},
while diagrams with one and two circles denote contributions suppressed by one 
or two powers of $Q/M_{QCD}$ in Eq. \eqref{strong}.
Similarly, on the heavy-baryon side of Fig. \ref{Fig:Match.1b} the diagrams 
with zero and two circles 
denote contributions from, respectively, the leading $T$-violating Lagrangian 
$\mathcal L^{(1)}_{\slashchi, f =2}$ \eqref{eq:lagr.3} 
and the subleading $T$-violating Lagrangian 
$\mathcal L^{(3)}_{\slashchi^1, \, f =2}$ \eqref{eq:2deriv.1}. 
Equating the relativistic and non-relativistic Green's functions we find
\begin{equation}
c_1^{(1)} = \frac{2 g_A}{m_N},
\qquad c^{(2)}_1 
= c_2^{(2)}
= \frac{g_A}{2 m^2_N},
\label{rel0}
\end{equation}
and, for the $T$-violating Green's function,
\begin{equation}
\rho\, \delta m_N 
= \bar g_0, 
\qquad \rho\, \zeta_{1}
= \rho\, \zeta_{6}
= \frac{\bar g_0}{2 m^2_N}.
\label{rel1}
\end{equation}

\begin{figure}[tb]
\includegraphics[width=2.5cm]{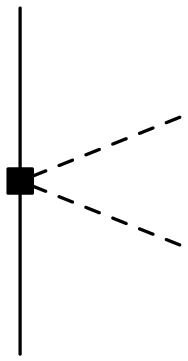}
\includegraphics[width=2.5cm]{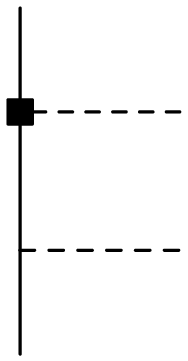}
\includegraphics[width=2.5cm]{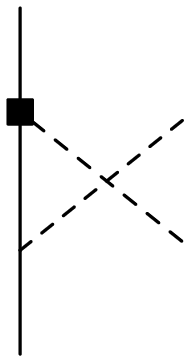}
\includegraphics[width=2.5cm]{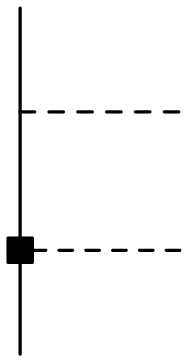}
\includegraphics[width=2.5cm]{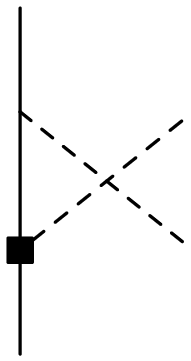}
\vspace{0.5cm}

\includegraphics[width=2.5cm]{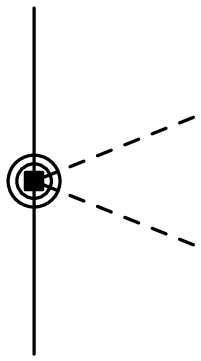}
\includegraphics[width=2.5cm]{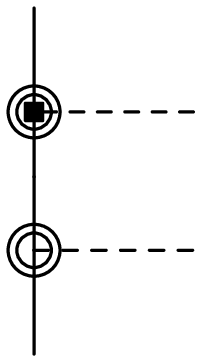}
\includegraphics[width=2.5cm]{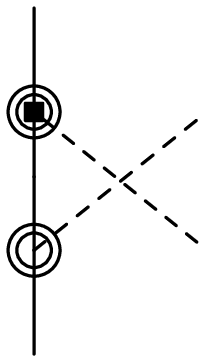}
\includegraphics[width=2.5cm]{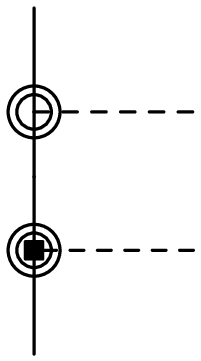}
\includegraphics[width=2.5cm]{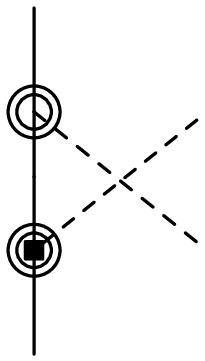}

\includegraphics[width=2.5cm]{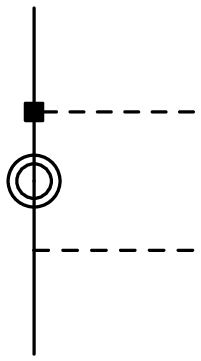}
\includegraphics[width=2.5cm]{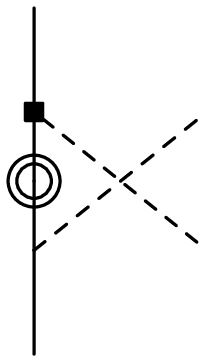}
\includegraphics[width=2.5cm]{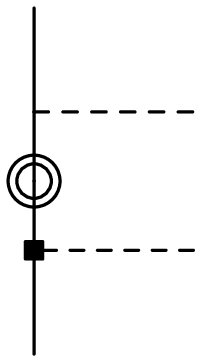}
\includegraphics[width=2.5cm]{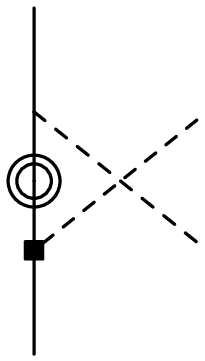}
\caption{Matching of the $T$-violating two-pion Green's function. 
In the top row, the nucleon is relativistic.
In the bottom row, 
the nucleon is described by the heavy-baryon Lagrangian. 
The double circle indicates that in each heavy-baryon diagram 
we consider corrections to the heavy-baryon propagator and vertices 
with up to two powers of $Q/M_{QCD}$ 
with respect to the leading $T$-conserving and $T$-violating diagrams.}
\label{Fig:Match.2}
\end{figure}

The Feynman diagrams for the matching of the two-pion $T$-violating Green's 
function 
are shown in Fig. \ref{Fig:Match.2}.
The first row shows the relativistic diagrams. 
As before, the $T$-violating vertices from Eq. \eqref{eq:RelLagr.1}
are denoted by squares:
in the first diagram, the $T$-violating coupling is either $\bar h^{(2)}_1$ 
or $\bar \eta_4$, while the last four diagrams are proportional to 
$\bar g_0$ or $\bar\eta_3$.
The $T$-conserving vertices come from the Lagrangian \eqref{eq:RelLagr.0}
and are proportional to the axial coupling $g_A$. 
The second and third rows contain the diagrams evaluated in the heavy-baryon 
theory. 
The double circle indicates that we consider vertices and corrections to 
the heavy-baryon propagator in the $T$-conserving 
and $T$-violating chiral Lagrangians with up to two powers of $Q/M_{QCD}$ 
with respect to $\mathcal L^{(0)}_{\chi,\, f = 2}$ \eqref{eq:QCD.16}  
and $\mathcal L^{(1)}_{\slashchi,\, f =2}$ \eqref{eq:lagr.3}.
Equating the two-pion Green's functions we find
 \begin{equation}
\rho\, \beta_1 = \bar h^{(2)}_1, 
\qquad  
\rho\, \zeta_8 = \frac{g_A \bar g_0}{m^2_N} - \frac{\bar h^{(2)}_1}{m_N}.
\label{rel2}
\end{equation}

The relations for the subleading $T$-conserving operators in Eq. \eqref{rel0} 
and for the isospin-breaking coefficients in Eqs. \eqref{rel1} and \eqref{rel2}
reproduce those in Refs.
\cite{Bernard:1992qa,Fettes:2000gb},
obtained 
by integrating the antinucleon field out of a relativistic Lagrangian, 
once a field redefinition is used to eliminate the time derivatives acting 
on the nucleon field from the power-suppressed Lagrangians. We refer to 
\cite{jordy2} for more details.

Equations \eqref{rel1} and \eqref{rel2} lead to Eq. \eqref{eq:rpi}.
Equation \eqref{eq:RelLagr.1} and the matching above imply that 
the coefficients 
$\zeta_3$, $\zeta_4$, 
$\zeta_9$, $\zeta_{10}$, and $\zeta_{12}$ 
are new arbitrary low-energy constants, 
not linked to the couplings appearing in the $\Delta = 1$ and $\Delta =2$ 
$T$-violating Lagrangians. 
The operators proportional to $\zeta_{11}$ and $\zeta_{13}$ do not appear in 
the relativistic Lagrangian, so their coefficient could be linked to 
$\delta m_N$ or $\beta_1$. In order to find the exact relation, 
we should match three-pion Green's functions. 
We refrain from doing this here because
these three-pion operators play no role in any foreseeable
phenomenological application.

The relations \eqref{eq:rpiEDM} can be obtained with the same method, 
by equating the relativistic and non-relativistic three-point Green's 
functions with two nucleon and one photon fields.

\section{Some $T$-Conserving Electromagnetic Terms}
\label{TconsEM}

In Sec. \ref{EMinteractions}
we constructed the $T$-violating interactions
stemming from the $\bar \theta$ term, a $P_4$, and their $T$-conserving 
partners from the associated quark mass splitting, a $P_3$. 
At the same orders as the terms considered there, there exist
$T$-conserving interactions that have no $T$-violating
partners, which come from the chiral-breaking average nucleon mass,
an $S_4$.
In this appendix we display these terms
and see how
such unpaired interactions impair our ability to extract
information about the $T$-violating operators from
$T$-conserving quantities.
Implications are discussed in Sec. \ref{discon}.
The same pattern is repeated at higher orders.

\noindent
{\bf Pion sector.}
In addition to the $P_a \otimes T_{34} $
terms discussed in Sec. \ref{pionalpha},
at the same order
the other structures $S_4$
and $S_4 \otimes T_{34} \otimes T_{34}$ give
\begin{equation}\label{eq:lagrsuppr.9prime}
\mathcal L^{(3)}_{\slashchi, f = 0, \rm{em}} = 
- \frac{1}{2D} 
\left\{\delta^{(3)}_{1, \textrm{em}} m^2_{\pi} \; \boldpi^2
+\frac{\delta^{(3)}_{2, \rm{em}} m^2_{\pi}}{F^2_{\pi} D^2}  
\left(1-\frac{\boldpi^2}{F^2_{\pi}}\right) \left(\boldpi^2 - \pi_3^2\right)
\right\}.
\end{equation}
These are corrections to the pion mass and pion mass difference,
and associated interactions. The coefficients
are of order 
\begin{equation}
\delta^{(3)}_{1, 2, \textrm{em}} m^2_{\pi} 
= \mathcal O\left(\frac{\alpha_{\textrm{em}}}{\pi} m^2_{\pi}\right),
\end{equation}
and therefore much smaller than the leading pion-mass splitting 
\cite{vanKolck,isoviolphen}
$\delta m^2_{\pi, \textrm{em}} =\mathcal O(\alpha_{\textrm{em}}M_{QCD}^2/\pi)$
(see App. \ref{appB}).

\noindent
{\bf Pion-nucleon sector.}
The other structure that we have at the
same order as the leading $T$-violating pion-nucleon interactions,
Sec. \ref{piNalpha},
is $S_4 \otimes \, e A_{\mu} \left(I^{\mu}/6 + T^{\mu}_{34}\right) 
\otimes \, e A_{\nu} \left(I^{\nu}/6 + T^{\nu}_{34}\right) $.
It would exist in the absence of any $P$ vector in the QCD Lagrangian,
and it gives rise to no $T$ violation:
\begin{eqnarray}\label{eq:lagrsuppr.11}
\mathcal L^{(4)}_{\slashchi,\, f = 2,\, \textrm{em}} & = &
c^{(4)}_{27,\, \rm{em}} 
\left(1- \frac{ 2 \boldpi^2}{F^2_{\pi}D} \right) \bar N N 
 + c^{(4)}_{28, \, \rm{em}} 
\bar N \left( t_3 - \frac{2 \pi_3}{F^2_{\pi} D } \boldt \cdot \boldpi \right) N
\nonumber \\ 
&& + \frac{4 c^{(4)}_{29,\, \rm{em}}}{F^2_{\pi} D^2}   
\bar N \left(\boldpi^2 t_{3} - \pi_3 \boldpi \cdot \boldt \right) N 
+ \frac{4 c^{(4)}_{30, \, \rm{em}}}{F^2_{\pi} D^3}
\left(1-\frac{\boldpi^2}{F^2_{\pi}}\right) \left(\boldpi^2 - \pi_3^2\right)
\bar N N.
\end{eqnarray}
The $c^{(4)}_{27, \rm{em}}$
operator 
corresponds to $S_4$ and to the fourth component of the vector in 
$S_4 \otimes T_{34} \otimes T_{34}$.
The $c^{(4)}_{28, \rm{em}}$
operator, with the properties of $P_3$, is generated by the 
vector in $S_4 \otimes T_{34}$. 
The 
$c^{(4)}_{29, \rm{em}}$ and $c^{(4)}_{30, \rm{em}}$
operators realize a three-index and a five-index tensor 
in $S_4 \otimes T_{34}$ and $S_{4} \otimes T_{34} \otimes T_{34}$,
respectively.
The coefficients scale as
\begin{equation}
c^{(4)}_{27-30, \rm{em}} = \mathcal O\left(
\frac{\alpha_{\rm{em}}}{\pi}\frac{m^2_{\pi}}{M_{QCD}}\right).
\end{equation}
The $T$-conserving, isospin-breaking operators with coefficients 
$c^{(4)}_{1, \rm{em}}$ and $c^{(4)}_{28, \rm{em}}$ in 
Eqs. \eqref{eq:lagrsuppr.10} and \eqref{eq:lagrsuppr.11} have exactly the same 
structure and the same transformation properties under the chiral group,
which are also present in Eq. \eqref{eq:lagr.3}.
They are corrections to the nucleon mass splitting and
CSB pion-nucleon interactions, and cannot be separated experimentally from
$\delta m_N$ or each other. 
This is not enough to constrain the ${\cal O}(\alpha_{\rm{em}}/\pi)$ 
correction 
to  the $T$-violating pion-nucleon coupling $\bar N \boldpi \cdot \boldt N$, 
which is proportional only to $c^{(4)}_{1, \rm{em}}$. 
The same argument can be repeated for the operators $c^{(4)}_{3, \rm{em}}$ 
and $c^{(4)}_{27, \rm{em}}$ {\it vis \`a vis} Eq. \eqref{sigma},
so that also the ${\cal O}(\alpha_{\rm{em}} M^2_{QCD}/\pi m^2_{\pi})$ 
correction 
to the coupling $\pi_3 \bar N N$ cannot be constrained by 
$T$-conserving observables. 

{\bf Photon-nucleon sector.}
At $\Delta =3$ the important $T$-violating interactions appear
in Sec. \ref{photonnucleon},
which contribute to the nucleon EDM at short distances.
Further $T$-conserving interactions at the same order are 
\begin{eqnarray}\label{eq:lagr.6}
\mathcal L^{(3)}_{\slashchi, f =2, \rm{em}}  &=&   
- \frac{2 c^{(3)}_{6, \rm{em}} }{F_{\pi} D} \bar{N} \boldpi \cdot \boldt  
\left(S^{\mu} v^{\nu} - S^{\nu} v^{\mu}\right) N\, e F_{\mu \nu} 
+ c^{(3)}_{7, \rm{em}} \left(1-\frac{2\boldpi^2}{F^2_{\pi}}\right)  
\bar{N}   i \left[S^{\mu} , S^{\nu} \right] N\, e F_{\mu \nu}
\nonumber\\
&&+ \frac{2 c^{(3)}_{8, \rm{em}} \pi_3}{F_{\pi} D} 
\bar N \left( S^{\mu}v^{\nu} - S^{\nu} v^{\mu} \right) N\, e F_{\mu \nu}   
+c^{(3)}_{9, \rm{em}}   
\bar N \left( t_{3} - \frac{2 \pi_3}{F^2_{\pi} D}  \boldpi \cdot \boldt \right)
i \left[ S^{\mu},  S^{\nu} \right] N\, e F_{\mu \nu}    
\nonumber\\ 
&& +c^{(3)}_{10, \rm{em}}
\left(1- \frac{2\boldpi^2}{F^2_{\pi}D}\right) 
\bar N \left[ \left( 1-\frac{2\boldpi^2}{F^2_{\pi}D}\right) t_3 
+ \frac{2\pi_3 }{F^2_{\pi}D} \boldpi \cdot \boldt \right] 
i \left[ S^{\mu},  S^{\nu} \right] N\, e F_{\mu \nu}.
\end{eqnarray}
Here the first two operators realize
the fourth component of the vector $S$;
the others come from the tensor product $T_{34} \otimes S_4$,
$c^{(3)}_{8, \rm{em}}$  and $c^{(3)}_{9, \rm{em}}$ transforming as the third 
component of a vector with the same properties as $P$, 
$c^{(3)}_{10, \rm{em}}$ representing a three-index antisymmetric tensor.
These operators 
are 
$P$- and $T$-conserving contributions to the nucleon magnetic dipole moment
and to 
pion photoproduction. 
The coefficients 
scale as
\begin{equation}
c^{(3)}_{6-10, \rm{em}} =
\mathcal O\left( \frac{m^2_{\pi}}{ M^3_{QCD}}\right). 
\end{equation}
Once again, 
electromagnetic interactions and the chiral symmetry breaking due to the 
quark masses conspire to destroy the relation \eqref{eq:lagr.2} between 
$T$-violating couplings and isospin-breaking interactions already in the 
leading-order realization of operators with $f=2$ and explicit photons. 
This emerges in Eqs. \eqref{eq:lagr.5} and \eqref{eq:lagr.6}: 
in order to constrain the short-distance contributions to the nucleon EDM 
it would be necessary to extract the coefficients 
$c^{(3)}_{1,\, \rm{em}}$, $c^{(3)}_{3,\, \rm{em}}$ 
and $c^{(3)}_{5,\, \rm{em}}$ from $T$-conserving observables. 
However, it is impossible to disentangle the coefficients 
$c^{(3)}_{1,\rm{em}}$ and $c^{(3)}_{8, \rm{em}}$ 
or $c^{(3)}_{3, \rm{em}}$ and $c^{(3)}_{6, \rm{em}}$ 
in the measurement of any such observable because the 
$T$-conserving operators they multiply have exactly the same structure 
and the same chiral properties.

\section{Pion-Nucleon Form Factor with Tadpoles}
\label{piNFFwtad}

\begin{figure}
\centering
\includegraphics[width=2.5cm]{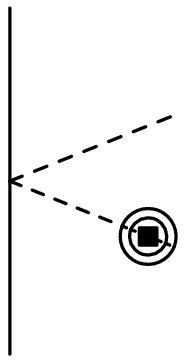} 
\includegraphics[width=2.5cm]{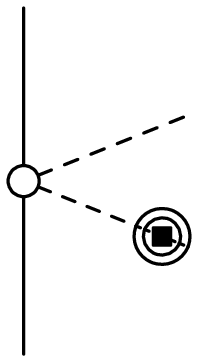} 
\includegraphics[width=2.5cm]{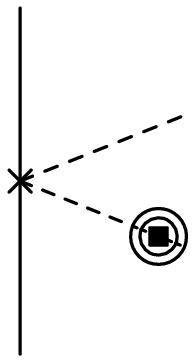} 
\caption{Tadpole contributions 
to the pion-nucleon form factors $F_i(q, K)$, $i=1,2,3$. 
The $T$-violating vertex from Eq. \eqref{eq:lagrtad.1}
is indicated by a twice-circled square. 
The unmarked vertex is the Weinberg-Tomozawa term in Eq. \eqref{eq:QCD.16}.
The circle denotes both the 
nucleon sigma term from Eq. \eqref{sigma} 
and a recoil correction to the Weinberg-Tomozawa term from Eq. \eqref{strong}, 
while the cross represents the isospin-breaking operator in 
Eq. \eqref{eq:lagr.3}.
} 
\label{Fig.6a}
\end{figure}

Here we show how one gets the results of Sec. \ref{FF}
for the 
pion-nucleon form factor without rotating away tadpoles.
We use the same Lagrangian \eqref{eq:piN},
but with $\delta^{(2)}m_\pi^2\to 0$. Instead, we have to include explicitly
the tadpole in Eq. \eqref{eq:lagrtad.1}.
It generates 
tadpole trees, shown in Fig. \ref{Fig.6a}, which
contribute to all three form factors.
The $T$-violating tadpole \eqref{eq:lagrtad.1}
connects to the outgoing pion via seagulls
from the nucleon covariant derivative in Eq. \eqref{eq:QCD.16}
(the so-called Weinberg-Tomozawa term),
from a recoil correction to it found in Eq. \eqref{strong}, 
from the nucleon sigma term \eqref{sigma},
and from the isospin-breaking operator in Eq. \eqref{eq:lagr.3}.

In this case we get also an additional term in the form factor,
\begin{equation}
V_a(q, K) = \frac{2i}{F_{\pi}}
\left[F_1(q, K)  t_a + F_2(q, K) \delta_{a 3} 
+ F_{3}(q,K) \delta_{a3} t_3 \right] 
+ V_{a, \textrm{tad}},
\label{FFformwtad}
\end{equation}
where 
\begin{eqnarray}
 F_1(q, K) & =&  - \bar g_0 
\left[ 1 + \frac{\delta^{(2)}m^2_{\pi}}{2 m^2_{\pi}} 
+ \frac{m_\pi^2}{(2\pi F_\pi)^2}
f\left(\frac{v \cdot q}{2m_\pi}, \frac{v \cdot K}{m_\pi}\right)\right] 
+ 2 \bar{\bar{h}}^{(3)}_2 
- \frac{1}{2} \left(\bar{\bar{\eta}}_2 
+ \bar \eta_3\right) \left(v \cdot q \right)^2    
\nonumber\\
&& - \bar{\bar{\eta}}_5 \left(v \cdot K\right)^2
 + \frac{\bar \eta_3}{2}  \vec q^{\;2} + \frac{\bar g_0}{2 m_N^2} \vec K^2 
+ i \frac{\bar g_0}{2 m_N^2} \vec S\cdot \left(\vec K \times \vec q\right),
\label{formfactorwtad}\\
F_2(q,K) &= & 2 \bar h^{(3)}_{1}  
- \rho\Delta m_N \frac{\delta^{(2)} m^2_{\pi}}{m^2_{\pi}} ,
\label{formfactor2wtad} \\
F_3(q,K) &= & -  \bar g_0 \frac{\delta^{(2)} m^2_{\pi}}{2 m^2_{\pi}}, 
\label{formfactor3wtad}
\end{eqnarray}
and
\begin{equation}
V_{a, \rm{tad}}(q, K) =  
\varepsilon^{3 a b} t_b \frac{\rho }{F_{\pi}} 
\frac{\delta^{(2)} m^2_{\pi}}{m^2_{\pi}}
\left[v\cdot q - \frac{\vec K \cdot \vec q}{m_N}\right].
\label{strange}
\end{equation}

These relations are slightly different than in the case
of the field redefinitions, Eqs. \eqref{formfactor},
\eqref{formfactor2}, and \eqref{formfactor3}.
This is not surprising because in general
a field redefinition changes quantities off-shell.
When the nucleons are on-shell, Eqs. \eqref{onshell1} and \eqref{onshell2} 
hold.
As in the main text, the function $f(v\cdot q/2m_\pi, v\cdot K/m_\pi)$
becomes higher order.
More care has to be taken, however, with Eq. \eqref{strange}, which gives
\begin{equation}
V_{a, \textrm{tad}} (q,K)= 
- \frac{i \bar g_0}{F_{\pi}} 
\frac{\delta^{(2)} m_\pi}{m^2_{\pi}} (t_a - \delta_{a 3} t_3),
\end{equation}
so that the on-shell form factors become
exactly Eqs. \eqref{formfactorshell},
\eqref{formfactorshell2}, and \eqref{formfactorshell3}.

\end{document}